\documentclass[aps,prx,reprint,superscriptaddress,twocolumn]{revtex4-2}
\usepackage{graphicx}
\usepackage{color}
\usepackage[dvipsnames]{xcolor}
\definecolor{DarkGreen}{rgb}{0.0, 0.5, 0.0}

\usepackage[T1]{fontenc}
\usepackage{textcomp}
\usepackage{bm}
\usepackage{amsmath,amssymb,amsthm}
\usepackage{comment}
\usepackage{grffile}
\usepackage{algpseudocode}

\usepackage[
  colorlinks=true,
  citecolor=DarkGreen,
  linkcolor=blue,
  urlcolor=blue,
  linktocpage=true
]{hyperref}

\newcommand{\cket}[1]{\left|#1\right\rangle}
\newcommand{\bra}[1]{\left\langle#1\right|}
\newcommand{\bracket}[2]{\left\langle#1|#2\right\rangle}
\theoremstyle{plain}
\newtheorem{theorem}{Theorem}

\begin{document}


\title{Correspondence between Asymptotic Quantum Many-Body Scars in Closed Systems and Diffusive Nambu-Goldstone Modes in Open Systems}

\author{Masaya Kunimi}
\email{kunimi@rs.tus.ac.jp}
\affiliation{Department of Physics, Tokyo University of Science, 1-3 Kagurazaka, Tokyo 162-8601, Japan}
\author{Taiki Haga}
\email{taiki.haga@omu.ac.jp}
\affiliation{Department of Physics and Electronics, Osaka Metropolitan University, Sakai-shi, Osaka 599-8531, Japan}
\author{Masaya Nakagawa}
\email{nakagawa@cat.phys.s.u-tokyo.ac.jp}
\affiliation{Department of Physics, The University of Tokyo, 7-3-1 Hongo, Tokyo 113-0033, Japan}


\date{\today}

\begin{abstract}
Hydrodynamic relaxation in open quantum many-body systems can be understood in terms of diffusive Nambu--Goldstone (NG) modes associated with strong-to-weak spontaneous symmetry breaking (SWSSB). Here, we find a common structure underlying weak ergodicity breaking in closed quantum systems and diffusive hydrodynamics in open quantum systems. We establish a spectral correspondence between the parent Hamiltonian of rainbow quantum many-body scar (RQMBS) states in closed systems and the effective Lindbladian of strongly dissipative open systems with strong $U(1)$ symmetry and local charge-dephasing Lindblad operators. For broad classes of spin, fermionic, and bosonic systems, we construct asymptotic RQMBS (ARQMBS) states. Within the enlarged scar subspace, the parent Hamiltonian admits a local Rokhsar--Kivelson-type frustration-free decomposition. A single-mode variational construction gives an $O(k^2)$ upper bound on its lowest excitation energy and yields states satisfying the defining ARQMBS criteria: orthogonality, vanishing energy variance, and characteristic entanglement scaling. Under vectorization, RQMBS states are mapped to maximally mixed infinite-temperature states in fixed-charge sectors, which exhibit long-range SWSSB order under finite-density conditions. Assuming uniformly bounded positive transition rates, the effective Lindbladian has the same local projectors as the parent Hamiltonian but with positive transition-dependent weights, yielding two-sided spectral bounds and the same system-size scaling of their gaps. For uniform rates, the two operators are proportional to each other and their eigenmodes coincide. Therefore, an exact quadratic parent-Hamiltonian branch maps directly to a diffusive NG mode. Our results connect different mechanisms of slow relaxation in closed and open quantum many-body systems.
\end{abstract}
\maketitle
\section{Introduction}\label{sec:Introduction}
Symmetry and spontaneous symmetry breaking (SSB) play an essential role in various fields of physics~\cite{Anderson1972more,Sachdev2011quantum}. In isolated quantum systems, if the system has a continuous symmetry and it is spontaneously broken, Nambu--Goldstone (NG) modes appear as gapless collective excitations~\cite{Nambu1960quasi,Goldstone1961field}, which determine the low-energy physics~\cite{Watanabe2012unified}. Examples of such NG modes include magnons associated with spontaneously broken spin-rotation symmetry~\cite{Holstein1940field}, acoustic phonons associated with broken  translational symmetry in crystals~\cite{Leutwyler1996phonons}, and Bogoliubov modes associated with broken global $U(1)$ symmetry in superfluids~\cite{Bogoliubov1947theory}.

In open quantum systems, the notion of symmetry is refined into strong and weak symmetries~\cite{Buca2012note,Albert2014symmetries,Lieu2020symmetry}. A strong symmetry is respected separately by the Hamiltonian and every Lindblad jump operator, so that the associated charge is conserved along each quantum trajectory. By contrast, a weak symmetry requires only covariance of the Lindbladian under the symmetry transformation. Individual trajectories need not conserve the charge, whereas an initially symmetric density matrix remains symmetry invariant under the ensemble-averaged evolution. At the level of mixed states, strong $U(1)$ symmetry means that the density matrix is supported within a fixed-charge sector, whereas weak $U(1)$ symmetry requires only invariance under the diagonal symmetry transformation, $e^{-i\theta\hat{Q}}\hat{\rho} e^{i\theta\hat{Q}}=\hat{\rho}$, and a weakly but not strongly symmetric state may contain an incoherent mixture of different charge sectors~\cite{Sala2024spontaneous,Lu2024bilayer,Lessa2025strong}.

Recently, in open quantum systems, a novel type of SSB, dubbed strong-to-weak SSB (SWSSB), has been proposed and extensively studied~\cite{Lee2023quantum,Sala2024spontaneous,Kuno2024strong,Lu2024bilayer,Lessa2025strong,Huang2025hydrodynamics,Gu2025spontaneous,Hauser2026strong,Guo2025quantum,Luo2025topological,Shah2025instability,Orito2025strong,Zhang2025strong,Weinstein2025efficient,Sun2025scheme,Feng2025hardness,Liu2025diagnosing,Chen2025strong,Kuno2025intrinsic,Kuno2025strongtoweak,Guo2025strongtoweak,Ziereis2025strong,Ando2026Gauge,Singh2026mixed,Shu2026universal,Ding2026strong,Thorngren2026translation,Lessa2026mixed,Lee2026charge,Liu2026local,Divi2026local,Haga2026discontinuous,Kuno2026hierarchy,Kawamoto2026strong,Wang2026kardar,Wang2026strong,Tang2026mean,Wang2026observation}. In isolated systems, SSB can be detected by the long-range behavior of a correlation function~\cite{Tasaki2020physics}. Because conventional expectation values and correlation functions are linear in the density matrix, they generally cannot distinguish strong symmetry from weak symmetry. SWSSB is instead detected by nonlinear functions of the density matrix, such as fidelity and R\'enyi-2 correlators~\cite{Lessa2025strong}. A notable example of SWSSB is the maximally mixed infinite-temperature state within a fixed $U(1)$-charge sector at finite charge density~\cite{Sala2024spontaneous,Lessa2025strong,Ziereis2025strong}. For generic open quantum systems that exhibit diffusive hydrodynamic relaxation, the NG mode associated with SWSSB can be identified with the diffusion mode of the conserved charge density~\cite{Huang2025hydrodynamics}. The decay rate of the diffusive NG mode behaves as $\lambda(k)\simeq Dk^2+o(k^2)$, where $D$ is the diffusion constant and $k$ is the wave number. The decay rate of this mode vanishes in the long-wavelength limit.

In this work, we find a connection between the slow modes associated with SWSSB and long-lived nonthermal states in closed systems under appropriate conditions. We establish a spectral correspondence between the parent Hamiltonian underlying the construction of asymptotic rainbow quantum many-body scar states and the effective Lindbladian of strongly dissipative systems with local charge dephasing. This correspondence links weak ergodicity breaking in closed systems to dissipative hydrodynamics in open systems.

To explain this correspondence, we first briefly review quantum many-body scarring and its asymptotic extension. Quantum many-body scar (QMBS) states~\cite{Bernien2017probing,Turner2018weak,Turner2018quantum,Serbyn2021quantum,Papic2022weak,Moudgalya2022quantum,Chandran2023quantum}  constitute a class of nonergodic quantum many-body states, which violate the eigenstate thermalization hypothesis~\cite{Deutsch1991quantum,Srednicki1994chaos,Rigol2008thermalization}. QMBS states are atypical energy eigenstates embedded in an otherwise thermal spectrum and exhibit nonthermal entanglement structures. They often have subvolume-law entanglement, although volume-law scar states are also known~\cite{Turner2018quantum,Moudgalya2022quantum,Langlett2022rainbow,Mukherjee2026symmetric}. In many scarred models, QMBS states form an approximately or exactly equally spaced tower~\cite{Turner2018weak,Turner2018quantum,Choi2019emergent}. An appropriate superposition of QMBS states with exactly equally spaced energies exhibits perfect revivals, with a revival period inversely proportional to the spacing of the QMBS tower. Extensive previous works have shown that various Hamiltonians host QMBS states and have investigated the properties of these states~\cite{Shiraishi2017systematic,Moudgalya2018exact,Moudgalya2018entanglement,Turner2018weak,Turner2018quantum,Choi2019emergent,Ho2019periodic,Lin2019exact,Bull2019systematic,Khemani2019signatures,Schecter2019weak,Shiraishi2019connection,Ok2019topological,Iadecola2020quantum,Mark2020eta,Moudgalya2020large,Moudgalya2020eta,Mark2020exact,Bull2020quantum,Chattopadhyay2020quantum,Mark2020unified,Alhambra2020revivals,Lin2020quantum,Lee2020exact,Kuno2020flat,Lin2020slow,Mizuta2020exact,ODea2020from,Shibata2020onsager,Pakrouski2020many,Hudomal2020quantum,Surace2020weak,Sugiura2021many,Pakrouski2021group,Chertkov2021motif,Ren2021quasisymmetry,Ren2022deformed,Langlett2022rainbow,Tamura2022quantum,Desaules2022extensive,Gotta2022exact,Tang2022multimagnon,Wildeboer2022quantum,Omiya2023fractionalization,Omiya2023quantum,Desaules2023weak,Desaules2023prominent,Sanada2023quantum,Iversen2023tower,Halimeh2023robust,Kaneko2024quantum,Wang2024embedding,Kunimi2024proposal,Matsui2024exactly,Teretenkov2024duality,Nakagawa2024exact,Moudgalya2024exhaustive,Moudgalya2024symmetries,Imai2025quantum,Calajo2025quantum,Marche2025exceptional,Morettini2025transport,Ivanov2025volume,Ma2025liuvillian,Huang2025strongly,Gotta2025open,Gioia2025distinct,Matsui2025symmetry,Perciavalle2025local,Dooley2026parent,Mohapatra2026additional,Faugno2026nonequilibrium,Miao2026exact,Garcia2026lindblad,Banerjee2026quantum,ODea2026locality,Omiya2026any,Mukherjee2026symmetric,Gehrmann2026exact,Kerschbaumer2026coherent}. Some of the QMBS states have been experimentally observed~\cite{Bernien2017probing,Bluvstein2021controlling,Chen2022error,Su2023observation,Zhang2023many,Dong2023disorder,Desaules2024robust,Zhao2025observation}.

Recently, asymptotic quantum many-body scar (AQMBS) states have been proposed by Gotta {\it et al}.~\cite{Gotta2023Asymptotic} and the first experimental signatures of AQMBS states were reported on a trapped-ion quantum processor~\cite{Logaric2026dynamical}. Although these states are related to QMBS states, they have several distinct properties, which are summarized as follows~\cite{Gotta2023Asymptotic,Moudgalya2024symmetries,Ren2024quasi,Kunimi2025systematic,Hashimoto2026construction}: (i) orthogonality to the corresponding QMBS tower, (ii) an atypical or nonthermal entanglement structure, and (iii) vanishing energy variance in the thermodynamic limit. Property (iii), together with the Mandelstam--Tamm bound~\cite{Mandelstam1945uncertainty}, implies a diverging lower bound on the characteristic evolution timescale of an AQMBS state. This result indicates asymptotically long-lived nonthermal dynamics. Because the energy variance of the AQMBS states is nonzero in finite-size systems, the AQMBS states are not energy eigenstates of the Hamiltonian, in contrast to the QMBS states. This means that AQMBS states cannot be identified as individual energy eigenstates by numerical exact diagonalization. Very recently, two works have proposed systematic construction methods for AQMBS states~\cite{Kunimi2025systematic,Logaric2026dynamical}. Although these works focus on different systems, the common ingredient is to consider a parent Hamiltonian whose ground-state manifold contains QMBS states and whose gapless excitations are candidates for AQMBS states of the original Hamiltonian.

A particular class of QMBS states is closely related to open quantum systems. In open quantum systems, the infinite-temperature state in a fixed $U(1)$ sector is given by $\hat{\rho}_{\infty}\propto \sideset{}{'}\sum_{\bm{n}}\cket{\bm{n}}\bra{\bm{n}}$, where $|\bm{n}\rangle$ denotes a computational-basis state in the fixed $U(1)$ sector and the primed sum runs over that sector. Under vectorization, which maps operators to vectors in a doubled Hilbert space, the state $\hat{\rho}_{\infty}$ is mapped to the rainbow state:
\begin{align}
\hat{\rho}_{\infty}\to \cket{\psi_{\rm rainbow}}\propto \sideset{}{'}\sum_{\bm{n}}\cket{\bm{n}}\otimes\cket{\bm{n}}.\label{eq:mapping_infinite_temperature_state_to_rainbow_state}
\end{align}
When this state is an atypical exact eigenstate of an appropriate doubled Hamiltonian, it constitutes a rainbow quantum many-body scar (RQMBS) state (or an Einstein--Podolsky--Rosen scar state)~\cite{Langlett2022rainbow,Wildeboer2022quantum}. Teretenkov and Lychkovskiy demonstrated a duality between rainbow scar states in closed systems and infinite-temperature states in open quantum systems under detailed-balance conditions~\cite{Teretenkov2024duality}. 

These observations suggest a relation between asymptotic RQMBS (ARQMBS) states and diffusive NG modes since both describe excitations from the states related by the correspondence in Eq.~(\ref{eq:mapping_infinite_temperature_state_to_rainbow_state}). In this paper, we rigorously prove the existence of ARQMBS states in broad classes of spin, fermionic, and bosonic systems using the systematic construction method for AQMBS states proposed in Ref.~\cite{Kunimi2025systematic}. We then establish an explicit correspondence between the spectrum of the RQMBS parent Hamiltonian in closed systems and that of the second-order effective Lindbladian in the strongly dissipative regime for open systems with strong $U(1)$ symmetry. A key feature of this correspondence is that, when restricted to the enlarged scar subspace, the RQMBS parent Hamiltonian admits a local Rokhsar--Kivelson-type frustration-free decomposition~\cite{Rokhsar1988superconductivity} and is gapless under appropriate conditions. For local gapless frustration-free Hamiltonians satisfying the assumptions of Refs.~\cite{Watanabe2024critical,Masaoka2024quadratic,Masaoka2025rigorous}, if the long-wavelength dispersion is characterized by $\epsilon(k)\sim k^z$, then $z\geq2$.

Our results provide a structural origin of the gaplessness and slow hydrodynamic scaling in open quantum many-body systems. In particular, when the parent Hamiltonian underlying the ARQMBS states has a quadratic low-energy dispersion, the corresponding low-decay-rate mode also has a quadratic dispersion and, under appropriate conditions, can be identified with a diffusive NG mode associated with SWSSB.

\section{Overview}\label{sec:overview}
Before presenting the precise formulation and proofs, we provide an overview of the main results and outline the structure of the rest of the paper. The detailed proofs and assumptions are presented in subsequent sections. Here, we focus on one-dimensional spin-1/2 systems. In the subsequent sections, the results will be generalized to higher-spin and multicomponent fermionic and bosonic systems. 
\subsection{Setting}\label{subsec:setting}
First, we consider isolated systems. The two-leg ladder Hamiltonian (rainbow scar Hamiltonian) is defined as
\begin{align}
\hat{H}_{\rm tot}&=\hat{H}\otimes\hat{1}-\hat{1}\otimes\hat{H}^{\ast}-\frac{\hbar}{2}\sum_j(\hat{L}_j\otimes\hat{1}-\hat{1}\otimes\hat{L}_j)^2,\label{eq:definition_of_Htot_overview}
\end{align}
where $\hat{H}$ acts on ladder $1$ and $-\hat{H}^{\ast}$ acts on ladder 2, $\hat{1}$ is the identity operator, and $\hat{L}_j$ is a local Hermitian function of $\{\hat{S}_i^z\}$. Here, $\hat{S}_i^z$ is the $z$ component of the spin-1/2 operator at the $i$th site. We assume that $\hat{H}_{\rm tot}$ possesses $U(1)\times U(1)$ symmetry, meaning that the total magnetization of each ladder is conserved. In addition, we assume that the total Hamiltonian can be written as a sum of local operators. 

We can show that the RQMBS state is an exact eigenstate of the total Hamiltonian~\cite{Langlett2022rainbow,Wildeboer2022quantum}:
\begin{align}
\cket{S_n}\propto \sideset{}{'}\sum_{\bm{n}}\cket{\bm{n}}\otimes\cket{\bm{n}},\quad \hat{H}_{\rm tot}\cket{S_n}=0,\label{eq:RQMBS_state_overview}
\end{align}
where $n$ labels a fixed $U(1)$ sector, $\cket{\bm{n}}$ is a product-basis state, and the primed sum runs over that sector.

Next, we consider the corresponding open quantum systems. The Lindblad master equation is given by
\begin{align}
\frac{d}{dt}\hat{\rho}(t)&=-\frac{i}{\hbar}[\hat{H}, \hat{\rho}(t)]+\sum_j\left[\hat{L}_j\hat{\rho}(t)\hat{L}_j-\frac{1}{2}\{\hat{L}_j^2, \hat{\rho}(t)\}\right],\label{eq:Lindblad_equation_overview}
\end{align}
where $\hat{\rho}(t)$ is the density matrix at time $t$, and $\hat{H}$ and $\hat{L}_j$ are the same as those in Eq.~(\ref{eq:definition_of_Htot_overview}). The interaction between the ladders in closed systems corresponds to the dissipative term generated by the Lindblad operators in open quantum systems. Under the assumptions on $\hat{H}$ and $\hat{L}_j$, the system possesses strong $U(1)$ symmetry~\cite{Buca2012note,Albert2014symmetries,Lieu2020symmetry}:
\begin{align}
[\hat{H}, \hat{S}_{\rm tot}^z]=0,\quad [\hat{L}_j, \hat{S}_{\rm tot}^z]=0,\quad \text{for all }j,\label{eq:definition_of_Strong_U(1)_symmetry_overview}
\end{align}
where $\hat{S}_{\rm tot}^z\equiv \sum_{i=1}^M\hat{S}_i^z$. In the vectorized form, the Lindblad master equation~\cite{Medvedyeva2016exact,Haga2023quasiparticles,Teretenkov2024duality} is written as 
\begin{align}
\frac{d}{dt}\cket{\rho(t)}&=\hat{\mathcal{L}}\cket{\rho(t)},\label{eq:vectrized_form_Lindblad_overview}\\
\hat{\mathcal{L}}&\equiv -\frac{i}{\hbar}(\hat{H}\otimes\hat{1}-\hat{1}\otimes\hat{H}^{\ast})+\sum_j\left(\hat{L}_j\otimes\hat{L}_j\right.\notag \\
&\left.\quad -\frac{1}{2}\hat{L}_j^2\otimes\hat{1}-\frac{1}{2}\hat{1}\otimes\hat{L}_j^2\right).\label{eq:vectorized_form_of_Lindblad_master_equation_overview}
\end{align}
We find that the Lindbladian $\hat{\mathcal{L}}$ has a form similar to that of the Hamiltonian in Eq.~(\ref{eq:definition_of_Htot_overview}):
\begin{align}
i\hbar\hat{\mathcal{L}}&=\hat{H}\otimes\hat{1}-\hat{1}\otimes\hat{H}^{\ast}-\frac{i\hbar}{2}\sum_j(\hat{L}_j\otimes\hat{1}-\hat{1}\otimes\hat{L}_j)^2.\label{eq:Hamiltonian-like_form_Lindbladian_overview}
\end{align}
The only difference is the presence of the imaginary unit in the interaction between ladders. 

We can show that the infinite-temperature state in each fixed $U(1)$ sector is a steady state of the Lindblad equation:
\begin{align}
\hat{\rho}_{{\rm s}, n}\propto \sideset{}{'}\sum_{\bm{n}}\cket{\bm{n}}\bra{\bm{n}}\leftrightarrow\cket{\rho_{{\rm s},n}}\propto \sideset{}{'}\sum_{\bm{n}}\cket{\bm{n}}\otimes\cket{\bm{n}}.\label{eq:infinit_temperature_state_overview}
\end{align}
The vectorized form of the infinite-temperature state is the same as the RQMBS state (\ref{eq:RQMBS_state_overview}). This state-level correspondence follows directly from vectorization. Under suitable detailed-balance conditions, Teretenkov and Lychkovskiy~\cite{Teretenkov2024duality} extended this state-level correspondence to a dynamical duality between open systems and closed bilayer systems.

In addition, we can show that the infinite-temperature state under the finite-density conditions exhibits SWSSB order, i.e., conventional correlation functions, such as ${\rm Tr}(\hat{\rho}_{{\rm s},n}\hat{S}_i^+\hat{S}_j^-)$, do not exhibit long-range order, whereas the R\'enyi-2~\cite{Lessa2025strong} and the R\'enyi-1 correlators~\cite{Liu2025diagnosing,Weinstein2025efficient} exhibit long-range order.

\begin{figure}[t]
\centering
\includegraphics[width=8.6cm,clip]{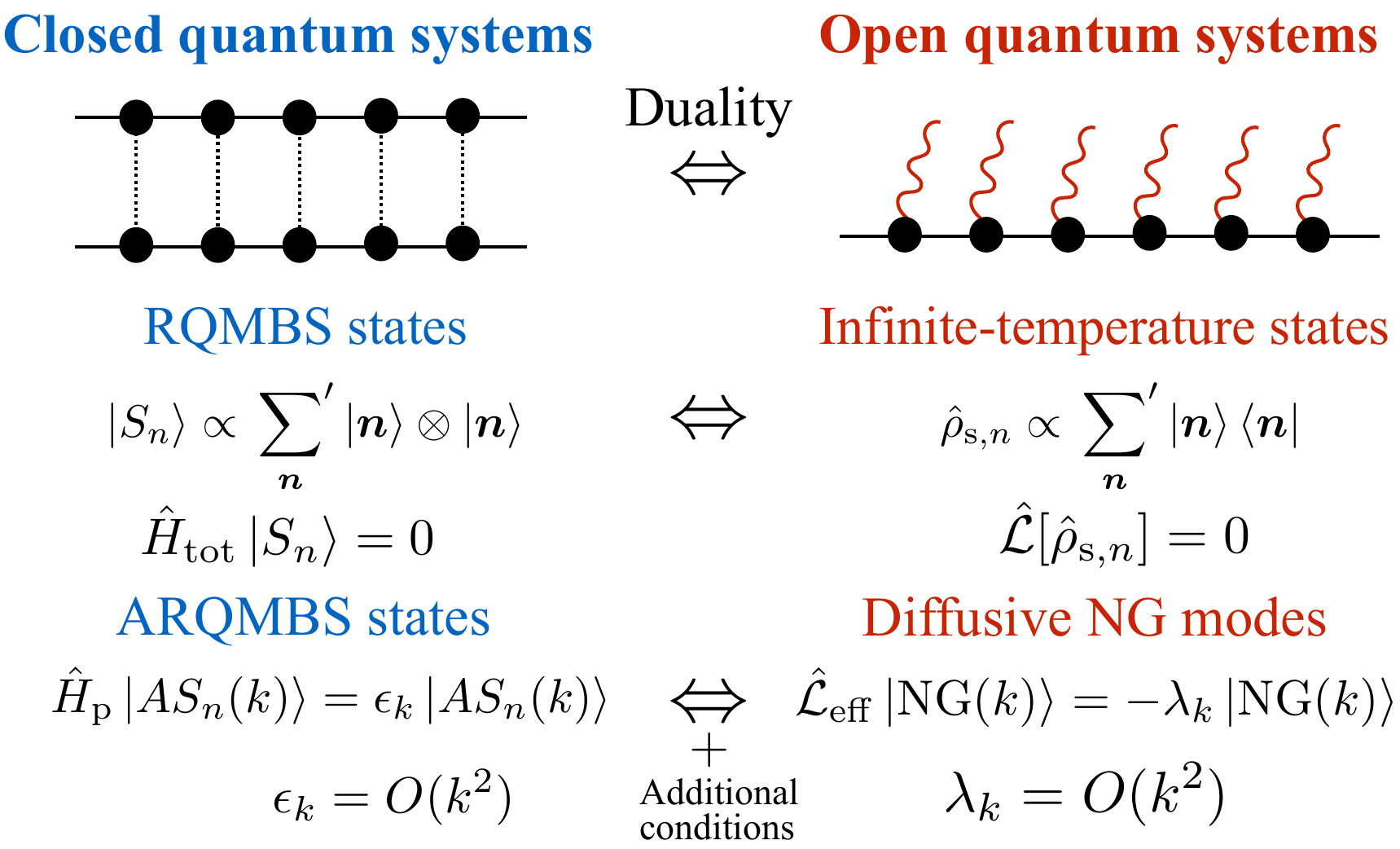}
\caption{Schematic of the correspondence between closed quantum systems and open quantum systems for the spin-$1/2$ example. 
}
\label{fig:schematic_duality}
\vspace{-0.75em}
\end{figure}%
\subsection{Main results}\label{subsec:main_results}

In this subsection, we state two main theorems without providing their proofs. 

For isolated systems, we show the following theorem.
\begin{theorem}[Informal statement: existence of ARQMBS states]\label{thm:ARQMBS}
Let $\hat{\mathcal{P}}$ be a projector onto the Hilbert subspace called the enlarged scar subspace spanned by $\{\cket{\bm{n}}\otimes\cket{\bm{n}}\}$. The parent Hamiltonian whose zero-energy ground states are the RQMBS states can be written as
\begin{align}
\hat{H}_{\rm p}\equiv \hat{\mathcal{P}}(\hat{H}\otimes\hat{1}-\hat{1}\otimes\hat{H}^{\ast})^2\hat{\mathcal{P}}=\sum_j\hat{h}_{{\rm p},j},\label{eq:parent_Hamiltonian_overview}
\end{align}
where $\hat{h}_{{\rm p},j}$ is a local Hermitian operator and $\hat{h}_{{\rm p},j}\ge 0$. The parent Hamiltonian is frustration-free $(\text{i.e.,~} \hat{h}_{{\rm p},j}\cket{S_n}=0 \text{ for all }j)$. In addition, under the assumptions summarized in Sec.~\ref{subsec:summary_construction_ARQMBS}, and within a fixed-charge sector satisfying the finite-density condition,  the finite-size gap of the parent Hamiltonian in the fixed $U(1)$ sector admits an $O(1/M^2)$ upper bound, where $M$ is the number of lattice sites.
\end{theorem}
Theorem \ref{thm:ARQMBS} relies on the systematic construction of the AQMBS states in systems satisfying a restricted spectrum-generating algebra~\cite{Kunimi2025systematic}. The upper bound on the finite-size gap implies that the parent Hamiltonian is gapless in the thermodynamic limit. Following Ref.~\cite{Kunimi2025systematic}, the corresponding low-energy variational state is a candidate AQMBS state of the original Hamiltonian. In this case, we can show that the corresponding single-mode variational state is orthogonal to the RQMBS states, exhibits the same characteristic entanglement scaling as the corresponding RQMBS state, and has vanishing energy variance in the thermodynamic limit. Therefore, we prove the existence of the ARQMBS states in the rainbow scar Hamiltonian. 

For open quantum systems, we prove the following theorem.
\begin{theorem}[Informal statement: correspondence between the parent-Hamiltonian spectrum and slow modes associated with SWSSB in the strongly dissipative regime]\label{thm:dissipative_NG}
In the strongly dissipative regime, the second-order effective Lindbladian can be written as
\begin{align}
\hat{\mathcal{L}}_{\rm eff}&=-\frac{2}{\hbar^2}\sum_{\bm{n},\bm{m},\bm{n}\not=\bm{m}}\frac{|H_{\bm{n},\bm{m}}|^2}{\Gamma_{\bm{n},\bm{m}}}\hat{P}_{\bm{n},\bm{m}},\label{eq:effective_Lindbladian_overview}
\end{align}
where $H_{\bm{n},\bm{m}}$,  $\hat{P}_{\bm{n},\bm{m}}$, and $\Gamma_{\bm{n},\bm{m}}$ are defined by Eqs.~(\ref{eq:definition_of_H_n_m_spin-S}), (\ref{eq:definition_of_operator_P_spin-S}), and (\ref{eq:definition_of_Gamma_n_m}), respectively. The eigenvalue equation for the effective Lindbladian is given by
\begin{align}
\hat{\mathcal{L}}_{\rm eff}\cket{\psi_i}=-\lambda_i\cket{\psi_i},\label{eq:eigenvalue_problem_effective_Lindbladian_overview}
\end{align}
where $\lambda_i$ is real and nonnegative and satisfies $0=\lambda_1\le \lambda_2\le\lambda_3\le \cdots$. If $\Gamma_{\bm{n},\bm{m}}$ satisfies the inequality $0<\Gamma_{\rm min}\le \Gamma_{\bm{n},\bm{m}}\le \Gamma_{{\rm max}}<\infty$, where $\Gamma_{\rm min,max}$ are positive system-size-independent constants, the following inequality holds:
\begin{align}
\frac{\epsilon_i}{\Gamma_{\rm max}}\le \hbar^2\lambda_i\le \frac{\epsilon_i}{\Gamma_{\rm min}},\label{eq:inequality_Lindbladian_spectra_overview}
\end{align}
where $\epsilon_i$ is the $i$th eigenvalue of the parent Hamiltonian and satisfies $0=\epsilon_1<\epsilon_2\le \epsilon_3\le \ldots$. Combining this inequality with Theorem~\ref{thm:ARQMBS} yields $\lambda_2=O(1/M^2)$.   
\end{theorem}
When $\Gamma_{\bm{n},\bm{m}}=\Gamma>0$ for all transitions with $H_{\bm{n},\bm{m}}\not=0$, the effective Lindbladian becomes $\hat{\mathcal{L}}_{\rm eff}=-\hat{H}_{\rm p}/(\hbar^2\Gamma)$. Thus, the two operators have identical eigenmodes and proportional spectra. In addition, as discussed in the next subsection, we present an example in which the parent-Hamiltonian spectrum has an exact quadratic low-energy branch. In this case, the slow modes represent diffusive modes. We summarize this correspondence in Fig.~\ref{fig:schematic_duality}.

\subsection{Example}\label{subsec:Example_overview}
In this subsection, we consider an example of the correspondence between closed systems and open quantum systems. We consider a spin-$1/2$ {\it XXZ} Hamiltonian and the following $\{\hat{L}_j\}$:
\begin{align}
\hat{H}&\equiv J\sum_{j=1}^{M-1}(\hat{S}_j^x\hat{S}_{j+1}^x+\hat{S}_j^y\hat{S}_{j+1}^y+\Delta\hat{S}_j^z\hat{S}_{j+1}^z),\label{eq:definition_of_XXZ_Hamiltonian_overview}\\
\hat{L}_j&\equiv \sqrt{\gamma}\hat{S}_j^z,\label{eq:definition_of_Lj_overview}
\end{align}
where $J>0$ is the interaction strength, $\Delta$ is the anisotropy parameter, and $\gamma>0$ is the inter-ladder coupling strength in the closed systems and the dephasing rate in the open systems. Here, we impose open boundary conditions. 

First, we consider isolated systems. The parent Hamiltonian for the RQMBS states is represented as
\begin{align}
\hat{H}_{\rm p}&=J^2\sum_{j=1}^{M-1}\left(\frac{1}{4}-\hat{\bm{s}}_j\cdot\hat{\bm{s}}_{j+1}\right)\label{eq:parent_Hamiltonian_S=1/2_XXZ_oveview}
\end{align}
in the enlarged scar subspace, where we have used the single-ladder mapping $\cket{\bm{n}}\otimes\cket{\bm{n}}\to \cket{\bm{n}}$, and $\hat{\bm{s}}_j$ is the spin-1/2 operator at site $j$ after mapping. Because the parent Hamiltonian is the ferromagnetic Heisenberg model, we obtain low-energy gapless magnon excitations:
\begin{align}
&\hat{H}_{\rm p}\cket{AS_n(k)}=\epsilon_k\cket{AS_n(k)},\label{eq:eigenvalue_problem_parent_Hamiltonian_overview}\\
&\epsilon_k\equiv J^2\left[1-\cos\left(\frac{\pi k}{M}\right)\right],\quad k=1,2,\ldots,M-1,\label{eq:disperion_parent_Hamiltonian_overview}
\end{align}
where $\cket{AS_n(k)}$ is the ARQMBS state and the explicit expression is given by Eq.~(\ref{eq:definition_of_magnon_ex_spin-1/2_ex}). For fixed $k=O(1)$, we obtain $\epsilon_k=O(1/M^2)$.

Next, we consider the open quantum systems. In this case, the effective Lindbladian becomes $\hat{\mathcal{L}}_{\rm eff}=-\hat{H}_{\rm p}/(\hbar^2\gamma)$. Therefore, for fixed $k=O(1)$, the decay rate $\lambda_k$ can be written as $\lambda_k=\epsilon_k/(\hbar^2\gamma)=O(1/M^2)$. 

\begin{figure}[t]
\centering
\includegraphics[width=8.6cm,clip]{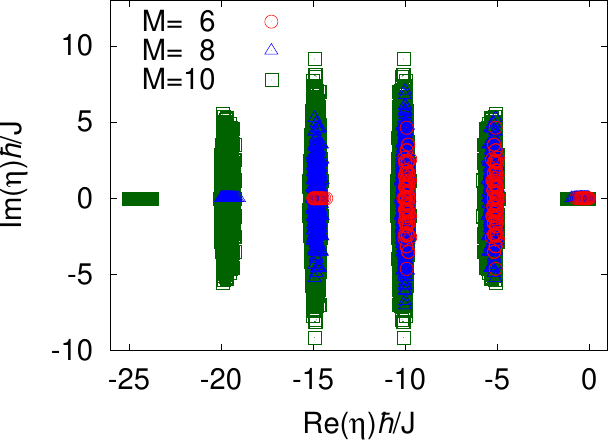}
\caption{Full Lindbladian spectrum $\eta$ for $\Delta=2$ and $\gamma=5J/\hbar$ in the sector $\hat{S}_{\rm tot}^z\otimes\hat{1}=\hat{1}\otimes\hat{S}_{\rm tot}^z=0$, where $\hat{S}_{\rm tot}^z\equiv \sum_{j=1}^M\hat{S}_j^z$. The system sizes are $M=6, 8$, and $10$.}
\label{fig:Lindbladian_spectra}
\vspace{-0.75em}
\end{figure}%

\begin{figure}[t]
\centering
\includegraphics[width=8.6cm,clip]{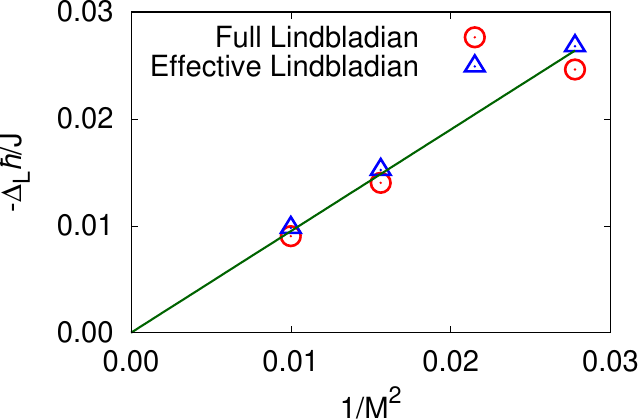}
\caption{Comparison of the slowest decay rates obtained from the full and effective Lindbladians. The parameters are the same as those in Fig.~\ref{fig:Lindbladian_spectra}. The green solid line represents $1/M^2$ scaling as a guide to the eye.}
\label{fig:slowest_decaying_mode}
\vspace{-0.75em}
\end{figure}

\begin{figure}[t]
\centering
\includegraphics[width=8.6cm,clip]{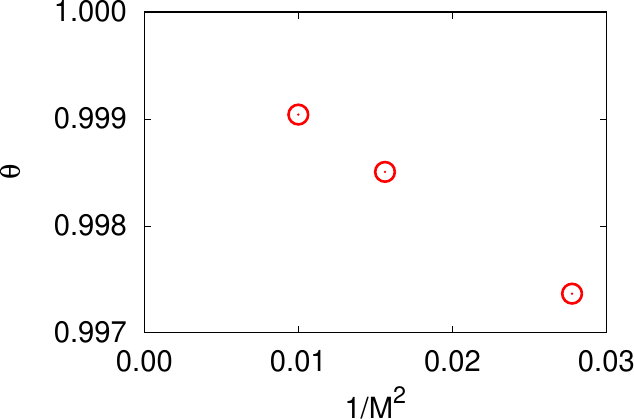}
\caption{Overlap between the ARQMBS state $|AS_{M/2}(1)\rangle$ and the slowest-decaying mode. }
\label{fig:overlap_slowest_decaying_mode}
\vspace{-0.75em}
\end{figure}

To numerically test the predicted diffusive scaling of the slowest mode in the full Lindbladian $\hat{\mathcal{L}}$, we perform numerical diagonalization of $\hat{\mathcal{L}}$. Figure \ref{fig:Lindbladian_spectra} shows the spectrum of the full Lindbladian: $\hat{\mathcal{L}}\cket{\varphi_i}=\eta_i\cket{\varphi_i}$. Here, we focus on the slowest-decaying mode, which has the nonzero eigenvalue closest to the imaginary axis. We define $\Delta_{\rm L}\equiv \max_{{\rm Re}(\eta_j)<0}[{\rm Re}(\eta_j)]$. Figure \ref{fig:slowest_decaying_mode} shows the system-size dependence of $\Delta_{\rm L}$ of the full Lindbladian and the effective Lindbladian. For the effective Lindbladian, $\Delta_{{\rm L,eff}}=-\epsilon_{k=1}/(\hbar^2\gamma)$. The results indicate $M^{-2}$ scaling and are consistent with the analytical results. We also investigate the overlap between the ARQMBS state and the slowest-decaying mode, which is defined by~\cite{Marche2025exceptional}
\begin{align}
\theta\equiv \frac{\left|\langle AS_n(1)|\varphi_{\rm slow} \rangle\right|}{\sqrt{\langle AS_n(1)|AS_n(1)\rangle \langle \varphi_{\rm slow}|\varphi_{\rm slow} \rangle}},\label{eq:definition_of_overlap_overview}
\end{align}
where $\cket{\varphi_{\rm slow}}$ is the eigenvector corresponding to the slowest-decaying mode obtained by the numerical diagonalization. Figure \ref{fig:overlap_slowest_decaying_mode} shows the overlap as a function of the system size. We find that the overlap is close to unity and that the ARQMBS state well approximates the slowest-decaying mode of the full Lindbladian. These results provide numerical evidence that the correspondence remains accurate for the full Lindbladian in the parameter regime and system sizes studied here.

\subsection{Structure of the paper}\label{subsec:structure_of_paper}

The remainder of this paper is organized as follows. In Sec.~\ref{sec:framework}, we review the formalism for the systematic construction of AQMBS states. In Sec.~\ref{sec:Proof_of_ARQMBS}, we rigorously demonstrate the existence of ARQMBS states in various systems, including quantum spin systems with $U(1)\times U(1)$ symmetry, multicomponent fermionic systems, and multicomponent bosonic systems with $U(1)^f\times U(1)^f$ symmetry. In Sec.~\ref{sec:Proof_of_correspondence}, under the assumptions specified therein, we establish a spectral correspondence between the parent Hamiltonian underlying the ARQMBS construction and the second-order effective Lindbladian for systems with local Hermitian charge-dephasing Lindblad operators. When the parent Hamiltonian has a quadratic low-energy dispersion, we identify the corresponding low-decay-rate modes with diffusive NG modes. In Sec.~\ref{sec:Example}, we consider several examples of closed and open systems and explicitly show the correspondence. In Sec.~\ref{sec:summary}, we summarize our results and discuss the future outlook. Technical details are provided in the appendices.

\section{General framework}\label{sec:framework}
In this section, we review the systematic construction of AQMBS states in isolated quantum systems proposed by Kunimi {\it et al}.~\cite{Kunimi2025systematic}. This construction plays an essential role in the present work. The construction combines the restricted spectrum-generating algebra (RSGA)~\cite{Batista2009canted,Mark2020eta,Moudgalya2020eta} with the symmetry-based formalism for QMBS states~\cite{ODea2020from}. We also review the recent extension of the RSGA to systems with multiple ladder operators~\cite{Hashimoto2026construction}.

\subsection{Restricted spectrum generating algebra}\label{subsec:RSGA}
\subsubsection{Single-ladder case}\label{subsec:sigle_ladder_operator}
We first consider a system in which a single ladder operator $\hat{Q}^{\dagger}$ is present~\cite{Batista2009canted,Mark2020eta,Moudgalya2020eta}. We assume that the ladder operator $\hat{Q}^{\dagger}$ is non-Hermitian and can be written as a sum of local operators $\hat{q}_j^{\dagger}$, i.e., $\hat{Q}^{\dagger}\equiv \sum_j\hat{q}_j^{\dagger}$. Let $\hat{H}$ be a nonintegrable Hamiltonian and $\cket{S_0}$ a normalized root state satisfying $\hat{Q}^{\dagger}\cket{S_0}\not=0$ and $\bracket{S_0}{S_0}=1$. We further assume that $\cket{S_0}$ cannot be written in the form $\hat{Q}^{\dagger}\cket{\phi}$ for any eigenstate $\cket{\phi}$ of $\hat{H}$. The RSGA-$p~(p=1,2,\ldots)$ is defined as
\begin{subequations}
\begin{align}
\hat{H}\cket{S_0}&=E_0\cket{S_0},\label{eq:RSGA-n_condition1}\\
[\hat{H}, \hat{Q}^{\dagger}]\cket{S_0}&=\mathcal{E}\hat{Q}^{\dagger}\cket{S_0},\label{eq:RSGA-n_condition2}\\
\underbrace{[[[[\hat{H},\hat{Q}^{\dagger}],\hat{Q}^{\dagger}],\ldots],\ldots]}_{r\text{ times}}\cket{S_0}&=0,\quad \text{for } 2\le r \le p,\label{eq:RSGA-n_condition3}\\
\underbrace{[[[[\hat{H},\hat{Q}^{\dagger}],\hat{Q}^{\dagger}],\ldots],\ldots]}_{p+1\text{ times}}&=0,\label{eq:RSGA-n_condition4}
\end{align}
\end{subequations}
where $E_0$ and $\mathcal{E}$ are real numbers.  For $p=1$, the condition in Eq.~(\ref{eq:RSGA-n_condition3}) is absent. These conditions imply that, for every nonnegative integer $n$, the unnormalized state
\begin{align}
|\tilde{S}_n\rangle\equiv (\hat{Q}^{\dagger})^n\cket{S_0},\label{eq:definition_of_un_normalized_Sn}
\end{align}
either vanishes or is an exact eigenstate satisfying
\begin{align}
\hat{H}|\tilde{S}_n\rangle&=(E_0+n\mathcal{E})|\tilde{S}_n\rangle.\label{eq:eigenstate_Sn}
\end{align}
Therefore, the RSGA conditions generate an equally spaced tower of exact energy eigenstates. When these eigenstates are nonthermal states embedded in an otherwise thermal spectrum, they constitute a tower of QMBS states. The normalized QMBS state is defined by $\cket{S_n}\equiv |\tilde{S}_n\rangle/\sqrt{\langle\tilde{S}_n|\tilde{S}_n\rangle}$. The scar subspace is defined as
\begin{align}
\mathcal{H}_{\rm scar}&\equiv{\rm Span}\{\cket{S_n} ~|~  n \in\mathcal{I}\},\label{eq:definition_of_Scar_subspace}\\
\mathcal{I}&\equiv \{ n\in \mathbb{Z}_{\ge 0} ~|~ (\hat{Q}^{\dagger})^n\cket{S_0}\not=0\}.\label{eq:setof_nonzero_QMBS_states}
\end{align}

\subsubsection{Multiladder case}\label{subsec:multi_ladder_operator}

Next, we consider the case with multiple ladder operators~\cite{Hashimoto2026construction}. Let us consider the operators $\{\hat{Q}^{\dagger}_1, \hat{Q}^{\dagger}_2,\ldots,\hat{Q}^{\dagger}_{N_{\rm L}}\}$, where $N_{\rm L}$ is a positive integer. We assume that these operators are non-Hermitian and can each be written as a sum of local operators: $\hat{Q}_l^{\dagger}=\sum_j\hat{q}^{\dagger}_{l,j}$. The normalized root state is denoted by $\cket{S_{\bm{0}}}$. As in the single-ladder case, we assume that $\cket{S_{\bm{0}}}$ cannot be written in the form $\hat{Q}_l^{\dagger}\cket{\phi}$ for any $l=1,\ldots,N_{\rm L}$ and for any eigenstate $\cket{\phi}$ of $\hat{H}$. The multiladder RSGA-$p$ (MLRSGA-$p$) is defined as
\begin{subequations}
\begin{align}
\hat{H}\cket{S_{\bm{0}}}&=E_0\cket{S_{\bm{0}}},\label{eq:MLRSGA_relation1}\\
[\hat{H}, \hat{Q}_l^{\dagger}]\cket{S_{\bm{0}}}&=\mathcal{E}_l\hat{Q}^{\dagger}_l\cket{S_{\bm{0}}},\notag \\
&\hspace{3.0em}\text{for }l=1,\ldots,N_{\rm L},\label{eq:MLRSGA_relation2}\\
\underbrace{[[[[\hat{H}, \hat{Q}_l^{\dagger}],\hat{Q}^{\dagger}_l],\ldots],\ldots]}_{r \text{ times}}\cket{S_{\bm{0}}}&=0,~\text{for }2\le r\le p,\notag \\
&\hspace{3.0em}\text{ and }l=1,\ldots,N_{\rm L}, \label{eq:MLRSGA_relation3}\\
\underbrace{[[[[\hat{H}, \hat{Q}_l^{\dagger}],\hat{Q}^{\dagger}_l],\ldots],\ldots]}_{p+1 \text{ times}}&=0,~\text{for }l=1,\ldots,N_{\rm L}, \label{eq:MLRSGA_relation4}\\
[\hat{Q}_l^{\dagger},\hat{Q}_k^{\dagger}]&=0,\;\text{for all }l, k,\label{eq:MLRSGA_relation5}\\
[[\hat{H}, \hat{Q}^{\dagger}_l],\hat{Q}_k^{\dagger}]&=0,\; \text{for all }l,k \text{ with }l\not=k,\label{eq:MLRSGA_relation6}
\end{align}
\end{subequations}
where $E_0$ and $\mathcal{E}_l$ are real constants. When the above relations hold, we obtain
\begin{align}
\hat{H}|\tilde{S}_{\bm{n}}\rangle&=\left(E_0+\sum_{l=1}^{N_{\rm L}}n_l\mathcal{E}_l\right)|\tilde{S}_{\bm{n}}\rangle,\label{eq:consequence_MLRSGA}\\
|\tilde{S}_{\bm{n}}\rangle&\equiv \prod_{l=1}^{N_{\rm L}}(\hat{Q}_l^{\dagger})^{n_l}\cket{S_{\bm{0}}},\label{eq:definition_of_QMBS_ML_case}
\end{align}
where $\bm{n}\equiv (n_1,n_2,\ldots,n_{N_{\rm L}})$, and each $n_l$ is a nonnegative integer. Thus, the MLRSGA generates a multidimensional tower of exact energy eigenstates with an energy increment $\mathcal{E}_l$ along the $l$th ladder direction. The nonzero states $|\tilde{S}_{\bm{n}}\rangle$ form a tower of unnormalized exact eigenstates. They constitute QMBS states when they are atypical nonthermal states embedded in an otherwise thermal spectrum. The normalized QMBS state for the multiladder case is also defined by $\cket{S_{\bm{n}}}\equiv |\tilde{S}_{\bm{n}}\rangle/\sqrt{\langle \tilde{S}_{\bm{n}}|\tilde{S}_{\bm{n}}\rangle}$. In the multiladder case, the scar subspace can be defined as
\begin{align}
\mathcal{H}_{\rm scar}&\equiv {\rm Span}\{|S_{\bm{n}}\rangle~|~ \bm{n}\in \mathcal{I}_{\rm L}\},\label{eq:definition_of_scar_subspace_multiladder_case}\\
\mathcal{I}_{\rm L}&\equiv \{\bm{n}\in\mathbb{Z}^{N_{\rm L}}_{\ge 0}~|~ |\tilde{S}_{\bm{n}}\rangle\not=0 \}.\label{eq:definition_of_set_I_L}
\end{align}

The RSGA guarantees the existence of the tower of exact eigenstates, but does not make explicit the structure of a local Hamiltonian supporting that tower. In the next subsection, we review the symmetry-based formalism, which provides such a constructive decomposition of the Hamiltonian.

\subsection{Symmetry-based formalism}\label{subsec:Symmetry-based formalism}
The RSGA reviewed above provides algebraic conditions under which a set of states generated by ladder operators becomes an exact tower of energy eigenstates. The symmetry-based formalism~\cite{ODea2020from} gives a complementary, constructive viewpoint: it decomposes the Hamiltonian into terms with distinct roles, thereby making explicit how a scar subspace can be embedded in an otherwise thermal many-body spectrum. In this subsection, we review this formalism for both the single-ladder and multiladder cases.

\subsubsection{Single-ladder case}\label{subsubsec:Symmetry-based formalism_single_ladder}
We first consider the single-ladder case. In the symmetry-based
formalism, the Hamiltonian is decomposed as
\begin{align}
\hat{H}=\hat{H}_{\rm A}+\hat{H}_{\rm SG}+\hat{H}_{\rm sym}.\label{eq:symmetry_based_formalism_Hamiltonian}
\end{align}
The first term $\hat{H}_{\rm A}$ annihilates all QMBS states:
\begin{align}
\hat{H}_{\rm A}\cket{S_n}=0,\quad \text{for }n\in\mathcal{I}.\label{eq:definition_of_annihilation_part}
\end{align}
The second term, \(\hat H_{\rm SG}\), is the spectrum-generating part. We introduce a Hermitian onsite symmetry generator
\begin{align}
\hat{Q}^z\equiv \sum_j\hat{q}_j^z,\label{eq:definition_of_Qz}
\end{align}
where $\hat{q}_j^z$ is a Hermitian single-site operator with
\(\mathrm{supp}(\hat q_j^z)=\{j\}\). We assume
\begin{align}
[\hat{Q}^{z}, \hat{Q}^{\dagger}]=\hat{Q}^{\dagger}.\label{eq:commutation_relation_Qz_and_Qdagger_single_ladder}
\end{align}
If the root state is an eigenstate of $\hat Q^z$,
\begin{align}
\hat{Q}^z\cket{S_0}=Q_0^z\cket{S_0},\label{eq:S0_is_eigenstate_of_Sz}
\end{align}
then the tower states satisfy
\begin{align}
\hat Q^z\cket{S_n}=(Q_0^z+n)\cket{S_n}.\label{eq:Qz_eigenvalue_single}
\end{align}
The spectrum-generating term is chosen as~\footnote{As shown in Refs.~\cite{Kunimi2025systematic,Hashimoto2026construction}, the term $\hat H_{\rm SG}$ may include an additional operator $\hat C$. We do not consider this possibility here, because $\hat{C}$ is not used in this paper.}
\begin{align}
\hat{H}_{\rm SG}=\mathcal{E} \hat{Q}^z,\label{eq:expression_H_SG}
\end{align}
which gives
\begin{align}
[\hat{H}_{\rm SG},\hat{Q}^\dagger]=\mathcal{E}\hat{Q}^\dagger.\label{eq:SGA_for_single_ladder_operator}
\end{align}

The third term, $\hat{H}_{\rm sym}$, preserves the scar multiplet. We assume
\begin{align}
[\hat{H}_{\rm sym},\hat{Q}^z]=0,\qquad
  [\hat{H}_{\rm sym},\hat{Q}^{\dagger}]=0.\label{eq:Conditions_for_Hsym}
\end{align}
If $\cket{S_0}$ is an eigenstate of $\hat{H}_{\rm sym}$, then all states in the tower have the same contribution from $\hat{H}_{\rm sym}$. Therefore, after absorbing all $n$-independent contributions into $E_0$, the full Hamiltonian acts on the tower as
\begin{align}
\hat{H}\cket{S_n}=(E_0+n\mathcal{E})\cket{S_n}.\label{eq:reproduce_QMBS_from_symmetry_based_formalism}
\end{align}
Thus, the symmetry-based decomposition provides a constructive realization of the single-ladder RSGA within the scar subspace.

\subsubsection{Multiladder case}\label{subsubsec:Symmetry-based formalism_multi_ladder}
Next, we consider the multiladder case. In this case, the Hamiltonian is decomposed as in Eq.~(\ref{eq:symmetry_based_formalism_Hamiltonian}). The annihilating part of the Hamiltonian satisfies
\begin{align}
\hat{H}_{\rm A}\cket{S_{\bm{n}}}=0,\quad \text{for }\bm{n}\in \mathcal{I}_{\rm L}.\label{eq:H_A_for_multiladder_operator_case}
\end{align}
We introduce mutually commuting Hermitian charge operators $\{\hat{Q}_{\mu}^z\}_{\mu=1}^{N_{\rm C}}$:
\begin{align}
[\hat{Q}^z_{\mu}, \hat{Q}_{\nu}^z]=0,\quad \mu,\nu=1,\ldots,N_{\rm C}.\label{eq:mutually_commuting_charges}
\end{align}
When the ladder operators originate from a Lie-algebraic structure, the charge operators may be chosen as Cartan generators and the ladder operators as weight operators~\cite{Hall2015LieGroups}. In the charge-resolving cases considered here, different multi-indices $\bm{n}$ label distinct weight sectors. We assume that the following relations hold:
\begin{align}
[\hat{Q}_{\mu}^z, \hat{Q}_l^{\dagger}]=q_{\mu l}\hat{Q}_l^\dagger, \quad \mu=1,\ldots, N_{\rm C},\quad l=1,\ldots, N_{\rm L},\label{eq:charge_matrix_multiladder}
\end{align}
where $q_{\mu l}\in\mathbb{R}$ is the charge carried by $\hat{Q}_l^{\dagger}$ under $\hat{Q}_{\mu}^z$. The diagonal case $N_{\rm C}=N_{\rm L}$ and $q_{\mu l}=\delta_{\mu l}$ gives a one-to-one correspondence between charge and ladder directions.

We further assume that the root state is a simultaneous eigenstate of all charge operators,
\begin{align}
\hat{Q}_{\mu}^z\cket{S_{\bm 0}}=Q_{\mu,0}^z\cket{S_{\bm 0}}, \quad \mu=1,\ldots, N_{\rm C}.\label{eq:relation_Qz_multiladder_case_for_S0}
\end{align}
Then the multiladder tower states satisfy
\begin{align}
\hat{Q}_{\mu}^z\cket{S_{\bm n}}=\left(Q_{\mu,0}^z+\sum_{l=1}^{N_L}n_l q_{\mu l}\right)\cket{S_{\bm n}}.\label{eq:relation_Qz_multiladder_case_for_Sn}
\end{align}
We denote the corresponding weight vector by
\begin{align}
\bm{Q}^z(\bm{n})&\equiv \left(Q_1^z(\bm{n}), \ldots, Q_{N_{\rm C}}^z(\bm{n})\right),\notag \\
Q_{\mu}^z(\bm{n})&=Q^z_{\mu,0}+\sum_{l=1}^{N_{\rm L}}n_lq_{\mu l}.\label{eq:definition_of_weight_vector}
\end{align}
We further assume that the map $\bm{n}\to \bm{Q}^z(\bm{n})$ is injective on $\mathcal{I}_{\rm L}$. Under this charge-resolving condition, different multi-indices label distinct weight sectors. The spectrum-generating part is chosen as
\begin{align}
\hat{H}_{\rm SG}=\sum_{\mu=1}^{N_{\rm C}}h_{\mu} \hat{Q}_{\mu}^z,\quad h_{\mu}\in\mathbb{R}.\label{eq:HSG_multiladder}
\end{align}
Using Eq.~(\ref{eq:charge_matrix_multiladder}), we obtain
\begin{align}
[\hat{H}_{\rm SG},\hat{Q}_l^{\dagger}]=\mathcal{E}_l\hat{Q}_l^{\dagger},\quad \mathcal{E}_l=\sum_{\mu=1}^{N_{\rm C}}h_{\mu} q_{\mu l}.\label{eq:HSG_MLRSGA_increment}
\end{align}
The third term, $\hat{H}_{\rm sym}$, is assumed to satisfy
\begin{align}
[\hat{H}_{\rm sym},\hat{Q}_{\mu}^z]=0,\quad [\hat{H}_{\rm sym},\hat{Q}_l^{\dagger}]=0,\label{eq:conditions_Hsym_for_multiladder_case}
\end{align}
for all $\mu$ and $l$. If $\hat{H}_{\rm sym}$ contributes only an $\bm{n}$-independent energy shift within the scar subspace, the full Hamiltonian satisfies
\begin{align}
\hat{H}\cket{S_{\bm n}}=\left(E_0+\sum_{l=1}^{N_{\rm L}}n_l\mathcal{E}_l\right)\cket{S_{\bm n}}.\label{eq:reproduce_tower_eigenstate_multiladder}
\end{align}
Thus, the symmetry-based formalism gives a constructive realization of the MLRSGA whenever the MLRSGA energy increments can be represented as $\mathcal{E}_l=\sum_{\mu} h_{\mu} q_{\mu l}$.

\subsection{Systematic construction of asymptotic quantum many-body scar states}\label{subsec:systematic_construction}
In this subsection, we review a systematic construction method for AQMBS states~\cite{Kunimi2025systematic,Hashimoto2026construction}. This method is based on constructing a parent Hamiltonian whose ground-state manifold contains the QMBS states. If the parent Hamiltonian is gapless, its low-energy excitations provide candidates for AQMBS states of the original Hamiltonian. In the following, we formulate the systematic construction of AQMBS states in the multiladder case. The single-ladder case is recovered by setting $N_{\rm L}=1$.

The key ingredient of the construction is to introduce an enlarged scar subspace $\mathcal{H}_P$ as~\cite{Kunimi2025systematic,Hashimoto2026construction}
\begin{align}
\mathcal{H}_P&\equiv \bigoplus_{\bm{n}\in \mathcal{I}_{\rm L}}\mathcal{H}_{P_{\bm{n}}},\label{eq:definition_enlarged_scar_subspace}\\
\mathcal{H}_{P_{\bm{n}}}&\equiv {\rm Span}\{\cket{\bm{\eta}}\in\mathcal{H} ~|~ \hat{Q}^z_{\mu}\cket{\bm{\eta}}=Q_{\mu}^z(\bm{n})\cket{\bm{\eta}}\text{ for all }\mu,\notag \\
&\hspace{16.0em}\bracket{\bm{\eta}}{S_{\bm{n}}}\not=0\},\label{eq:definition_of_enlarged_subspace_H_P_n}
\end{align}
where $\mathcal{H}$ is the full Hilbert space, and $\cket{\bm{\eta}}$ denotes a product-basis state that is a simultaneous eigenstate of all $\{\hat{Q}^z_{\mu}\}_{\mu=1}^{N_{\rm C}}$. Because different $\bm{n}$'s correspond to distinct weight sectors, the subspaces $\mathcal{H}_{P_{\bm{n}}}$ are mutually orthogonal. We note that the scar subspace defined in Eq.~(\ref{eq:definition_of_scar_subspace_multiladder_case}) is a subspace of the enlarged scar subspace. We also introduce the complement of the enlarged scar subspace $\mathcal{H}_{\rm Q}\equiv \mathcal{H}_P^{\perp}$.

Here, we introduce the projector onto $\mathcal{H}_{P_{\bm{n}}}$ as $\hat{\mathcal{P}}_{\bm{n}}$. The projector onto the total enlarged scar subspace is defined by
\begin{align}
\hat{\mathcal{P}}\equiv \sum_{\bm{n}\in\mathcal{I}_{\rm L}}\hat{\mathcal{P}}_{\bm{n}}.\label{eq:definition_of_projection_operator_onto_total_enlarged_scar_subspace}
\end{align}
The projector onto $\mathcal{H}_Q$ is defined as $\hat{\mathcal{Q}}\equiv \hat{1}-\hat{\mathcal{P}}$, where $\hat{1}$ is the identity operator in $\mathcal{H}$. 

In the systematic construction of AQMBS states, we impose the following structural assumptions on the annihilating part of the Hamiltonian. We decompose $\hat{H}_{\rm A}$ as
\begin{align}
\hat{H}_{\rm A}&\equiv \hat{H}_0+\hat{H}_{\rm p}',\label{eq:decompose_HA_into_H0_and_Hp'}\\
\hat{H}_0&\equiv \sum_j\hat{h}_j,\label{eq:H0_as_a_sum_of_local_Hamiltonian}\\
\hat{H}_{\rm p}'&\equiv \sum_{\bm{n}\in\mathcal{I}_{\rm L}}c_{\bm{n}}\hat{\mathcal{P}}_{\bm{n}}\hat{H}_0^2\hat{\mathcal{P}}_{\bm{n}},\label{eq:definition_of_Hp'}
\end{align}
where $\hat{h}_j$ is a local Hermitian operator and $c_{\bm{n}}$ is a real constant with dimensions of inverse energy. We further assume 
\begin{align}
\hat{\mathcal{P}}\hat{h}_j\hat{\mathcal{P}}=0,\quad \text{for all }j.\label{eq:assumptions_hj}
\end{align}

We define the parent Hamiltonian $\hat{H}_{\rm p}$ as
\begin{align}
\hat{H}_{\rm p}&\equiv \hat{\mathcal{P}}\hat{H}_0^2\hat{\mathcal{P}}.\label{eq:definition_of_the_parent_Hamiltonian}
\end{align}
From the definition, $\hat{H}_{\rm p}$ is positive semidefinite ($\hat{H}_{\rm p}\ge0$). Together with the annihilation condition $\hat{H}_{\rm A}\cket{S_{\bm n}}=0$ and the orthogonal decomposition $\mathcal{H}=\mathcal{H}_P\oplus\mathcal{H}_Q$, the condition $\hat{\mathcal{P}}\hat{H}_0\hat{\mathcal{P}}=0$ implies $\hat{H}_0\cket{S_{\bm n}}=0$. We obtain 
\begin{align}
\hat{H}_{\rm p}\cket{S_{\bm{n}}}=0,\quad \text{for }\bm{n}\in\mathcal{I}_{\rm L}.\label{eq:scar_states_are_the_ground_state_of_Hp}
\end{align}
Therefore, the QMBS states are zero-energy ground states of $\hat{H}_{\rm p}$. We further assume that the off-diagonal blocks of the parent Hamiltonian are zero:
\begin{align}
\hat{\mathcal{P}}_{\bm{n}}\hat{H}_0^2\hat{\mathcal{P}}_{\bm{m}}=0,\quad \text{for all }\bm{n}\not=\bm{m}.\label{eq:assumptions_off-diagonal_elements_of_the_parent_Hamiltonian}
\end{align}
The parent Hamiltonian then takes the block-diagonal form:
\begin{align}
\hat{H}_{\rm p}=\sum_{\bm{n}\in\mathcal{I}_{\rm L}}\hat{\mathcal{P}}_{\bm{n}}\hat{H}_0^2\hat{\mathcal{P}}_{\bm{n}}\equiv \sum_{\bm{n}\in\mathcal{I}_{\rm L}}\hat{H}_{\rm p}^{(\bm{n})}.\label{eq:diagonal_form_of_parant_Hamiltonian}
\end{align}

We now evaluate the energy variance. Let $\cket{\varphi_{\bm{n}}}$ be a normalized state in $\mathcal{H}_{P_{\bm{n}}}$. In deriving the energy variance below, we assume that $\hat{H}_{\rm SG}$ and $\hat{H}_{\rm sym}$ act as constants within each enlarged scar subspace $\mathcal{H}_{P_{\bm{n}}}$. The energy variance for $\cket{\varphi_{\bm{n}}}$ becomes
\begin{align}
&\Delta E_{\bm{n}}^2\notag \\
&=\bra{\varphi_{\bm{n}}}\hat{H}^2\cket{\varphi_{\bm{n}}}-(\bra{\varphi_{\bm{n}}}\hat{H}\cket{\varphi_{\bm{n}}})^2\notag \\
&=\bra{\varphi_{\bm{n}}}\hat{H}^2_0\cket{\varphi_{\bm{n}}}\notag \\
&+c_{\bm{n}}^2\left[\bra{\varphi_{\bm{n}}}(\hat{\mathcal{P}}_{\bm{n}}\hat{H}_0^2\hat{\mathcal{P}}_{\bm{n}})^2\cket{\varphi_{\bm{n}}}-(\bra{\varphi_{\bm{n}}}\hat{\mathcal{P}}_{\bm{n}}\hat{H}_0^2\hat{\mathcal{P}}_{\bm{n}}\cket{\varphi_{\bm{n}}})^2\right],\label{eq:energy_variance_for_H_Pn}
\end{align}
where we have used the assumption (\ref{eq:assumptions_hj}). If we choose $\cket{\varphi_{\bm{n}}}$ to be an eigenstate of $\hat{\mathcal{P}}_{\bm{n}}\hat{H}_0^2\hat{\mathcal{P}}_{\bm{n}}$, the bracketed term in Eq.~(\ref{eq:energy_variance_for_H_Pn}) vanishes. Therefore, we obtain
\begin{align}
\Delta E_{\bm{n}}^2=\bra{AS_{\bm{n}}}\hat{\mathcal{P}}_{\bm{n}}\hat{H}_0^2\hat{\mathcal{P}}_{\bm{n}}\cket{AS_{\bm{n}}},\label{eq:energy_variance_optimal_form}
\end{align}
where $\cket{AS_{\bm{n}}}$ is a normalized eigenstate of $\hat{\mathcal{P}}_{\bm{n}}\hat{H}_0^2\hat{\mathcal{P}}_{\bm{n}}$. 
This result shows that, if $\hat{H}_{\rm p}^{(\bm{n})}$ has a sequence of normalized excited eigenstates whose eigenvalues vanish in the thermodynamic limit, then their energy variances with respect to the original Hamiltonian also vanish. Provided that these states are orthogonal to the QMBS states and remain nonthermal, they can be identified as AQMBS states.

\section{Proof of the existence of asymptotic rainbow quantum many-body scar states}\label{sec:Proof_of_ARQMBS}
In this section, we rigorously prove the existence of ARQMBS states using the systematic construction reviewed in the previous section. We begin by introducing RQMBS states~\cite{Langlett2022rainbow,Wildeboer2022quantum}, a special class of QMBS states that arise in systems with a doubled Hilbert-space structure, such as two-leg ladders and bilayer systems. Their defining feature is maximal entanglement across the leg, or layer, bipartition: after tracing out one leg or layer, the reduced density matrix of the remaining subsystem is maximally mixed on its support. We then establish the existence of ARQMBS states in spin-$S$ systems, multicomponent fermionic systems, and multicomponent bosonic systems. Concrete examples will be presented in Sec.~\ref{sec:Example}.

\subsection{Setup}\label{subsec:setup}
Here, we introduce rainbow QMBS states. Throughout this section, the total Hilbert space is given by $\mathcal{H}_{\rm tot}=\mathcal{H}\otimes \mathcal{H}$, where $\mathcal{H}$ is the Hilbert space of each subsystem. The subsystems are labeled by $1$ and $2$. The sites of each subsystem form a $d$-dimensional Bravais lattice. The position of the $j$th lattice site of the subsystem $\alpha$ is given by $\bm{R}_{j}\equiv \sum_{l=1}^ds_{j}^{(l)}\bm{a}_l$, where $s_{j}^{(l)}\in\mathbb{Z}$ and $\bm{a}_{l}$ are the primitive lattice vectors. We also introduce reciprocal lattice vectors $\{\bm{b}_l\}$ satisfying $\bm{a}_l\cdot\bm{b}_{l'}=2\pi\delta_{l,l'}$. Let $M$ denote the total number of lattice sites in each subsystem. Let $M_l$ denote the number of lattice sites along the $l$th direction. The total number of lattice sites is then given by $M=\prod_{l=1}^dM_l$. 

We consider a Hamiltonian of the form~\cite{Langlett2022rainbow,Wildeboer2022quantum}
\begin{align}
\hat{H}_{\rm tot}&\equiv \hat{H}\otimes\hat{1}-\hat{1}\otimes\hat{H}^{\ast}+\hat{H}_{\rm int}\notag \\
&\equiv \hat{H}_0+\hat{H}_{\rm int}.\label{eq:Hamiltonian_rainbow_QMBS}
\end{align}
Here, $\hat{H}$ acts on one subsystem, whereas $-\hat{H}^{\ast}$ acts on the other subsystem. The symbol ${}^{\ast}$ denotes complex conjugation in the product basis $\{|\bm{n}\rangle\}$, $\hat{H}_{\rm int}$ is the interaction Hamiltonian between the two subsystems, and $\hat{1}$ is the identity operator on $\mathcal{H}$. We assume that the system has a $U(1)\times U(1)$ symmetry associated with charge conservation in the two subsystems. We set $\hat{U}=e^{-i\hat{N}\theta}$, where $\hat{N}$ is the charge operator and $\theta$ is a real number. The $U(1)\times U(1)$ symmetry is equivalent to
\begin{align}
[\hat{H}, \hat{U}]&=0,\quad [\hat{H}^{\ast}, \hat{U}]=0,\label{eq:commutation_relation_H_and_U}\\
[\hat{H}_{\rm int}, \hat{U}\otimes\hat{1}]&=0,\quad [\hat{H}_{\rm int}, \hat{1}\otimes\hat{U}]=0.\label{eq:commutation_relation_Hint_and_U}
\end{align}
This means that the charge in each subsystem is conserved separately. In Sec.~\ref{subsec:spin-S}, we consider $U(1)\times U(1)$ symmetry, and in Secs.~\ref{subsec:multicomponent_Fermion} and \ref{subsec:multicomponent_Boson}, we treat higher symmetry such as $U(1)^f\times U(1)^f$ symmetry for $f\in\mathbb{Z}_{>0}$. 

In this paper, we restrict the functional form of the interaction Hamiltonian $\hat{H}_{\rm int}$ to
\begin{align}
\hat{H}_{\rm int}&\equiv -\frac{\hbar}{2}\sum_j(\hat{L}_j\otimes\hat{1}-\hat{1}\otimes\hat{L}_j)^2,\label{eq:interaction_Hamiltonian_between_ladders}
\end{align}
where $\hat{L}_j$ is a function of local charge-density operators and has dimensions of $[T]^{-1/2}$. For example, in the $U(1)$-symmetric spin system, $\hat{L}_j$ is a function of $\{\hat{S}_j^z\}$, where $\hat{S}_j^z$ is the spin operator along the $z$ direction at site $j$, and for Hubbard-like systems, $\hat{L}_j$ is a function of $\{\hat{n}_{j}\}$, where $\hat{n}_{j}$ is the number operator at site $j$. This means that the interaction terms have only diagonal elements. 

We can show that the Hamiltonian (\ref{eq:Hamiltonian_rainbow_QMBS}) has exact energy eigenstates~\cite{Wildeboer2022quantum}:
\begin{align}
\hat{H}_{\rm tot}|\tilde{S}_n\rangle&=0,\label{eq:RQMBS_for_general_case}\\
|\tilde{S}_n\rangle&\equiv n!\sideset{}{'}\sum_{\bm{n}}\cket{\bm{n}}\otimes\cket{\bm{n}},\label{eq:expression_of_RQMBS}
\end{align}
where $|\bm{n}\rangle$ is a product-basis state and the primed sum in Eq.~(\ref{eq:expression_of_RQMBS}) runs over the basis states satisfying the condition $\hat{N}|\bm{n}\rangle=N_n|\bm{n}\rangle$. The state $|\tilde{S}_n\rangle$ is an unnormalized rainbow QMBS state. A proof is given in Appendix~\ref{app:proof_of_RQMBS}.

We also define the locality of operators following Ref.~\cite{ODea2026locality}, which will be used throughout the rest of this paper. An operator
\begin{align}
\hat{O}=\sum_j \hat{o}_j,\label{eq:definition_of_hat_O}
\end{align}
is called $r$-local if each local term $\hat o_j$ has support $X_j=\mathrm{supp}(\hat{o}_j)$ containing at most $r$ sites, i.e.,
\begin{align}
|X_j|\le r,\quad \text{for all }j.\label{eq:definition_of_r-local}
\end{align}

We next define the range of an operator. Let $d_{\rm lat}(\bm{R}_j,\bm{R}_k)$ be the graph-theoretic distance on a single subsystem, namely the minimum number of bonds connecting the two sites. In the doubled system, we define
\begin{align}
d_{\rm G}\left((\bm{R}_{j},\alpha),(\bm{R}_{k},\beta)\right)=d_{\rm lat}(\bm{R}_j,\bm{R}_k)+(1-\delta_{\alpha,\beta}),\label{eq:defimition_of_graph_theoretic_distance_in_doubled_state}
\end{align}
where $\alpha,\beta=1,2$ are the labels of subsystems. For a set of sites
\begin{align}
X=\{(\bm{R}_{j_1},\alpha_1),\ldots,(\bm{R}_{j_x},\alpha_x)\},\label{eq:definition_of_set_of_sites}
\end{align}
we define its diameter by
\begin{align}
\mathrm{diam}(X)=1+\max_{(\bm R_{j},\alpha),(\bm R_{k},\beta)\in X}
  d_G\left((\bm{R}_j,\alpha),(\bm{R}_k,\beta)\right).
\end{align}
We say that $\hat{O}=\sum_j\hat{o}_j$ has range $R$ if
\begin{align}
\mathrm{diam}\left(\mathrm{supp}(\hat{o}_j)\right)\le R,\quad\text{for all }j.\label{eq:definition_of_Range_R}
\end{align}
For example, we consider one-dimensional systems. The operator $\hat{S}_{1,1}^z\hat{S}_{2,1}^z$ is a $2$-local operator of range $2$. Another example is $\hat{S}_{1,1}^z\hat{S}_{100,1}^z$. In this case, the operator is $2$-local and has range $100$.

We assume that the Hamiltonian $\hat{H}_{\rm tot}$ is $r_0$-local and has range $R_0$ and both $r_0$ and $R_0$ are $O(1)$. Then, $\hat{H}_{\rm tot}$ can be written as the sum of local Hermitian operators
\begin{align}
\hat{H}_{\rm tot}=\sum_j\hat{h}_{{\rm tot},j}.\label{eq:Hamiltonian_sum_of_local_operators}
\end{align}

\subsection{Spin-$S$ systems}\label{subsec:spin-S}
In this subsection, we prove the existence of ARQMBS states in spin-$S$ systems.

\subsubsection{Rainbow quantum many-body scar states}\label{subsubsec:RQMBS_spin-S}
Here, we show the existence of RQMBS states in spin-$S$ systems. Let $\hat{S}_j^{\mu}\;(\mu=x,y,z)$ be a spin-$S$ operator at site $j$. We define the total magnetization operator $\hat{S}_{\rm tot}^z\equiv \sum_j\hat{S}_j^z$ of the subsystem. The Hamiltonian satisfies
\begin{align}
[\hat{H}, \hat{S}_{\rm tot}^z]&=0,\label{eq:assumption_Hamiltonian_spin-S}\\
[\hat{L}_j, \hat{S}_{\rm tot}^z]&=0,\quad\text{for all }j.\label{eq:assumption_L_spin^S}
\end{align}
For convenience, we introduce the charge-density operator $\hat{n}_j\equiv \hat{S}_j^z+S$ and
\begin{align}
\hat{N}_{\rm tot}&\equiv \frac{1}{2}(\hat{N}\otimes\hat{1}+\hat{1}\otimes\hat{N}),\label{eq:definition_of_charge_operator_spin-S}\\
\hat{N}&\equiv \sum_{j=1}^M\hat{n}_j.\label{eq:definition_of_charge_subsystem}
\end{align}
In this subsection, we use the product basis $\cket{\bm{n}}\equiv |n_1,n_2,\ldots, n_M\rangle$ satisfying $\hat{S}_j^z\cket{\bm{n}}=n_j\cket{\bm{n}}$ for $n_j=-S,-S+1,\ldots, S$. The Hamiltonian $\hat{H}$ commutes with $\hat{N}$ and each $\hat{L}_j$ is a function of the local $\{\hat{S}_l^z\}$ operators.

As shown in Sec.~\ref{subsec:setup}, the rainbow QMBS state $|\tilde{S}_n\rangle$ is an energy eigenstate of the system. In the spin-$S$ case, the rainbow QMBS state takes the form
\begin{align}
|\tilde{S}_n\rangle\equiv n!\sum_{\bm{n}:\hat{N}\cket{\bm{n}}=n\cket{\bm{n}}}\cket{\bm{n}}\otimes\cket{\bm{n}},\label{eq:rainbow_QMBS_for_spin-S}
\end{align}
where the sum runs over all product-basis states satisfying $\hat{N}\cket{\bm{n}}=n\cket{\bm{n}}$. Here, we show that the rainbow QMBS states $|\tilde{S}_n\rangle$ can be generated by repeatedly applying a ladder operator $\hat{Q}^{\dagger}$ to a root state $\cket{S_0}$ and that they satisfy the RSGA. We define
\begin{align}
\cket{S_0}\equiv \cket{-S,-S,\ldots,-S}\otimes\cket{-S,-S,\ldots,-S}.\label{eq:definition_of_root_state_spin-S}
\end{align}
Direct calculations show that $\hat{H}_{\rm tot}\cket{S_0}=0$ and $\hat{N}_{\rm tot}\cket{S_0}=0$. We choose $\cket{S_0}$ as the root state.

We define the operator
\begin{align}
\hat{Q}^{\dagger}&\equiv \sum_{j=1}^M\hat{q}_j^{\dagger},\label{eq:definition_of_Qdagger_spin-S}\\
\hat{q}_j^{\dagger}&\equiv \sum_{m=-S}^{S-1}(m+S+1)(\hat{F}_j^{m+1,m}\otimes\hat{F}_j^{m+1,m}),\label{eq:definition_of_small_qj_spin-S}\\
\hat{F}_j^{\mu,\nu}&\equiv \cket{\mu}\bra{\nu}_j,\label{eq:definition_of_F_spin-S}
\end{align}
where $\cket{\mu}\bra{\nu}_j$ represents the operator $\cket{\mu}\bra{\nu}$ acting on site $j$. Direct calculations show that
\begin{align}
\hat{q}_j^{\dagger}\cket{m}_j\otimes\cket{m}_j&=(m+S+1)\cket{m+1}_j\otimes\cket{m+1}_j.\label{eq:qj_dagger_acting_on_m_m_state_spin-S}
\end{align}
Using the above operator, we obtain
\begin{align}
\hat{Q}^{\dagger}\cket{S_0}&=|\tilde{S}_1\rangle.\label{eq:relation_Qdagger_to_S0_spin-S}
\end{align}
We can prove $(\hat{Q}^{\dagger})^n\cket{S_0}=|\tilde{S}_n\rangle$ by induction. The details of the proof are given in Appendix~\ref{app:detail_of_Qdagger_to_S0}. Therefore, we obtain 
\begin{align}
\hat{H}_{\rm tot}|\tilde{S}_n\rangle=0,\quad \text{for }0\le n\le 2SM.\label{eq:RQMBS_for_spin-S_case}
\end{align}
For $n>2SM$, $|\tilde{S}_n\rangle=0$ because $(\hat{Q}^{\dagger})^{2SM}|S_0\rangle\propto\cket{S,S,\ldots, S}\otimes\cket{S,S,\ldots, S}$ and $(\hat{Q}^{\dagger})^{2SM+1}|S_0\rangle=0$.

From Eq.~(\ref{eq:RQMBS_for_spin-S_case}), we can show that the RSGA conditions (\ref{eq:RSGA-n_condition1})--(\ref{eq:RSGA-n_condition3}) are satisfied. The remaining relation is the operator identity (\ref{eq:RSGA-n_condition4}). Using the $r_0$-locality of the Hamiltonian and the nilpotency of the ladder operator, we show that the operator identity holds at order at most $4Sr_0$:
\begin{align}
\underbrace{[[[[\hat{H}_{\rm tot},\hat{Q}^{\dagger}],\hat{Q}^{\dagger}],\ldots],\ldots]}_{4Sr_0+1\text{ times}}&=0.\label{eq:RSGA-n_condition4_spin-S}
\end{align}
The proof, which follows the method of Ref.~\cite{ODea2026locality}, is given in Appendix~\ref{app:proof_of_operator_identity_in_RSGA}.

\subsubsection{Parent Hamiltonian}\label{subsubsec:parent_Hamiltonian_spin-S}

The above results show that the rainbow-scar construction satisfies the RSGA conditions, which is a necessary condition for applying the systematic construction of AQMBS reviewed in Sec.~\ref{subsec:systematic_construction}. The enlarged scar subspace for the spin-$S$ systems is
\begin{align}
\mathcal{H}_{P_n}&\equiv {\rm Span}\{\cket{\bm{n}}\otimes\cket{\bm{n}}~|~ \cket{\bm{n}}\in\mathcal{H}, \hat{N}\cket{\bm{n}}=n\cket{\bm{n}} \},\label{eq:definition_of_enlarged_scar_subspace_Hn_spin-S}\\
\mathcal{H}_{P}&\equiv \bigoplus_{n\in\mathcal{I}}\mathcal{H}_{P_n}.\label{eq:definition_of_total_enlarged_scar_subspace_spin-S}
\end{align}
We introduce the projectors onto $\mathcal{H}_{P_n}$ and $\mathcal{H}_P$ as $\hat{\mathcal{P}}_n$ and $\hat{\mathcal{P}}$, respectively. Direct calculations show $\hat{H}_{\rm int}\hat{\mathcal{P}}=0$. Thus, $\hat{H}_{\rm int}$ annihilates the entire enlarged subspace and does not contribute to the energy expectation value or variance of states in this subspace. We can also show that
\begin{align}
\hat{\mathcal{P}}(\hat{h}_{j}\otimes \hat{1}-\hat{1}\otimes\hat{h}_j^{\ast})\hat{\mathcal{P}}=0.\label{eq:condition_PhjP=0_for_spin-S}
\end{align}
This means that the part of the Hamiltonian $\hat{H}\otimes\hat{1}-\hat{1}\otimes\hat{H}^{\ast}$ can be regarded as $\hat{H}_0$ in the symmetry-based formalism. Therefore, we can construct the parent Hamiltonian as
\begin{align}
\hat{H}_{\rm p}=\hat{\mathcal{P}}(\hat{H}\otimes\hat{1}-\hat{1}\otimes\hat{H}^{\ast})^2\hat{\mathcal{P}}.\label{eq:parent_Hamiltonian_Spin-S_case}
\end{align}

We can show that the parent Hamiltonian in Eq.~(\ref{eq:parent_Hamiltonian_Spin-S_case}) is a Rokhsar--Kivelson-type frustration-free Hamiltonian~\cite{Rokhsar1988superconductivity}. Using the projection operator
\begin{align}
\hat{\mathcal{P}}=\sum_{\bm{n}}(\cket{\bm{n}}\otimes\cket{\bm{n}})(\bra{\bm{n}}\otimes\bra{\bm{n}}),\label{eq:expression_of_projection_operator_spin-S}
\end{align}
we obtain
\begin{align}
\hat{H}_{\rm p}&=2\sum_{\bm{n},\bm{m}}|H_{\bm{n},\bm{m}}|^2\hat{P}_{\bm{n},\bm{m}},\label{eq:parent_Hamiltonian_bracket_form_spin-S}\\
H_{\bm{n},\bm{m}}&\equiv \bra{\bm{n}}\hat{H}\cket{\bm{m}},\label{eq:definition_of_H_n_m_spin-S}\\
\hat{P}_{\bm{n},\bm{m}}&\equiv \frac{1}{2}(\cket{\bm{n}}\otimes\cket{\bm{n}}-\cket{\bm{m}}\otimes\cket{\bm{m}})\notag \\
&\quad \times (\bra{\bm{n}}\otimes\bra{\bm{n}}-\bra{\bm{m}}\otimes\bra{\bm{m}}).\label{eq:definition_of_operator_P_spin-S}
\end{align}
Because $\hat{P}_{\bm{n},\bm{m}}|\tilde{S}_n\rangle=0$ for every pair with $H_{\bm{n},\bm{m}}\not=0$ and $\hat{P}_{\bm{n},\bm{m}}$ is positive semidefinite, the parent Hamiltonian is frustration-free. 
When regarded as an operator on the enlarged scar subspace $\mathcal{H}_{P}$, the parent Hamiltonian admits the local decomposition:
\begin{align}
\hat{H}_{\rm p}=\sum_j\hat{h}_{{\rm p},j},\label{eq:parent_Hamiltonian_local_sum_form}
\end{align}
where $\hat{h}_{{\rm p},j}$ is a Hermitian $r'$-local operator of range $R'$ $[r',R'=O(1)]$ and satisfies $\hat{h}_{{\rm p},j}|\tilde{S}_n\rangle=0$ for any $j$ and $n$. The proof is given in Appendix~\ref{app:proof_locality_parent_Hamitonian_spin-S}.

\subsubsection{Asymptotic rainbow quantum many-body scar states}\label{subsubsec:ARQMBS_spin-S}

We next show that the parent Hamiltonian is gapless. As discussed in Sec.~\ref{subsec:systematic_construction}, gapless excited states of the parent Hamiltonian are candidates for AQMBS states. To prove this, we further assume that the restriction of $\hat{H}_{\mathrm{p}}$ to each enlarged charge sector $\mathcal{H}_{P_{n}}$ has a unique zero-energy ground state, given by the normalized RQMBS state $|S_{n}\rangle$. We emphasize that the uniqueness assumption is needed only to establish the gaplessness of the parent Hamiltonian. The ARQMBS properties of the variational states themselves do not rely on this assumption. In addition, we assume that the number $N_{\rm p}(M)$ of local terms in Eq.~(\ref{eq:parent_Hamiltonian_local_sum_form}) and the local terms of the parent Hamiltonian satisfy
\begin{align}
N_{\rm p}(M)\le c_{\rm p}M, \text{ with }c_{\rm p}=O(1),\label{eq:assumption_number_of_hpj}\\
\| \hat{h}_{{\rm p},j}\|_{\rm op}\le J_{\rm p}=O(1),\label{eq:assumption_local_term_parent_Hamiltonian}
\end{align}
where the operator norm is defined by~\cite{Tasaki2020physics}
\begin{align}
\|\hat{A}\|_{\rm op}\equiv \sup_{\substack{\cket{\psi}\in\mathcal{H}_{\rm tot} \\ \cket{\psi}\not=0}}\frac{\|\hat{A}\cket{\psi}\|}{\|\cket{\psi}\|}.\label{eq:definition_of_operator_norm_Spin-S}
\end{align}

The proof is based on the single-mode approximation (SMA)~\cite{Bijl1940lowest,Feynman1954atomic,Girvin1986magneto,Arovas1988extended,Han2025models}. In the SMA, we obtain a variational estimate of low-energy excitations. We define a variational state for $1\le n\le 2SM-1$~\cite{Gotta2023Asymptotic,Kunimi2024proposal,Kunimi2025systematic}:
\begin{align}
|\widetilde{AS}_n\rangle&\equiv \hat{Q}^{\dagger}(\bm{k})|\tilde{S}_{n-1}\rangle=\frac{1}{n}\hat{\rho}(\bm{k})|\tilde{S}_n\rangle,\label{eq:definition_of_AS_n_spin-S}\\
\hat{Q}^{\dagger}(\bm{k})&\equiv \sum_{j=1}^Me^{i\bm{k}\cdot\bm{R}_j}\hat{q}_j^{\dagger},\label{eq:definition_of_Qdagger_of_k_spin-S}\\
\hat{\rho}(\bm{k})&\equiv \sum_{j=1}^Me^{i\bm{k}\cdot\bm{R}_j}(\hat{n}_j\otimes \hat{1}),\label{eq:definition_of_rho_of_k_spin-S}
\end{align}
where $\bm{k}$ is a $d$-dimensional wave vector. Here, the second equality in Eq.~(\ref{eq:definition_of_AS_n_spin-S}) can be derived using an argument similar to that used for Eq.~(\ref{eq:proof_of_Qdagger_to_n_to_S0_calculation}). We take the finite Bravais lattice to be $\bm{R}_j=\sum_{l=1}^ds_j^{(l)}\bm{a}_l$ for $s_j^{(l)}=0,1,\ldots, M_l-1$, where $M=\prod_{l=1}^dM_l$. We choose the wave vector $\bm{k}$ as
\begin{align}
\bm{k}\equiv \sum_{l=1}^d\frac{\nu_l}{M_l}\bm{b}_l,\label{eq:definition_of_crystal_momentum_spin-S}
\end{align}
where $\nu_l$ is an integer. We assume that at least one $\nu_{l}$ is not divisible by $M_{l}$ unless stated otherwise.

The energy variance of the corresponding normalized variational state $|AS_n\rangle\equiv |\widetilde{AS}_n\rangle/\sqrt{\langle \widetilde{AS}_n|\widetilde{AS}_n\rangle}$ can be written as
\begin{align}
\Delta E_n^2&=\frac{\langle\widetilde{AS}_n| \hat{H}_{\rm p}|\widetilde{AS}_n\rangle}{\langle\widetilde{AS}_n|\widetilde{AS}_n\rangle}\equiv \frac{f_{\rm n}(\bm{k})}{f_{\rm d}(\bm{k})},\label{eq:expression_energy_variance_for_AS_n}
\end{align}
where we used $|\widetilde{AS}_n\rangle\in \mathcal{H}_{P_n}$, $\hat{\mathcal{P}}\hat{H}_0\hat{\mathcal{P}}=0$, and $\hat{H}_{\rm int}\hat{\mathcal{P}}=0$. 

Let $\Delta_{{\rm p},n}$ denote the energy gap, i.e., the smallest positive eigenvalue of the parent Hamiltonian within $\mathcal{H}_{P_n}$. We evaluate $\Delta_{{\rm p},n}$ using the SMA, which is a variational method for estimating the lowest excitation energy. The variational state must be orthogonal to the ground state. As shown in Appendix~\ref{app:proof_gaplessness_parent_Hamitonian_spin-S}, $|\widetilde{AS}_n\rangle$ is orthogonal to $|\tilde{S}_n\rangle$. Under the uniqueness assumptions stated above, the variational principle gives
\begin{align}
\Delta_{{\rm p},n}\le \frac{f_{\rm n}(\bm{k})}{f_{\rm d}(\bm{k})}=\Delta E_n^2.\label{eq:bound_gap_and_energy_variance}
\end{align}
An upper bound on the parent-Hamiltonian gap is given by
\begin{align}
\Delta_{{\rm p},n}\le \Delta E_n^2&\le C_4|\bm{k}|^2,\label{eq:inequality_for_energy_variance}
\end{align}
where $C_4=O(1)$ under the assumption $\epsilon\le n/M\le 2S-\epsilon$ for a positive system-size-independent constant $\epsilon$. The proof is given in Appendix \ref{app:proof_gaplessness_parent_Hamitonian_spin-S}. Choosing a primitive direction $a$ such that $M_{a}\to\infty$ and setting
\begin{align}
\bm{k}_{\rm min}=\frac{\bm{b}_{a}}{M_{a}},\label{eq:definition_of_kmin}
\end{align}
we obtain
\begin{align}
\Delta_{\mathrm{p},n} \leq C_{4}|\bm{k}_{\min}|^{2}\to 0.\label{eq:bound_Delta_p_n_by_kmin}
\end{align}
This proves that the parent Hamiltonian is gapless. At the same time, the energy variance of the variational state vanishes in the thermodynamic limit.

\begin{figure}[t]
\centering
\includegraphics[width=8.6cm,clip]{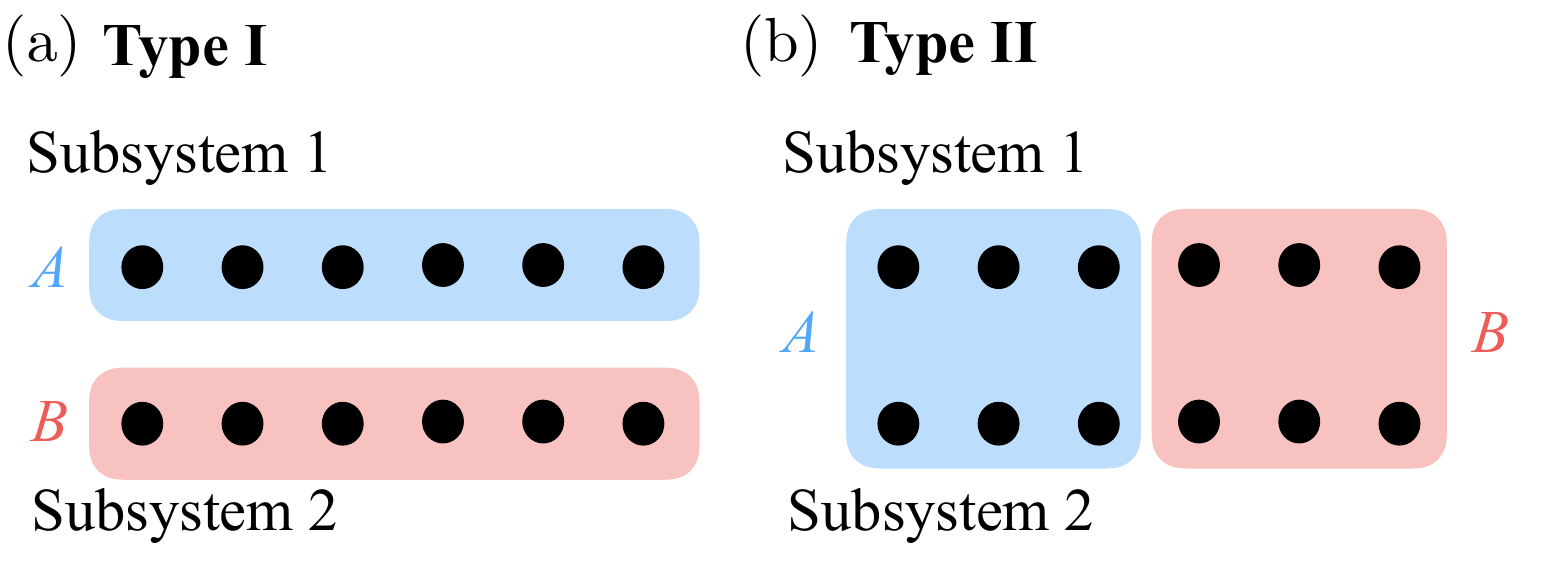}
\caption{Schematic of the two bipartitions of the system. The upper and lower black circles represent the lattice sites of the subsystems 1 and 2, respectively. (a) Type-I bipartition. The blue and red regions represent subsystems $A$ and $B$, respectively, which coincide with subsystems 1 and 2. (b) Type-II bipartition. Subsystem $A$ contains half of the sites in each of subsystems 1 and 2, and subsystem $B$ contains the remaining half. See also Fig.~1 of Ref.~\cite{Wildeboer2022quantum}.
}
\label{fig:schematic_bipartition}
\vspace{-0.75em}
\end{figure}

Finally, we discuss the entanglement entropy of the variational state $|\widetilde{AS}_n\rangle$. As discussed in Ref.~\cite{Wildeboer2022quantum}, the scaling law of entanglement entropy (EE) of RQMBS states depends on the choice of bipartition into subsystems $A$ and $B$. For the type-I bipartition shown in Fig.~\ref{fig:schematic_bipartition}(a), the EE obeys a volume law, whereas it obeys a subvolume law for the type-II bipartition shown in Fig.~\ref{fig:schematic_bipartition}(b). We next evaluate the EE of the normalized variational state $|AS_{n}\rangle$. As shown in Appendix~\ref{app:entanglement_entropy_of_the_variational_states}, it obeys a volume law for the type-I bipartition and a subvolume law for the type-II spatial bipartition, in agreement with the corresponding RQMBS state. Together with its orthogonality to the RQMBS tower and its vanishing energy variance, this shows that the sequence of variational states constitutes ARQMBS states in spin-$S$ systems.

\subsection{Multicomponent fermionic systems}\label{subsec:multicomponent_Fermion}
In this subsection, we consider multicomponent fermionic systems. The proof follows essentially the same steps as in the spin-$S$ case. The main differences are the larger symmetry group and the need to treat fermionic signs carefully. 

\subsubsection{Rainbow quantum many-body scar states}\label{subsubsec:RQMBS_fermions}
Here, we show the existence of RQMBS states in $N_{\rm F}$-component fermionic systems. Let $\hat{c}_{j,\sigma}$ and $\hat{c}^{\dagger}_{j,\sigma}$ denote the annihilation and creation operators, respectively, for a fermion of component $\sigma\in\{1,2,\ldots,N_{\rm F}\}$ at site $j$. Throughout this subsection, we assume that the fermion operators satisfy the usual canonical anticommutation relations within each subsystem, whereas fermionic operators acting on different subsystems commute~\footnote{Following Ref.~\cite{Wildeboer2022quantum}, this convention is equivalent to the conventional one in which fermionic operators acting on different subsystems anticommute. When we define $\hat{d}_{j,\sigma,1}\equiv \hat{c}_{j,\sigma}\otimes\hat{1}$ and $\hat{d}_{j,\sigma,2}\equiv (-1)^{\hat{N}}\otimes\hat{c}_{j,\sigma}$, we recover the usual anticommutation relations $\{\hat{d}_{j,\sigma,\gamma}, \hat{d}_{j',\sigma',\gamma'}\}=0$ and $\{\hat{d}_{j,\sigma,\gamma}, \hat{d}^{\dagger}_{j',\sigma',\gamma'}\}=\delta_{j,j'}\delta_{\sigma,\sigma'}\delta_{\gamma,\gamma'}$, where $\gamma,\gamma'\in\{1,2\}$.
}:
\begin{align}
[\hat{c}_{j,\sigma}\otimes\hat{1}, \hat{1}\otimes\hat{c}_{j',\sigma'}]=0,\; [\hat{c}_{j,\sigma}^{\dagger}\otimes\hat{1}, \hat{1}\otimes\hat{c}_{j',\sigma'}^{\dagger}]=0,\label{eq:commutation_relation1_in_different_subsystems}\\
[\hat{c}_{j,\sigma}\otimes\hat{1}, \hat{1}\otimes\hat{c}_{j',\sigma'}^{\dagger}]=0.\label{eq:commutation_relation2_in_different_subsystems}
\end{align}

In this subsection, we use the following basis states:
\begin{align}
\cket{\bm{n}}\otimes\cket{\bm{m}}&\equiv \left[\prod_{j=1}^M\prod_{\sigma=1}^{N_{\rm F}}(\hat{c}_{j,\sigma}^{\dagger})^{n_{j,\sigma}}\right]\cket{\rm vac}\notag \\
&\quad \otimes\left[\prod_{j=1}^M\prod_{\sigma=1}^{N_{\rm F}}(\hat{c}^{\dagger}_{j,\sigma})^{m_{j,\sigma}}\right]\cket{\rm vac},\label{eq:definition_of_fermion_basis}
\end{align}
where $\cket{\rm vac}$ denotes the normalized vacuum state of each subsystem, $n_{j,\sigma},m_{j,\sigma}\in\{0,1\}$, and the products of fermionic creation operators are ordered according to
\begin{align}
\prod_{j=1}^M\prod_{\sigma=1}^{N_{\rm F}}\hat{A}_{j,\sigma}\equiv \hat{A}_{1,1}\hat{A}_{1,2}\cdots\hat{A}_{1,N_{\rm F}}\hat{A}_{2,1}\hat{A}_{2,2}\cdots\hat{A}_{M,N_{\rm F}}.\label{eq:definition_of_order_of_fermion_operators}
\end{align}
We define the total fermion-number operators for component $\sigma$ in a subsystem and in the doubled system, respectively, as
\begin{align}
\hat{N}_{\sigma}&\equiv \sum_{j=1}^M\hat{c}^{\dagger}_{j,\sigma}\hat{c}_{j,\sigma}\equiv \sum_{j=1}^M\hat{n}_{j,\sigma},\label{eq:definition_of_Fermion_number_operator_sigma}\\
\hat{N}_{{\rm tot},\sigma}&\equiv \frac{1}{2}(\hat{N}_{\sigma}\otimes\hat{1}+\hat{1}\otimes\hat{N}_{\sigma}),\label{eq:definition_of_Ntot_sigma_fermion}
\end{align}
where $\hat{n}_{j,\sigma}$ is the number operator for component $\sigma$ at site $j$. The total fermion-number operators in a subsystem and in the doubled system are defined by
\begin{align}
\hat{N}&\equiv \sum_{\sigma=1}^{N_{\rm F}}\hat{N}_{\sigma},\label{eq:definition_of_total_fermion_number_operator}\\
\hat{N}_{\rm tot}&\equiv \frac{1}{2}(\hat{N}\otimes\hat{1}+\hat{1}\otimes\hat{N}).\label{eq:definition_of_Ntot_fermion}
\end{align}

Here, we consider a system with $U(1)^f\times U(1)^f$ symmetry, where $1\le f\le N_{\rm F}$. This means that the $f$ independent particle-number charges are conserved in each subsystem. For $f=1$, the total particle number in each subsystem is conserved, and for $f=N_{\rm F}$, the particle number of each component is conserved separately in each subsystem. Throughout this subsection, $N_{\mathrm{F}}$, $f$, and the group sizes $\{|g_s|\}_{s=1}^{f}$ are kept fixed in the thermodynamic limit. To specify the conserved particle numbers, we partition the set of fermionic components into $f$ nonempty disjoint subsets $g_s$:
\begin{align}
\{1,2,\ldots,N_{\rm F}\}\equiv \bigcup_{s=1}^fg_s.\label{eq:definition_of_subset_gs}
\end{align}
The corresponding conserved particle-number operators are defined by
\begin{align}
\hat{\mathcal{N}}_s\equiv \sum_{\sigma \in g_s}\hat{N}_{\sigma},\quad s=1,2,\ldots, f.\label{eq:definition_of_conserved_particle_number}
\end{align}
We assume that the Hamiltonian and the operators $\hat{L}_j$ satisfy
\begin{align}
[\hat{H}, \hat{\mathcal{N}}_s]&=0,\quad \text{ for all }s,\label{eq:commutation_relation_H_and_Nl_fermion}\\
[\hat{L}_j, \hat{\mathcal{N}}_s]&=0,\quad \text{for all }j, s.\label{eq:commutation_relation_Lj_and_Nl_fermion}
\end{align}
We further assume that no additional conserved particle-number charges, linearly independent of $\{\hat{N}_{s}\}_{s=1}^{f}$, can be formed as linear combinations of $\{\hat{N}_{\sigma}\}_{\sigma=1}^{N_{\mathrm F}}$. Here, $\hat{L}_j$ is a function of the local $\{\hat{n}_{x,\sigma}\}$ operators.

We now construct RQMBS states in multicomponent fermionic systems. We choose the root state to be
\begin{align}
\cket{S_{\bm{0}}}&\equiv \cket{\rm vac}\otimes\cket{\rm vac}.\label{eq:root_state_fermionic_systems}
\end{align}
For each $\bm{l}$, we define the unnormalized RQMBS state by
\begin{align}
|\tilde{S}_{\bm{l}}\rangle\equiv \left(\prod_{s=1}^fl_s!\right)\sum_{\bm{n}:\hat{\mathcal{N}}_s\cket{\bm{n}}=l_s\cket{\bm{n}} }\cket{\bm{n}}\otimes\cket{\bm{n}},\label{eq:definition_of_RQMBS_fermion_systems}
\end{align}
where $\bm{l}=(l_1,l_2,\ldots, l_f)$. We define the ladder operator
\begin{align}
\hat{Q}^{\dagger}_s&\equiv \sum_{j=1}^M\hat{q}_{s,j}^{\dagger},\label{eq:definition_of_Ql_dagger_fermions}\\
\hat{q}_{s,j}^{\dagger}&\equiv \sum_{\sigma\in g_s}\hat{c}^{\dagger}_{j,\sigma}\otimes\hat{c}^{\dagger}_{j,\sigma}.\label{eq:defintion_of_q_dagger_j_l_fermions}
\end{align}
Calculations similar to those in the spin-$S$ case (see Appendix \ref{app:detail_of_Qdagger_to_S0}) show that 
\begin{align}
|\tilde{S}_{\bm{l}}\rangle=\prod_{s=1}^f(\hat{Q}_s^{\dagger})^{l_s}\cket{S_{\bm{0}}},\quad\text{for } n=\sum_{s=1}^fl_s.\label{eq:RQMBS_ladder_operator_form_fermions}
\end{align}
Using the functional form of the Hamiltonian, we obtain
\begin{align}
\hat{H}_{\rm tot}|\tilde{S}_{\bm{l}}\rangle=0,\quad \text{for }\bm{l}\in \mathcal{I}_{\rm L},\label{eq:eigenstate_RQMBS_Htot_fermion_systems}
\end{align}
where $\mathcal{I}_{\rm L}$ is defined by Eq.~(\ref{eq:definition_of_set_I_L}).

Here, we investigate the MLRSGA conditions for fermionic systems. From Eq.~(\ref{eq:eigenstate_RQMBS_Htot_fermion_systems}), we can show that the MLRSGA conditions (\ref{eq:MLRSGA_relation1}), (\ref{eq:MLRSGA_relation2}), and (\ref{eq:MLRSGA_relation3}) are satisfied. From the definition of the ladder operators, we can also show that condition (\ref{eq:MLRSGA_relation5}) is satisfied. In the present case, condition (\ref{eq:MLRSGA_relation6}) is not needed for the subsequent construction because condition (\ref{eq:MLRSGA_relation6}) ensures that $|\tilde{S}_{\bm{l}}\rangle$ is an eigenstate of the Hamiltonian~\cite{Hashimoto2026construction} while Eq.~(\ref{eq:eigenstate_RQMBS_Htot_fermion_systems}) directly establishes the exact multiladder tower. The remaining conditions are operator identities in Eq.~(\ref{eq:MLRSGA_relation4}). Because the total Hamiltonian is $r_0$-local and has range $R_0$, with $r_0,R_0=O(1)$, and because $\hat{q}_{s,j}^{\dagger}$ is nilpotent, the operator identities hold at order at most $2r_0\max_s(|g_s|)$ (see Appendix \ref{app:proof_of_operator_identity_in_RSGA}):
\begin{align}
\underbrace{[[[[\hat{H}_{\rm tot}, \hat{Q}_s^{\dagger}], \hat{Q}^{\dagger}_s],\ldots],\ldots]}_{2|g_s|r_0+1}=0,\quad \text{for all }s.\label{eq:MLRSGA_operator_identity_for_fermions}
\end{align}
Therefore, the fermionic system possesses the exact multiladder RQMBS tower and satisfies all the algebraic conditions required for the subsequent construction.

\subsubsection{Parent Hamiltonian}\label{subsubsec:parent_Hamiltonian_fermions}
In the previous subsubsection, we have shown that multicomponent fermionic systems possess an exact multiladder RQMBS tower and satisfy all algebraic relations required for the subsequent construction. We can apply the systematic construction method for multiladder systems as reviewed in Sec.~\ref{subsec:systematic_construction}. The enlarged scar subspace for the fermionic systems is given by
\begin{align}
\mathcal{H}_{P_{\bm{l}}}&\equiv {\rm Span}\{\cket{\bm{n}}\otimes\cket{\bm{n}}~|~ \cket{\bm{n}}\in\mathcal{H},~\hat{\mathcal{N}}_s\cket{\bm{n}}=l_s\cket{\bm{n}},\notag \\
&\hspace{12.0em}\text{ for all }s=1,2,\ldots,f  \},\label{eq:definition_of_enlarged_scar_subspace_H_P_l_fermions}\\
\mathcal{H}_P&\equiv \bigoplus_{\bm{l}\in\mathcal{I}_{\rm L}}\mathcal{H}_{P_{\bm{l}}}.\label{eq:definition_of_total_enlarged_scar_subspace}
\end{align}
We introduce the projectors onto $\mathcal{H}_{P}$ and $\mathcal{H}_{P_{\bm{l}}}$ by $\hat{\mathcal{P}}$ and $\hat{\mathcal{P}}_{\bm{l}}$, respectively. As in the spin-$S$ case, we can show that
\begin{align}
\hat{\mathcal{P}}(\hat{h}_j\otimes\hat{1}-\hat{1}\otimes\hat{h}_j^{\ast})\hat{\mathcal{P}}&=0.\label{eq:PhjP=0_for_fermionic_systems}
\end{align}
Therefore, we can construct the parent Hamiltonian as
\begin{align}
\hat{H}_{\rm p}=\hat{\mathcal{P}}(\hat{H}\otimes\hat{1}-\hat{1}\otimes\hat{H}^{\ast})^2\hat{\mathcal{P}}.\label{eq:parent_Hamiltonian_for_fermionic_systems}
\end{align}

We can also show that the parent Hamiltonian is of Rokhsar--Kivelson type and admits a local frustration-free decomposition, i.e., we obtain
\begin{align}
\hat{H}_{\rm p}&=2\sum_{\bm{n},\bm{m}}|H_{\bm{n},\bm{m}}|^2\hat{P}_{\bm{n},\bm{m}},\label{eq:RK_form_of_parent_Hamiltonian_fermions}\\
&=\sum_j\hat{h}_{{\rm p},j},\label{eq:local_decompotion_parent_Hamiltonian_fermions}
\end{align}
where we regard $\hat{H}_{\rm p}$ as an operator restricted to the enlarged scar subspace $\mathcal{H}_P$. To prove Eq.~(\ref{eq:local_decompotion_parent_Hamiltonian_fermions}), we need an additional discussion of fermionic signs. The details are given in Appendix \ref{app:proof_local_FF_decomposition_parent_Hamitonian_fermions}.

\subsubsection{Asymptotic rainbow quantum many-body scar states}\label{subsubsec:ARQMBS_fermions}
We next show that the parent Hamiltonian is gapless in fermionic systems. As in the spin-$S$ case, we restrict $\hat{H}_{\rm p}$ to the enlarged scar subspace $\mathcal{H}_P$. We further assume that $\hat{H}_{\rm p}$ has a unique zero-energy ground state in each enlarged scar subspace $\mathcal{H}_{P_{\bm{l}}}$. We also assume that the number of local terms in the parent Hamiltonian satisfies Eq.~(\ref{eq:assumption_number_of_hpj}).

We consider a variational state for $\bm{l}\in \mathcal{I}_{\rm L}$ and $n=\sum_{s=1}^fl_s$
\begin{align}
|\widetilde{AS}_{\bm{l}}\rangle&\equiv \hat{Q}^{\dagger}_{s_0}(\bm{k})|\tilde{S}_{\bm{l}-\bm{e}_{s_0}}\rangle=\frac{1}{l_{s_0}}\hat{\rho}_{s_0}(\bm{k})|\tilde{S}_{\bm{l}}\rangle,\label{eq:definition_of_AS_l_fermions}\\
\hat{Q}^{\dagger}_{s_0}(\bm{k})&\equiv \sum_{j=1}^Me^{i\bm{k}\cdot\bm{R}_j}\hat{q}^{\dagger}_{s_0,j},\label{eq:definition_of_Qdagger_k_s0_fermions}\\
\hat{\rho}_{s_0}(\bm{k})&\equiv \sum_{j=1}^M\sum_{\sigma\in g_{s_0}}e^{i\bm{k}\cdot\bm{R}_j}(\hat{n}_{j,\sigma}\otimes\hat{1}),\label{eq:definition_of_rho_k_s0_fermions}
\end{align}
where $s_0\in\{1,2,\ldots,f\}$ and $\bm{e}_{s_0}$ denotes the unit vector whose $s_0$th component is one and whose remaining components are zero. 

The energy variance of the corresponding normalized variational state $|AS_{\bm{l}}\rangle\equiv |\widetilde{AS}_{\bm{l}}\rangle/\sqrt{\langle\widetilde{AS}_{\bm{l}} | \widetilde{AS}_{\bm{l}} \rangle}$ can be written as
\begin{align}
\Delta E_{\bm{l}}^2&=\frac{\langle \widetilde{AS}_{\bm{l}}| \hat{H}_{\rm p} |\widetilde{AS}_{\bm{l}}\rangle}{\langle \widetilde{AS}_{\bm{l}}| \widetilde{AS}_{\bm{l}}\rangle}\equiv \frac{f_{\rm n}(\bm{k})}{f_{\rm d}(\bm{k})},\label{eq:energy_variance_for_fermions}
\end{align}
where we used $|AS_{\bm{l}}\rangle\in\mathcal{H}_{P_{\bm{l}}}$, $\hat{\mathcal{P}}\hat{H}_0\hat{\mathcal{P}}=0$, and $\hat{H}_{\rm int}\hat{\mathcal{P}}=0$. As shown in Appendix \ref{app:proof_gaplessness_parent_Hamitonian_fermions}, the energy gap of the parent Hamiltonian $\Delta_{{\rm p},\bm{l}}$ and the energy variance satisfy the following inequality:
\begin{align}
\Delta_{{\rm p},\bm{l}}\le \Delta E_{\bm{l}}^2\le C_5\bm{k}^2,\label{eq:inequality_for_energy_gap_and_variance_fermions}
\end{align}
where $C_5$ is an $O(1)$ constant. Here, we assume that $\epsilon\le l_{s_0}/(|g_{s_0}|M)\le 1-\epsilon$ for a positive system-size-independent constant $\epsilon$, that $\|\hat{h}_{{\rm p},j}\|_{\rm op}\le J_{\rm p}=O(1)$, and that Eq.~(\ref{eq:assumption_number_of_hpj}) holds. Following the discussion in the spin-$S$ case, we choose the same sequence of nonzero lattice wave vectors as in Eq.~(\ref{eq:definition_of_kmin}), namely $\bm{k}=\bm{k}_{\min}$, for which $|\bm{k}_{\min}|\to0$ in the thermodynamic limit. This proves that the parent Hamiltonian is gapless. At the same time, the energy variance of the variational state vanishes in the thermodynamic limit.

Finally, we discuss the EE of the variational state $|AS_{\bm{l}}\rangle$. As shown in Appendix   \ref{app:entanglement_entropy_of_the_variational_states_fermion}, the EE of $|AS_{\bm{l}}\rangle$ exhibits the same qualitative scaling as that of the corresponding RQMBS state.

\subsection{Multicomponent bosonic systems}\label{subsec:multicomponent_Boson}
In this subsection, we consider multicomponent bosonic systems. The proof differs from those for the spin-$S$ and multicomponent fermionic cases because the local Hilbert space of soft-core bosons has infinite dimension. Consequently, additional assumptions are required to establish the existence of ARQMBS states.

\subsubsection{Rainbow quantum many-body scar states}\label{subsubsec:RQMBS_bosons}
Here, we show the existence of RQMBS states in $N_{\rm B}$-component bosonic systems. Let $\hat{b}_{j,\sigma}$ and $\hat{b}_{j,\sigma}^{\dagger}$ denote the annihilation and creation operators, respectively, for a boson of component $\sigma\in\{1,2,\ldots,N_{\rm B}\}$ at site $j$. 

We define the component-resolved and total boson-number operators for a single subsystem and for the doubled system analogously to the fermionic case:
\begin{align}
\hat{N}_{\sigma}&\equiv \sum_{j=1}^M\hat{b}^{\dagger}_{j,\sigma}\hat{b}_{j,\sigma}\equiv \sum_{j=1}^M\hat{n}_{j,\sigma},\label{eq:definition_of_number_operator_sigma_bosons}\\
\hat{N}_{{\rm tot},\sigma}&\equiv \frac{1}{2}(\hat{N}_{\sigma}\otimes\hat{1}+\hat{1}\otimes\hat{N}_{\sigma}),\label{eq:definition_ot_boson-number_operator_total_system_boson}\\
\hat{N}&\equiv \sum_{\sigma=1}^{N_{\rm B}}\hat{N}_{\sigma},\label{eq:definition_ot_total_number_of_bosons_in_subsystem_boson}\\
\hat{N}_{\rm tot}&\equiv \frac{1}{2}(\hat{N}\otimes\hat{1}+\hat{1}\otimes\hat{N}).\label{eq:definition_of_total_number_operator_in_doubled_system_boson}
\end{align}
As in the fermionic case, we consider a system with $U(1)^{f}\times U(1)^{f}$ symmetry, where $1\leq f\leq N_{\rm B}$. We partition the component labels into $f$ nonempty, pairwise disjoint subsets
$\{g_s\}_{s=1}^{f}$:
\begin{align}
\{1,2,\ldots, N_{\rm B}\}\equiv \bigcup_{s=1}^fg_s.\label{eq:definition_of_gs_bosons}
\end{align}
The corresponding conserved particle-number operators are defined by
\begin{align}
\hat{\mathcal{N}}_s&\equiv \sum_{\sigma\in g_s}\hat{N}_{\sigma}.\label{eq:definition_of_chages_bosons}
\end{align}
We assume that the Hamiltonian $\hat{H}$ and the operators $\{\hat{L}_{j}\}$ satisfy
\begin{align}
[\hat{H}, \hat{\mathcal{N}}_s]&=0,\quad \text{for all }s,\label{eq:commutation_relation_H_and_Ns_bosons}\\
[\hat{L}_j, \hat{\mathcal{N}}_s]&=0,\quad \text{for all }j, s.\label{eq:commutation_relation_Lj_and_Ns_bosons}
\end{align}
As in the fermionic case, we further assume that no additional conserved particle-number charges, linearly independent of $\{\hat{N}_{s}\}_{s=1}^{f}$, can be formed as linear combinations of $\{\hat{N}_{\sigma}\}_{\sigma=1}^{N_{{\rm B}}}$. Here, each $\hat{L}_{j}$ is a function of the local number operators $\{\hat{n}_{x,\sigma}\}$.

As in the fermionic case, we can construct exact eigenstates of the Hamiltonian. We choose the root state to be
\begin{align}
\cket{S_{\bm{0}}}\equiv \cket{\rm vac}\otimes\cket{\rm vac},\label{eq:root_state_bosonic_systems}
\end{align}
where $\cket{\rm vac}$ denotes the normalized vacuum state of each subsystem. For each $\bm{l}$, we define the unnormalized RQMBS state by
\begin{align}
|\tilde{S}_{\bm{l}}\rangle&\equiv \left(\prod_{s=1}^fl_s!\right)\sum_{\bm{n}: \hat{\mathcal{N}}_s\cket{\bm{n}}=l_s\cket{\bm{n}}}\cket{\bm{n}}\otimes\cket{\bm{n}},\label{eq:definition_of_RQMBS_boson}\\
\cket{\bm{n}}&\equiv \prod_{j=1}^M\prod_{\sigma=1}^{N_{\rm B}}\frac{(\hat{b}^{\dagger}_{j,\sigma})^{n_{j,\sigma}}}{\sqrt{n_{j,\sigma}!}}\cket{\rm vac},\label{eq:definition_of_basis_boson}
\end{align}
where $\bm{l}\equiv (l_1,l_2,\ldots,l_f)$. Direct calculations show that
\begin{align}
|\tilde{S}_{\bm{l}}\rangle&=\prod_{s=1}^f(\hat{Q}^{\dagger}_{s})^{l_s}\cket{S_{\bm{0}}},\label{eq:rewrite_RQMBS_for_bosons_by_ladder_operator}\\
\hat{Q}_s^{\dagger}&\equiv \sum_{j=1}^M\hat{q}_{s,j}^{\dagger},\label{eq:definition_of_ladder_operator_bosons}\\
\hat{q}_{s,j}^{\dagger}&\equiv \sum_{\sigma\in g_s}\hat{b}^{\dagger}_{j,\sigma}\otimes\hat{b}^{\dagger}_{j,\sigma}.\label{eq:definition_of_local_ladder_operator_bosons}
\end{align}
Using the assumed form of the Hamiltonian, we obtain
\begin{align}
\hat{H}_{\rm tot}|\tilde{S}_{\bm{l}}\rangle=0,\quad \text{for }\bm{l}\in\mathcal{I}_{\rm L},\label{eq:exact_eigenstate_RQMBS_bosons}
\end{align}
where $\mathcal{I}_{\rm L}$ is defined by Eq.~(\ref{eq:definition_of_set_I_L}).

Here, we investigate the MLRSGA conditions for bosonic systems. From Eq.~(\ref{eq:exact_eigenstate_RQMBS_bosons}), we can show that the MLRSGA conditions (\ref{eq:MLRSGA_relation1}), (\ref{eq:MLRSGA_relation2}), and (\ref{eq:MLRSGA_relation3}) are satisfied. From the definition of the ladder operators, we can also show that condition (\ref{eq:MLRSGA_relation5}) is satisfied. For the same reason as in the fermionic case, condition (\ref{eq:MLRSGA_relation6}) is not required for the subsequent construction. The remaining conditions are operator identities in Eq.~(\ref{eq:MLRSGA_relation4}). In contrast to the spin-$S$ and fermionic cases, we cannot establish the operator identities in Eq.~(\ref{eq:MLRSGA_relation4}) using locality and nilpotency, because the local ladder operators in Eq.~(\ref{eq:definition_of_local_ladder_operator_bosons}) are not nilpotent in bosonic systems: $(\hat{q}_{s,j}^{\dagger})^w\not=0$ for all $w\in\mathbb{Z}_{\ge 0}$.

To prove the operator identities in Eq.~(\ref{eq:MLRSGA_relation4}), we further assume that the total Hamiltonian is a normal-ordered polynomial of uniformly bounded degree in the bosonic creation and annihilation operators:
\begin{align}
\hat{H}_{\rm tot}&=\sum_{\substack{p,q\ge 0 \\ p+q\le r_{\rm B}}}\sum_{\substack{\bm{x},\bm{y}\in \mathcal{B}^p \\ \bm{x}',\bm{y}'\in\mathcal{B}^q}}h^{(p,q)}_{\bm{x},\bm{y};\bm{x}',\bm{y}'}\hat{b}^{\dagger}_{x_1}\cdots\hat{b}^{\dagger}_{x_p}\hat{b}_{y_1}\cdots\hat{b}_{y_p}\notag \\
&\otimes \hat{b}^{\dagger}_{x_1'}\cdots\hat{b}^{\dagger}_{x_q'}\hat{b}_{y_1'}\cdots\hat{b}_{y_q'},\label{eq:assumption_for_Htot_bosons}
\end{align}
where $r_{\rm B}\in\mathbb{Z}_{\geq 0}$ is independent of the system size and gives an upper bound on the total annihilation degree $p+q$ of each monomial. We define the set of single-particle bosonic modes by
\begin{align}
\mathcal{B}\equiv\left\{(j,\sigma)~|~ j=1,2,\ldots,M,\;\sigma=1,2,\ldots,N_{\rm B}\right\}.\label{eq:definition_of_set_B_bosons}
\end{align}
Here, $\mathcal{B}^{p}$ denotes the $p$-fold Cartesian product of $\mathcal{B}$. Thus, $\bm{x}=(x_{1},x_{2},\ldots,x_{p})\in\mathcal{B}^{p}$ is an ordered $p$-tuple of composite mode indices, and repeated mode indices are allowed. For $x_{a}=(j_{a},\sigma_{a})$, we use the shorthand notation $\hat{b}_{x_{a}}\equiv\hat{b}_{j_{a},\sigma_{a}}$. We set $\mathcal{B}^{0}\equiv\{\varnothing\}$, and an operator product over an empty tuple is understood as the identity operator. Without loss of generality, the coefficients $h^{(p,q)}_{\bm{x},\bm{y}; \bm{x}^{\prime},\bm{y}^{\prime}}$ are taken to be symmetric under independent permutations of the entries within each of $\bm{x}$, $\bm{y}$, $\bm{x}^{\prime}$, and $\bm{y}^{\prime}$. The Hermiticity of the total Hamiltonian requires $h^{(p,q)}_{\bm{x},\bm{y};\bm{x}',\bm{y}'}=[h^{(p,q)}_{\bm{y},\bm{x};\bm{y}',\bm{x}'}]^{\ast}$. To state the charge-neutrality conditions, for an arbitrary tuple $\bm{z}=(z_{1},z_{2},\ldots,z_{r})\in\mathcal{B}^{r}$, we define
\begin{align}
\nu_{s}(\bm{z})\equiv\#\left\{a\in\{1,2,\ldots,r\}\,\middle|\,z_{a}=(j_{a},\sigma_{a}),
\quad\sigma_{a}\in g_{s}\right\}.\label{eq:definition_of_number_of_group_bosons_s}
\end{align}
The coefficient $h^{(p,q)}_{\bm{x},\bm{y};\bm{x}',\bm{y}'}$ can be nonzero only if
\begin{align}
\nu_{s}(\bm{x})&=\nu_{s}(\bm{y}),&\nu_{s}(\bm{x}^{\prime})=\nu_{s}(\bm{y}^{\prime}),\quad s=1,2,\ldots,f,\label{eq:charge_neutral_conditions_boson}
\end{align}
which expresses the $U(1)^{f}\times U(1)^{f}$ symmetry of the total Hamiltonian. In addition, the coefficient vanishes unless the corresponding monomial satisfies the assumed $r_0$-locality and range $R_0$ conditions, where $r_0,R_0=O(1)$.

Under the above assumption, we can establish the operator identities in Eq.~(\ref{eq:MLRSGA_relation4}) for bosonic systems. The key ingredients are the following commutation relations:
\begin{align}
[\hat{b}_{j,\sigma}\otimes\hat{1}, \hat{Q}^{\dagger}_s]&=
\begin{cases}
\hat{1}\otimes \hat{b}^{\dagger}_{j,\sigma},\quad \sigma\in g_s,\\
0,\hspace{4.0em}\text{otherwise},
\end{cases}
\label{eq:commutation_relation_b1_and_Qdagger_bosons}\\
[\hat{1}\otimes\hat{b}_{j,\sigma}, \hat{Q}^{\dagger}_s]&=
\begin{cases}
\hat{b}^{\dagger}_{j,\sigma}\otimes\hat{1},\quad \sigma\in g_s,\\
0,\hspace{4.0em}\text{otherwise},
\end{cases}
\label{eq:commutation_relation_b2_and_Qdagger_bosons}\\
[\hat{b}_{j,\sigma}^{\dagger}\otimes\hat{1},\hat{Q}^{\dagger}_s]&=0,\quad [\hat{1}\otimes \hat{b}_{j,\sigma}^{\dagger},\hat{Q}^{\dagger}_s]=0.\label{eq:commutation_relation_bdagger12_and_Qdagger_bosons}
\end{align}
From these relations and the bounded-degree assumption on the Hamiltonian, the maximal annihilation degree in the commutator $[\hat{H}_{\rm tot},\hat{Q}_{s}^{\dagger}]$ is $r_{\rm B}-1$. Thus, the operator identities hold at order $r_{\rm B}$ or lower:
\begin{align}
\underbrace{[[[[\hat{H}_{\rm tot}, \hat{Q}_s^{\dagger}], \hat{Q}^{\dagger}_s],\ldots],\ldots]}_{r_{\rm B}+1}=0,\quad \text{for all }s.\label{eq:MLRSGA_operator_identity_for_bosons}
\end{align}
Therefore, multicomponent bosonic systems possess an exact multiladder RQMBS tower and satisfy all the algebraic conditions required for the subsequent construction under the additional assumption on the form of the Hamiltonian.

\subsubsection{Parent Hamiltonian}\label{subsubsec:parent_Hamiltonian_bosons}
In the previous subsubsection, we have shown that multicomponent bosonic systems possess an exact multiladder RQMBS tower and satisfy all algebraic relations required for the subsequent construction. We can apply the systematic construction method for multiladder systems as reviewed in Sec.~\ref{subsec:systematic_construction}. The enlarged scar subspace for bosonic systems is given by
\begin{align}
\mathcal{H}_{P_{\bm{l}}}&\equiv {\rm Span}\{\cket{\bm{n}}\otimes\cket{\bm{n}}~|~ \cket{\bm{n}}\in\mathcal{H},~\hat{\mathcal{N}}_s\cket{\bm{n}}=l_s\cket{\bm{n}},\notag \\
&\hspace{12.0em}\text{ for all }s=1,2,\ldots,f  \},\label{eq:definition_of_enlarged_scar_subspace_H_P_l_bosonss}\\
\mathcal{H}_P&\equiv \bigoplus_{\bm{l}\in\mathcal{I}_{\rm L}}\mathcal{H}_{P_{\bm{l}}}.\label{eq:definition_of_total_enlarged_scar_subspace_bosons}
\end{align}
We denote the projectors onto $\mathcal{H}_{P}$ and $\mathcal{H}_{P_{\bm l}}$ by $\hat{\mathcal P}$ and $\hat{P}_{\bm l}$, respectively. As in the spin-$S$ and fermionic cases, we can show that
\begin{align}
\hat{\mathcal{P}}(\hat{h}_j\otimes\hat{1}-\hat{1}\otimes\hat{h}_j^{\ast})\hat{\mathcal{P}}&=0.\label{eq:PhjP=0_for_bosonic_systems}
\end{align}
Therefore, we can construct the parent Hamiltonian as
\begin{align}
\hat{H}_{\rm p}=\hat{\mathcal{P}}(\hat{H}\otimes\hat{1}-\hat{1}\otimes\hat{H}^{\ast})^2\hat{\mathcal{P}}.\label{eq:parent_Hamiltonian_for_bosonic_systems}
\end{align}

We can also show that the parent Hamiltonian is of Rokhsar--Kivelson type and admits a local frustration-free decomposition, i.e., we obtain
\begin{align}
\hat{H}_{\rm p}&=2\sum_{\bm{n},\bm{m}}|H_{\bm{n},\bm{m}}|^2\hat{P}_{\bm{n},\bm{m}},\label{eq:RK_form_of_parent_Hamiltonian_bosons}\\
&=\sum_j\hat{h}_{{\rm p},j},\label{eq:local_decompotion_parent_Hamiltonian_bosons}
\end{align}
where $\hat{h}_{{\rm p},j}\ge0$ and we regard $\hat{H}_{\rm p}$ as an operator restricted to the enlarged scar subspace $\mathcal{H}_P$. The proof is essentially the same as in the spin-$S$ case (see Appendix \ref{app:proof_locality_parent_Hamitonian_spin-S}).

\subsubsection{Asymptotic rainbow quantum many-body scar states}\label{subsubsec:ARQMBS_bosonss}
We next show that the parent Hamiltonian is gapless in bosonic systems. As in the spin-$S$ and fermionic cases, we restrict $\hat{H}_{\rm p}$ to the enlarged scar subspace $\mathcal{H}_P$. We further assume that $\hat{H}_{\rm p}$ has a unique zero-energy ground state in each enlarged scar subspace $\mathcal{H}_{P_{\bm{l}}}$. We also assume that the number of local terms in the parent Hamiltonian satisfies Eq.~(\ref{eq:assumption_number_of_hpj}). 

The proof is almost identical to the corresponding proofs for the spin-$S$ and fermionic cases. However, an additional assumption is required because neither $\|\hat{h}_{{\rm p},j}\|_{\rm op}$ nor $\|\hat{n}_{j,\sigma}\|_{\rm op}$ can be bounded by an $O(1)$ constant in bosonic systems.

We consider a variational state for $\bm{l}\in \mathcal{I}_{\rm L}$
\begin{align}
|\widetilde{AS}_{\bm{l}}\rangle&\equiv \hat{Q}^{\dagger}_{s_0}(\bm{k})|\tilde{S}_{\bm{l}-\bm{e}_{s_0}}\rangle=\frac{1}{l_{s_0}}\hat{\rho}_{s_0}(\bm{k})|\tilde{S}_{\bm{l}}\rangle,\label{eq:definition_of_AS_l_bosons}\\
\hat{Q}^{\dagger}_{s_0}(\bm{k})&\equiv \sum_{j=1}^Me^{i\bm{k}\cdot\bm{R}_j}\hat{q}^{\dagger}_{s_0,j},\label{eq:definition_of_Qdagger_k_s0_bosons}\\
\hat{\rho}_{s_0}(\bm{k})&\equiv \sum_{j=1}^M\sum_{\sigma\in g_{s_0}}e^{i\bm{k}\cdot\bm{R}_j}(\hat{n}_{j,\sigma}\otimes\hat{1}),\label{eq:definition_of_rho_k_s0_bosons}
\end{align}
where $s_0\in\{1,2,\ldots,f\}$ and $\bm{e}_{s_0}$ denotes the unit vector whose $s_0$th component is one and whose remaining components are zero. 

The energy variance of the corresponding normalized variational state $|AS_{\bm{l}}\rangle\equiv |\widetilde{AS}_{\bm{l}}\rangle/\sqrt{\langle \widetilde{AS}_{\bm{l}}|\widetilde{AS}_{\bm{l}} \rangle}$ can be written as
\begin{align}
\Delta E_{\bm{l}}^2&=\frac{\langle \widetilde{AS}_{\bm{l}}| \hat{H}_{\rm p}| \widetilde{AS}_{\bm{l}} \rangle}{\langle\widetilde{AS}_{\bm{l}}|\widetilde{AS}_{\bm{l}}\rangle}\equiv \frac{f_{\rm n}(\bm{k})}{f_{\rm d}(\bm{k})},\label{eq:energy_variance_for_bosonic_systems}
\end{align}
where we used $|AS_{\bm{l}}\rangle\in \mathcal{H}_{P_{\bm{l}}}, \hat{\mathcal{P}}\hat{H}_0\hat{\mathcal{P}}=0$, and $\hat{H}_{\rm int}\hat{\mathcal{P}}=0$.  

As shown in Appendix \ref{app:proof_gaplessness_parent_Hamitonian_bosons}, the energy gap of the parent Hamiltonian $\Delta_{{\rm p},\bm{l}}$ and the energy variance satisfy the following inequality:
\begin{align}
\Delta_{{\rm p},\bm{l}}\le \Delta E_{\bm{l}}^2\le C_8\bm{k}^2,\label{eq:inequality_for_energy_gap_and_variance_bosons}
\end{align}
where $C_8$ is an $O(1)$ constant. Here, we assume that $0<\nu_-\le l_{s_0}/(|g_{s_0}|M)\le \nu_+<\infty$ and Eq.~(\ref{eq:assumption_number_of_hpj}) holds. To prove the bound in Eq.~(\ref{eq:inequality_for_energy_gap_and_variance_bosons}), we impose the following additional assumption:
\begin{align}
\frac{\| \hat{h}^{1/2}_{{\rm p},j}(\hat{n}_{x,\sigma}\otimes\hat{1})|\tilde{S}_{\bm{l}}\rangle \|^2}{\langle\tilde{S}_{\bm{l}}|\tilde{S}_{\bm{l}}\rangle}\le J_{\rm B},\text{ for all }j, x\in S_{{\rm p},j}^{(1)},\sigma\in g_{s_0}.\label{eq:assumptions_for_bosons}
\end{align}
Here, $J_{\rm B}>0$ is a system-size-independent constant. Following the discussion in the spin-$S$ and fermionic cases, we choose the same sequence of nonzero lattice wave vectors as in Eq.~(\ref{eq:definition_of_kmin}), namely $\bm{k}=\bm{k}_{\min}$, for which $|\bm{k}_{\min}|\to0$ in the thermodynamic limit. This proves that the parent Hamiltonian is gapless. At the same time, the energy variance of the variational state vanishes in the thermodynamic limit.

Finally, we discuss the EE of the variational state $|AS_{\bm l}\rangle$. As shown in Appendix \ref{app:entanglement_entropy_of_the_variational_states_boson}, its EE obeys a volume law for the type-I bipartition and a subvolume law for the type-II bipartition, in agreement with the corresponding RQMBS state.

\subsection{Summary of the construction of ARQMBS states}\label{subsec:summary_construction_ARQMBS}
Here, we summarize our construction of ARQMBS states. Our assumptions are as follows:
\begin{enumerate}
\item The total Hamiltonian has the form $\hat{H}_{\rm tot}=\hat{H}\otimes\hat{1}-\hat{1}\otimes\hat{H}^{\ast}-(\hbar/2)\sum_j(\hat{L}_j\otimes\hat{1}-\hat{1}\otimes\hat{L}_j)^2$.

\item The total Hamiltonian $\hat{H}_{\rm tot}$ is nonintegrable. 

\item The total Hamiltonian $\hat{H}_{\rm tot}$ is $r_0$-local and has range $R_0$, where $r_0,R_0=O(1)$. The subsystem Hamiltonian $\hat{H}$ is $r$-local and has range $R$, where $r, R=O(1)$.

\item For soft-core bosons, the total Hamiltonian is assumed to be a normal-ordered polynomial of uniformly bounded total annihilation degree $p+q\leq r_{\rm B}=O(1)$ and to satisfy the locality, range, Hermiticity, and charge-neutrality conditions stated in Eqs.~(\ref{eq:assumption_for_Htot_bosons})--(\ref{eq:charge_neutral_conditions_boson}).\label{eq:assumptions_soft_core_boson_Hamiltonian}

\item Spin-$S$ systems possess $U(1)\times U(1)$ symmetry, whereas multicomponent fermionic and bosonic systems possess $U(1)^{f}\times U(1)^{f}$ symmetry.

\item Each $\hat L_j$ is a local Hermitian operator that is a function of local charge-density operators.

\item The quantities $S$, $f$, $N_{\rm F}$, and $N_{\rm B}$, as
well as the group sizes $\{|g_s|\}$, are kept fixed in
the thermodynamic limit.

\item For multicomponent fermionic and bosonic systems, no additional conserved particle-number charge that is linearly independent of $\{\hat N_s\}_{s=1}^{f}$ can be formed as a linear combination of the component-resolved particle-number operators.

\item The corresponding normalized RQMBS state is the unique zero-energy ground state of the parent Hamiltonian restricted to each enlarged scar subspace $\mathcal H_{P_n}$ for spin systems or $\mathcal H_{P_{\bm l}}$ for multicomponent systems. This assumption is used only to establish the gaplessness of the parent Hamiltonian.

\item The number of local terms of the parent Hamiltonian is bounded by $c_{\rm p}M$ with $c_{\rm p}=O(1)$.

\item For spin-$S$ and fermionic systems, $\|\hat{h}_{{\rm p},j}\|_{\rm op}\le J_{\rm p}=O(1)$. For bosonic systems, we assume  Eq.~(\ref{eq:assumptions_for_bosons}).

\item The spatial dimension and the lattice coordination number are independent of the system size. The thermodynamic limit is taken along a sequence of finite Bravais lattices for which at least one linear size $M_a$ diverges. We assume that there exists a sequence of nonzero lattice wave vectors $\{\bm k_{\min}(M)\}$ satisfying $\sum_{j=1}^Me^{i\bm k_{\min}\cdot\bm R_j}=0$, $|\bm k_{\min}|\to0$. 

\item The sequence of charge sectors is taken at finite density. More precisely, we assume $\epsilon\leq n/M\leq2S-\epsilon$ for spin-$S$ systems, $\epsilon\leq l_{s_0}/(|g_{s_0}|M)\leq1-\epsilon$ for fermionic systems, and $0<\nu_-\leq l_{s_0}/(|g_{s_0}|M)\leq\nu_+<\infty$ for bosonic systems, where $\epsilon$, $\nu_{-}$, and $\nu_{+}$ are
positive constants independent of the system size.\label{eq:finite_density_assumptions}

\item For the EE scaling in bosonic systems, we assume $l_s=O(M)$ for every $s$.
\end{enumerate}

\section{Proof of the correspondence between ARQMBS states and diffusive Nambu--Goldstone modes}\label{sec:Proof_of_correspondence}
In this section, we rigorously prove the correspondence between ARQMBS states in closed systems and the low-decay-rate spectrum of the effective Lindbladian in the strongly dissipative regime. When the parent-Hamiltonian spectrum is quadratic, we identify the corresponding low-decay-rate modes with diffusive NG modes~\cite{Huang2025hydrodynamics}. To this end, we consider open quantum systems with strong $U(1)$ symmetry for spin-$S$ systems and strong $U(1)^f$ symmetry for multicomponent systems. The key feature is the duality between an open quantum system and a doubled system~\cite{Teretenkov2024duality}. First, we explain the details of this duality. Next, we derive the effective Lindbladian for the strongly dissipative regime. Finally, we establish the correspondence between the spectral problem underlying the ARQMBS construction and the low-decay-rate spectrum of the effective Lindbladian. 

\subsection{Setup}\label{subsec:setup_open_system}
Throughout this section, we consider the Lindblad master equation~\cite{Breuer2002theory,Rivas2012open}
\begin{align}
\frac{d}{dt}\hat{\rho}(t)&=-\frac{i}{\hbar}[\hat{H}, \hat{\rho}(t)]+\sum_j\left[\hat{L}_j\hat{\rho}(t)\hat{L}_j^{\dagger}-\frac{1}{2}\{\hat{L}_j^{\dagger}\hat{L}_j, \hat{\rho}(t)\}\right]\notag \\
&\equiv \hat{\mathcal{L}}[\hat{\rho}(t)],\label{eq:Lindblad_master_equation_open}
\end{align}
where $\hat{\rho}(t)$ is a density matrix at time $t$, $\{\hat{L}_j\}$ are Lindblad operators, and $\hat{\mathcal{L}}[\cdot]$ is a superoperator called the Lindbladian. We assume that the system has strong $U(1)$ symmetry for spin-$S$ systems and $U(1)^f$ symmetry for multicomponent systems, namely, the Hamiltonian and Lindblad operators satisfy~\cite{Buca2012note,Albert2014symmetries,Lieu2020symmetry}
\begin{align}
[\hat{H}, \hat{\mathcal{N}}_s]&=0,\quad \text{for all }s,\label{eq:condition_strong_U(1)_for_Hamiltonian}\\
[\hat{L}_j, \hat{\mathcal{N}}_s]&=0,\quad\text{for all }j, s,\label{eq:conditions_strong_U(1)_for_Lindblad}
\end{align}
where $\hat{\mathcal{N}}_s=\hat{S}_{\rm tot}^z$ for spin-$S$ systems, and $\hat{\mathcal{N}}_s$ is defined in Eqs.~(\ref{eq:definition_of_conserved_particle_number}) and (\ref{eq:definition_of_chages_bosons}) for fermionic and bosonic systems, respectively. In addition, we assume that the Hamiltonian $\hat{H}$ is $r$-local and has range $R$, with $r, R=O(1)$ as in the isolated systems. For soft-core boson Hamiltonians, we use assumption \ref{eq:assumptions_soft_core_boson_Hamiltonian} in Sec.~\ref{subsec:summary_construction_ARQMBS}. For the Lindblad operators, we assume that each $\hat{L}_j$ is a local Hermitian operator that is a function of local charge-density operators. 

From the strong-symmetry assumption and the Hermiticity of all $\hat{L}_j$, the Lindblad master equation has an infinite-temperature steady state in each $U(1)$-charge sector~\cite{Buca2012note,Yoshida2024uniqueness}:
\begin{align}
\hat{\mathcal{L}}[\hat{\rho}_{{\rm s},\bm{l}}]&=0,\label{eq:steady_state_each_U(1)_sector}\\
\hat{\rho}_{{\rm s},\bm{l}}&=\frac{1}{W_{\bm{l}}}\sum_{\bm{n}:\hat{\mathcal{N}}_s\cket{\bm{n}}=l_s\cket{\bm{n}}}\cket{\bm{n}}\bra{\bm{n}},\label{eq:infinite_temperature_state_in_each_U(1)_sector}
\end{align}
where $\hat{\rho}_{{\rm s},\bm{l}}$ is a steady state and $W_{\bm{l}}$ is the dimension of the $U(1)$-charge sector specified by charges $\bm{l}$. 

In what follows, we use the vectorized form of the Lindblad master equation~\cite{Medvedyeva2016exact,Haga2023quasiparticles,Teretenkov2024duality}. In this paper, we use the following convention:
\begin{align}
\hat{\rho}=\sum_{\bm{n},\bm{m}}\rho_{\bm{n},\bm{m}}\cket{\bm{n}}\bra{\bm{m}}~\leftrightarrow~ \cket{\rho}=\sum_{\bm{n},\bm{m}}\rho_{\bm{n},\bm{m}}\cket{\bm{n}}\otimes\cket{\bm{m}},\label{eq:convention_vectorization}
\end{align} 
where $\rho_{\bm{n},\bm{m}}\equiv \bra{\bm{n}}\hat{\rho}\cket{\bm{m}}$ and $\cket{\rho}$ is a state vector belonging to $\mathcal{H}\otimes\mathcal{H}$. Using this convention, the Lindblad master equation becomes
\begin{align}
\frac{d}{dt}|\rho(t)\rangle&=\hat{\mathcal{L}}\cket{\rho(t)},\label{eq:vectrization_form_of_Lindblad_master_equaiton}\\
\hat{\mathcal{L}}&\equiv -\frac{i}{\hbar}(\hat{H}\otimes\hat{1}-\hat{1}\otimes\hat{H}^{\ast})\notag \\
&\quad +\sum_j\left(\hat{L}_j\otimes\hat{L}_j-\frac{1}{2}\hat{L}_j^2\otimes\hat{1}-\frac{1}{2}\hat{1}\otimes \hat{L}_j^2\right)\notag \\
&=-\frac{i}{\hbar}(\hat{H}\otimes\hat{1}-\hat{1}\otimes\hat{H}^{\ast})\notag \\
&\quad -\frac{1}{2}\sum_j\left(\hat{L}_j\otimes\hat{1}-\hat{1}\otimes\hat{L}_j\right)^2.\label{eq:definition_of_Lindbladian_vectrization_form}
\end{align}
Here, the Lindbladian $\hat{\mathcal{L}}$ acts on the doubled Hilbert space $\mathcal{H}\otimes\mathcal{H}$. We have used the assumption that each Lindblad operator $\hat{L}_j$ is a Hermitian local function of the charge-density operators. Here, we introduce $\hat{\mathcal{H}}\equiv i\hbar\hat{\mathcal{L}}$. The Lindblad master equation takes the form of a Schr\"odinger-like equation:
\begin{align}
i\hbar\frac{d}{dt}\cket{\rho(t)}&=\hat{\mathcal{H}}\cket{\rho(t)},\label{eq:Schrodinger_like_equation_Lindblad}\\
\hat{\mathcal{H}}&=\hat{H}\otimes\hat{1}-\hat{1}\otimes\hat{H}^{\ast}\notag \\
&\quad-\frac{i\hbar}{2}\sum_j\left(\hat{L}_j\otimes\hat{1}-\hat{1}\otimes\hat{L}_j\right)^2.\label{eq:Hamiltonian_form_of_Lindbladian}
\end{align}
Comparing Eq.~(\ref{eq:Hamiltonian_form_of_Lindbladian}) with Eqs.~(\ref{eq:Hamiltonian_rainbow_QMBS}) and (\ref{eq:interaction_Hamiltonian_between_ladders}), we find that the only difference is the presence of the imaginary unit in the interaction term. This similarity plays an important role in the following analyses.

Here, we consider the vectorized form of the infinite-temperature state in Eq.~(\ref{eq:infinite_temperature_state_in_each_U(1)_sector}). Using the convention (\ref{eq:convention_vectorization}), we obtain
\begin{align}
\hat{\rho}_{{\rm s},\bm{l}}\leftrightarrow |\rho_{{\rm s},\bm{l}}\rangle\equiv \frac{1}{W_{\bm{l}}}\sum_{\bm{n}:\hat{\mathcal{N}}_s\cket{\bm{n}}=l_s\cket{\bm{n}}}\cket{\bm{n}}\otimes\cket{\bm{n}}.\label{eq:vectorization_form_of_infinite_temperature_state}
\end{align}
This state coincides with the RQMBS state [see Eqs.~(\ref{eq:rainbow_QMBS_for_spin-S}), (\ref{eq:definition_of_RQMBS_fermion_systems}), and (\ref{eq:definition_of_RQMBS_boson})] up to a normalization constant. This correspondence has been shown by Teretenkov and Lychkovskiy~\cite{Teretenkov2024duality}. 

We now show that the infinite-temperature state $\hat{\rho}_{{\rm s},\bm{l}}$ has SWSSB order under the finite-density assumptions summarized in item \ref{eq:finite_density_assumptions} of Sec.~\ref{subsec:summary_construction_ARQMBS}. To quantify this, we introduce the R\'enyi-2 correlator~\cite{Lessa2025strong}:
\begin{align}
R^{(2)}_{ij}&\equiv \frac{{\rm Tr}(\hat{\rho}\hat{O}_i\hat{O}_j^{\dagger}\hat{\rho}\hat{O}_j\hat{O}_i^{\dagger})}{{\rm Tr}(\hat{\rho}^2)},\label{eq:definition_of_Renyi2_correlator}
\end{align}
where $\hat{O}_i$ is an operator satisfying
\begin{align}
[\hat{\mathcal{N}}_s, \hat{O}_i]&=w_s\hat{O}_i,\quad s=1,\ldots,f, \quad\bm{w}\not=\bm{0}.\label{eq:condition_Oi_for_multicomponent_systems}
\end{align}
When $R_{ij}^{(2)}\to R^{(2)}\not=0$ for $|\bm{R}_i-\bm{R}_j|\to\infty$, the system exhibits SWSSB order. As shown in Appendix~\ref{app:details_of_Renyi2_correlator}, the infinite-temperature states of spin-$S$, multicomponent fermionic, and multicomponent bosonic systems exhibit SWSSB order. In Appendix~\ref{app:details_of_Renyi2_correlator}, we also show that the infinite-temperature states exhibit SWSSB order as diagnosed by the R\'enyi-1 correlator~\cite{Liu2025diagnosing,Weinstein2025efficient} and the fidelity correlator. 

\subsection{Effective Lindbladian}\label{subsec:Effective_Lindbladian}
Here, we derive the effective Lindbladian for the strongly dissipative regime following Ref.~\cite{Kessler2012generalized}. We decompose the Lindbladian into unperturbed and perturbative parts:
\begin{align}
\hat{\mathcal{L}}&\equiv \hat{\mathcal{L}}_0+\hat{\mathcal{V}},\label{eq:definition_of_nonperturbative_and_perturbative_part_Lindbladian}\\
\hat{\mathcal{L}}_0&\equiv -\frac{1}{2}\sum_j\left(\hat{L}_j\otimes\hat{1}-\hat{1}\otimes\hat{L}_j\right)^2,\label{eq:nonperturbative_part_of_Lindbladian}\\
\hat{\mathcal{V}}&\equiv -\frac{i}{\hbar}(\hat{H}\otimes\hat{1}-\hat{1}\otimes\hat{H}^{\ast}).\label{eq:perturbative_part_of_Lindbladian}
\end{align}

To apply the Schrieffer-Wolff transformation for open systems~\cite{Kessler2012generalized}, we consider ${\rm ker}\hat{\mathcal{L}}_0$, defined by $\hat{\mathcal{L}}_0\cket{\psi}=0$. Because each $\hat{L}_j$ is a local function of the charge-density operator, all paired product states $\cket{\bm{n}}\otimes\cket{\bm{n}}$ belong to the kernel of $\hat{\mathcal{L}}_0$. The Hilbert subspace spanned by $\cket{\bm{n}}\otimes\cket{\bm{n}}$ coincides with the enlarged scar subspace $\mathcal{H}_{P}$ [see Eq.~(\ref{eq:expression_of_projection_operator_spin-S})]. We define the projector onto the complement of $\mathcal{H}_P$ as
\begin{align}
\hat{\mathcal{Q}}\equiv \hat{1}-\hat{\mathcal{P}}.\label{eq:definition_of_projector_Q_Lindbladian}
\end{align}
We assume that these paired product states exhaust the zero eigenspace: ${\rm ker}\hat{\mathcal{L}}_0=\mathcal{H}_P$. Furthermore, we assume a system-size-independent dissipative gap:
\begin{align}
-\hat{\mathcal{Q}}\hat{\mathcal{L}}_0\hat{\mathcal{Q}}\ge \Delta\hat{\mathcal{Q}},\label{eq:assumption_denominator_of_effective_Lindbladian}
\end{align}
where $\Delta~(>0)$ is a system-size-independent constant with units of $[T]^{-1}$. To justify the perturbative treatment, we assume that $\hbar\Delta$ is much larger than other energy scales of the system. Note that we first take the strong-dissipation limit at fixed system size and subsequently take the thermodynamic limit.

We then derive the effective Lindbladian. Using the fact that $\hat{\mathcal{P}}\hat{\mathcal{L}}_0\hat{\mathcal{P}}=0$ and $\hat{\mathcal{P}}\hat{\mathcal{V}}\hat{\mathcal{P}}=0$, the second-order effective Lindbladian becomes
\begin{align}
\hat{\mathcal{L}}_{\rm eff}&\equiv \hat{\mathcal{P}}\hat{\mathcal{V}}\hat{\mathcal{Q}}(-\hat{\mathcal{Q}}\hat{\mathcal{L}}_0\hat{\mathcal{Q}})^{-1}\hat{\mathcal{Q}}\hat{\mathcal{V}}\hat{\mathcal{P}}.\label{eq:effective_Lindbladian_second_order}
\end{align}
Direct calculation shows that
\begin{align}
\hat{\mathcal{L}}_{\rm eff}&=-\frac{2}{\hbar^2}\sum_{\substack{\bm{n},\bm{m} \\ \bm{n}\not=\bm{m}}}\frac{|H_{\bm{n},\bm{m}}|^2}{\Gamma_{\bm{n},\bm{m}}}\hat{P}_{\bm{n},\bm{m}},\label{eq:effecive_Lindbladian_Rohksar-Kivelson_form}\\
\Gamma_{\bm{n},\bm{m}}&\equiv -\bra{\bm{n}}\otimes\bra{\bm{m}}\hat{\mathcal{Q}}\hat{\mathcal{L}}_0\hat{\mathcal{Q}}\cket{\bm{n}}\otimes\cket{\bm{m}}\notag \\
&=\frac{1}{2}\bra{\bm{n}}\otimes\bra{\bm{m}}\sum_j\left(\hat{L}_j\otimes\hat{1}-\hat{1}\otimes\hat{L}_j\right)^2\cket{\bm{n}}\otimes\cket{\bm{m}}.\label{eq:definition_of_Gamma_n_m}
\end{align}
Here, $H_{\bm{n},\bm{m}}$ and $\hat{P}_{\bm{n},\bm{m}}$ were defined by Eqs.~(\ref{eq:definition_of_H_n_m_spin-S}) and (\ref{eq:definition_of_operator_P_spin-S}), respectively. From the assumption in Eq.~(\ref{eq:assumption_denominator_of_effective_Lindbladian}), $\Gamma_{\bm{n},\bm{m}}>0$ for all $\bm{n}\not=\bm{m}$. This result implies that $-\hat{\mathcal{L}}_{\rm eff}$ has a Rokhsar--Kivelson-type frustration-free form [see Eq.~(\ref{eq:parent_Hamiltonian_bracket_form_spin-S})]. The only differences from the parent Hamiltonian are the additional factor $1/\Gamma_{\bm{n},\bm{m}}$, which originates from the dissipative terms, and an overall constant.

\subsection{Proof of gaplessness of the effective Lindbladian}
\label{subsec:proof_of_Ok_squared_spectra}

We now consider the eigenvalue problem of the effective Lindbladian $\hat{\mathcal{L}}_{\rm eff}$. In the following, we restrict both $\hat{\mathcal L}_{\rm eff}$ and $\hat H_{\rm p}$ to a fixed enlarged scar subspace $\mathcal H_{P_n}$ or $\mathcal H_{P_{\bm l}}$. From Eq.~(\ref{eq:effecive_Lindbladian_Rohksar-Kivelson_form}) and $\Gamma_{\bm{n},\bm{m}}>0$, the effective Lindbladian is negative semidefinite. We can write the eigenvalue equation as
\begin{align}
\hat{\mathcal{L}}_{\rm eff}\cket{\psi_i}&=-\lambda_i\cket{\psi_i},\label{eq:eigenvalue_problem_of_effective_Lindbladian}
\end{align}
where $\lambda_i\ge 0$ is the $i$th decay rate and $\cket{\psi_i}$ is the corresponding eigenvector. Here, we label the decay rates as $0=\lambda_1\le \lambda_2\le \lambda_3\le\ldots$, where $\cket{\psi_1}$ corresponds to the RQMBS state. 

Here, we prove that $\lambda_2=O(|\bm{k}_{\rm min}|^2)$. To prove this, we assume that
\begin{align}
&0<\Gamma_{\rm min}\le \Gamma_{\bm{n},\bm{m}}\le \Gamma_{\rm max}<\infty,\notag \\
&\hspace{5.0em}\text{for all } \bm{n}\not=\bm{m}\text{ such that }H_{\bm{n},\bm{m}}\not=0,\label{eq:assumption_for_Gamma}
\end{align}
where $\Gamma_{\rm min}$ and $\Gamma_{\rm max}$ are system-size-independent positive constants. From this assumption, we obtain
\begin{align}
\frac{\hat{H}_{\rm p}}{\Gamma_{\rm max}}\le-\hbar^2\hat{\mathcal{L}}_{\rm eff}\le\frac{\hat{H}_{\rm p}}{\Gamma_{\rm min}},\label{eq:inequality_for_parent_Hamiltonian_and_effective_Lindbladian}
\end{align}
where $\hat{H}_{\rm p}$ is the parent Hamiltonian of the RQMBS states [see Eq.~(\ref{eq:parent_Hamiltonian_bracket_form_spin-S})]. The eigenvalue equation for the parent Hamiltonian is written as
\begin{align}
\hat{H}_{\rm p}\cket{\varphi_i}&=\epsilon_i\cket{\varphi_i},\quad 0= \epsilon_1<\epsilon_2\le\cdots.\label{eq:eigenvalue_equation_parent_Hamiltonian}
\end{align}
Using the min-max principle~\cite{Tasaki2020physics,Masaoka2024quadratic}, we obtain the following eigenvalue bounds:
\begin{align}
\frac{\epsilon_i}{\Gamma_{\rm max}}\le \hbar^2\lambda_i \le \frac{\epsilon_i}{\Gamma_{\rm min}}.\label{eq:inequality_for_eivenvalues_parent_Hamiltonian_and_Lindblaidna}
\end{align}
Equation~(\ref{eq:inequality_for_eivenvalues_parent_Hamiltonian_and_Lindblaidna}) shows that the Lindbladian gap is bounded above and below by system-size-independent multiples of the parent-Hamiltonian gap. In particular, since $\epsilon_2=O(|\bm k_{\min}|^2)$, we obtain $\lambda_2=O(|\bm k_{\min}|^2)$, which proves that the effective Lindbladian is gapless. This result does not exclude a spectrum softer than quadratic.

As shown in Sec.~\ref{sec:Example}, the low-energy spectra of the parent Hamiltonians typically exhibit $\bm{k}^2$ dispersion. This corresponds to diffusion modes in open quantum systems. Reference \cite{Huang2025hydrodynamics} identified diffusive modes with $\bm{k}^2$ dispersion as NG modes in $U(1)$ SWSSB systems by field-theoretical methods. When $\Gamma_{\bm n,\bm m}=\Gamma$, an exact parent-Hamiltonian eigenmode with quadratic long-wavelength dispersion maps directly to a diffusive eigenmode of the effective Lindbladian.

Physically, the ladder operator $\hat{Q}^{\dagger}$ creates a pair across two subsystems in the doubled Hilbert space (dubbed an incoherenton in Ref.~\cite{Haga2023quasiparticles}), and the infinite-temperature state \eqref{eq:vectorization_form_of_infinite_temperature_state} is regarded as a Bose--Einstein condensate of such pairs. In fact, the long-range correlation in the R\'enyi-2 correlator \eqref{eq:definition_of_Renyi2_correlator} indicates off-diagonal long-range order in the doubled Hilbert space. The ARQMBS states create a long-wavelength density modulation on top of the infinite-temperature state, leading to diffusive modes with small relaxation rates. The quadratic or softer dispersion of the dissipative NG modes in $U(1)$ SWSSB shows a stark contrast with the Bogoliubov modes with linear dispersion in ordinary $U(1)$ SSB. From our results, the origin of the $O(\bm{k}^2)$ upper bound on the low-energy spectrum can be understood as a consequence of the locality and the frustration-free structure of $-\hat{\mathcal{L}}_{\rm eff}$.

Here, we remark on related works. Refs.~\cite{Ogunnaike2023unifying,Moudgalya2024symmetries} consider Brownian or noisy circuits whose averaged dynamics is governed by a superoperator with Hermitian local generators and no fixed coherent Hamiltonian. In contrast, our setup retains a deterministic Hamiltonian and restricts the Lindblad operators to local functions of charge-density operators. In addition, our setup excludes Lindblad operators with off-diagonal elements, such as $\hat{S}_j^x\hat{S}_{j+1}^x+\hat{S}_j^y\hat{S}_{j+1}^y$, whereas Refs.~\cite{Ogunnaike2023unifying,Moudgalya2024symmetries} allow such terms. Therefore, our results complement those of Refs.~\cite{Ogunnaike2023unifying,Moudgalya2024symmetries}.

\section{Examples}\label{sec:Example}
In this section, we consider several closed-system models exhibiting RQMBS and ARQMBS states, together with their open-system counterparts. For simplicity, we focus on two-leg ladders in closed systems and the corresponding one-dimensional chains in open systems. Some of these models were discussed in Ref.~\cite{Wildeboer2022quantum}.

Throughout this section, we assume that the full two-leg Hamiltonians are nonintegrable at the chosen generic parameter values in isolated systems. We exclude known integrable limits. In particular, the spin-$1/2$ model at $\Delta=0$ and the spinless-fermion model at $U=0$ are unitarily equivalent to the integrable one-dimensional Fermi--Hubbard model~\cite{Lieb1968absence,Medvedyeva2016exact}. We therefore exclude these parameter values when treating the models as scarred systems.

In all the examples below, interactions that are diagonal in the product basis do not contribute to the parent Hamiltonian. This follows from $\hat{H}_{\rm D}\hat{\mathcal P}=0$, where $\hat{H}_{\rm D}$ denotes the part of $\hat{H}_{\rm tot}$ that is diagonal in the product basis. These diagonal terms can change the integrability properties of the original two-leg Hamiltonian without modifying the parent Hamiltonian or its gapless excitations.

\subsection{Spin-1/2 {\it XXZ} model}\label{subsec:spin-1/2_XXZ}
The first example is a spin-$1/2$ {\it XXZ} model. We consider the following Hamiltonian and operators $\{\hat{L}_j\}$:
\begin{align}
\hat{H}&=J\sum_{j=1}^{M-1}(\hat{S}_{j}^x\hat{S}_{j+1}^x+\hat{S}_j^y\hat{S}_{j+1}^y+\Delta\hat{S}_j^z\hat{S}_{j+1}^z),\label{eq:definition_of_Hamiltonian_spin-1/2_XXZ_example}\\
\hat{L}_j&\equiv \sqrt{\gamma}\hat{S}_j^z,\label{eq:definition_of_Lj_spin-1/2_XXZ_ex}
\end{align}
where $\hat{S}_j^{\mu}\;(\mu=x,y,z)$ is the spin-1/2 operator at site $j$, $J$ is the interaction strength, $\Delta$ is the anisotropy parameter, and $\gamma(> 0)$ is the strength of the inter-ladder interaction and has units of $[T]^{-1}$. Here, we impose open boundary conditions. This model satisfies the following commutation relations $[\hat{H},\hat{S}_{\rm tot}^z]=0$ and $[\hat{L}_j, \hat{S}_{\rm tot}^z]=0$ for all $j$. Therefore, this model satisfies all the symmetry and locality conditions discussed in Sec.~\ref{subsec:spin-S}, allowing us to construct RQMBS and ARQMBS states. 

The ladder operator, the root state, and the RQMBS states are given by
\begin{align}
\hat{Q}^{\dagger}&=\sum_{j=1}^M\hat{S}_j^+\otimes\hat{S}_j^+,\label{eq:definition_of_ladder_operator_spin-1/2_XXZ_ex}\\
\cket{S_0}&\equiv \cket{-1/2 \ldots-1/2}\otimes \cket{-1/2 \ldots -1/2},\label{eq:definition_of_root_state_spin-1/2_XXZ_ex}\\
\cket{S_n}&=\frac{1}{\sqrt{W_{M,n}}}\sum_{\bm{n}: \hat{N}\cket{\bm{n}}=n\cket{\bm{n}}}\cket{\bm{n}}\otimes\cket{\bm{n}},\;n=0,1,\ldots,M,\label{eq:definition_of_RQMBS_state_spin-1/2_XXZ_ex}
\end{align}
where $\hat{S}_j^{\pm}\equiv \hat{S}_j^x\pm i\hat{S}_j^y$ and $W_{M,n}\equiv \binom{M}{n}$. The enlarged scar subspace can be written as
\begin{align}
\mathcal{H}_{P_n}&={\rm Span}\{\cket{\bm{n}}\otimes\cket{\bm{n}}~|~n_j=\pm1/2, \hat{N}\cket{\bm{n}}=n\cket{\bm{n}} \},\label{eq:enlarged_scar_subspace_spin-1/2_XXZ}\\
\mathcal{H}_P&\equiv \bigoplus_{n=0}^M\mathcal{H}_{P_n}.\label{eq:definition_ot_total_enlarged_scar_subspace_spin-1/2_XXZ}
\end{align}
Let $\hat{\mathcal{P}}_n$ and $\hat{\mathcal{P}}$ denote the projectors onto $\mathcal{H}_{P_n}$ and $\mathcal{H}_P$, respectively. To simplify the notation, we introduce $\hat{S}_{j,1}^{\mu}\equiv \hat{S}_j^{\mu}\otimes \hat{1}$ and $\hat{S}_{j,2}^{\mu}\equiv \hat{1}\otimes\hat{S}_{j}^{\mu}\;(\mu=x,y,z,+,-)$. Direct calculations show that the parent Hamiltonian of the system becomes
\begin{align}
&\hat{H}_{\rm p}\notag \\
&=\hat{\mathcal{P}}\hat{H}^2_0\hat{\mathcal{P}}\notag \\
&=J^2\hat{\mathcal{P}}\sum_{j=1}^{M-1}\left[\frac{1}{4}-\frac{1}{4}(\hat{S}_{j,1}^z+\hat{S}_{j,2}^z)(\hat{S}_{j+1,1}^z+\hat{S}_{j+1,2}^z)\right.\notag \\
&\left.\quad -\frac{1}{2}\hat{S}_{j,1}^+\hat{S}_{j+1,1}^-\hat{S}_{j,2}^+\hat{S}_{j+1,2}^--\frac{1}{2}\hat{S}_{j,1}^-\hat{S}_{j+1,1}^+\hat{S}_{j,2}^-\hat{S}_{j+1,2}^+\right]\hat{\mathcal{P}}.\label{eq:parent_Hamiltonian_S=1/2_XXZ_ex}
\end{align}
Since the enlarged scar subspace consists of paired states of the form $|\bm n\rangle\otimes|\bm n\rangle$, it can be mapped onto a single-ladder Hilbert space spanned by $\{|\bm n\rangle\in\mathcal H\}$. Under this mapping, we obtain $\hat{s}_j^z\equiv \hat{\mathcal{P}}(\hat{S}_{j,1}^z+\hat{S}_{j,2}^z)\hat{\mathcal{P}}/2$, $\hat{s}_j^+\equiv \hat{\mathcal{P}}\hat{S}_{j,1}^+\hat{S}_{j,2}^+\hat{\mathcal{P}}$, and $\hat{s}^-_j\equiv (\hat{s}_j^+)^{\dagger}$. The parent Hamiltonian reduces to
\begin{align}
\hat{H}_{\rm p}&=J^2\sum_{j=1}^{M-1}\left(\frac{1}{4}-\hat{\bm{s}}_j\cdot\hat{\bm{s}}_{j+1}\right).\label{eq:parent_Hamiltonian_spin-1/2_XXZ}
\end{align}
This is the spin-1/2 ferromagnetic Heisenberg model. 

As shown in Appendix C of Ref.~\cite{Kunimi2025systematic}, the low-energy excited states of the spin-1/2 ferromagnetic Heisenberg model are magnon excitations. In this case, exact energy eigenstates of Eq.~(\ref{eq:parent_Hamiltonian_spin-1/2_XXZ}) and their energy variances are given by
\begin{align}
\cket{AS_n(k)}&\propto \hat{s}^+(k)(\hat{s}_{\rm tot}^+)^{n-1}\cket{-1/2\ldots-1/2},\label{eq:definition_of_magnon_ex_spin-1/2_ex}\\
\Delta E_n^2&=J^2\left[1-\cos\left(\frac{\pi k}{M}\right)\right],\quad k=1,2,\ldots,M-1,\label{eq:exact_energy_variance_of_spin-1/2_XXZ_ex}
\end{align}
where $\hat{s}_{\rm tot}^+\equiv \sum_{j=1}^M\hat{s}_j^+$, $\hat{s}^+(k)\equiv\sum_{j=1}^Mf_j(k)\hat{s}_j^+ $, and $f_j(k)\equiv \cos[\pi k(2j-1)/(2M)]$. For a fixed positive integer $k=O(1)$ and $\epsilon\leq n/M\leq1-\epsilon$, where $\epsilon>0$ is a system-size-independent constant, we obtain $\Delta E_n^2=O(M^{-2})$. As shown in Appendix \ref{app:entanglement_entropy_of_the_variational_states}, the excited states have the same qualitative EE scaling as the corresponding RQMBS states. Therefore, these states constitute ARQMBS states.

Next, we consider the corresponding open quantum system. This model has been investigated in various contexts~\cite{Znidaric2010exact,Znidaric2010dephasing,Cai2013algebraic,Znidaric2014large,Medvedyeva2016influence,Gu2025spontaneous}. A similar but distinct setup was also investigated in Ref.~\cite{Hauser2026strong}.

In this case, the effective Lindbladian becomes
\begin{align}
\hat{\mathcal{L}}_{\rm eff}&=-\frac{J^2}{\hbar^2\gamma}\sum_{j=1}^{M-1}\left(\frac{1}{4}-\hat{\bm{s}}_j\cdot\hat{\bm{s}}_{j+1}\right)=-\frac{1}{\hbar^2\gamma}\hat{H}_{\rm p}.\label{eq:effective_Linabladian_S=1/2_XXZ}
\end{align}
Therefore, the effective Lindbladian is the spin-1/2 ferromagnetic Heisenberg model up to an overall constant. From Eq.~(\ref{eq:exact_energy_variance_of_spin-1/2_XXZ_ex}), we obtain the following Lindbladian decay rates:
\begin{align}
\lambda_k&=\frac{J^2}{\hbar^2 \gamma}\left[1-\cos\left(\frac{\pi k}{M}\right)\right],\;k=1,2,\ldots, M-1.\label{eq:Lindblad_spectra_spin-1/2_XXZ}
\end{align}
For $k=O(1)$, we obtain $\lambda_k=O(1/M^2)$, which is characteristic of diffusive scaling.

\subsection{Spin-1 {\it XXZ} model}\label{subsec:spin-1_XXZ}
The next example is a spin-1 XXZ model. The Hamiltonian and local operators $\{\hat L_j\}$ are defined by Eqs.~(\ref{eq:definition_of_Hamiltonian_spin-1/2_XXZ_example}) and (\ref{eq:definition_of_Lj_spin-1/2_XXZ_ex}), with $\hat S_j^\mu$ taken to be a spin-1 operator.

The ladder operator, the root state, and the RQMBS states are given by
\begin{align}
\hat{Q}^{\dagger}&\equiv \sum_{j=1}^M\left(\hat{F}_{j,1}^{0,-1}\hat{F}_{j,2}^{0,-1}+2\hat{F}_{j,1}^{1,0}\hat{F}_{j,2}^{1,0}\right),\label{eq:definition_of_Qdagger_S=1_system_ex}\\
\cket{S_0}&\equiv\cket{-,\ldots,-}\otimes \cket{-,\ldots,-},\label{eq:root_state_spin-1_XXZ_ex}\\
\cket{S_n}&\equiv  \frac{1}{\sqrt{W_{M,n}^{S=1}}}\sum_{\bm{n}:\hat{N}\cket{\bm{n}}=n\cket{\bm{n}}}\cket{\bm{n}}\otimes\cket{\bm{n}},\notag \\
&\hspace{10.0em}\text{for }\;n=0,1,\ldots,2M,\label{eq:definition_of_RQMBS_spin-1_XXZ_ex}
\end{align}
where $\hat{F}_{j,1}^{\mu,\nu}\equiv \hat{F}_j^{\mu,\nu}\otimes \hat{1}$ and $\hat{F}_{j,2}^{\mu,\nu}\equiv \hat{1}\otimes\hat{F}_j^{\mu,\nu}$. The normalization constant $W_{M,n}^{S=1}$ is determined by Eqs.~(\ref{eq:expression_of_W_appendix}) and (\ref{eq:reccurent_relation_W_appendix}) for the $S=1$ case. The enlarged scar subspace is defined by
\begin{align}
\mathcal{H}_{P_n}&={\rm Span}\{\cket{\bm{n}}\otimes\cket{\bm{n}}~|~n_j=0,\pm1, \hat{N}\cket{\bm{n}}=n\cket{\bm{n}} \},\label{eq:enlarged_scar_subspace_spin-1_XXZ}\\
\mathcal{H}_P&\equiv \bigoplus_{n=0}^{2M}\mathcal{H}_{P_n}.\label{eq:definition_ot_total_enlarged_scar_subspace_spin-1_XXZ}
\end{align}
After the single-ladder mapping used in Sec.~\ref{subsec:spin-1/2_XXZ}, the parent Hamiltonian becomes
\begin{align}
\hat{H}_{\rm p}&=4J^2\sum_{j=1}^{M-1}\hat{h}_{j,j+1}^{S=1},\label{eq:parent_Hamiltonian_spin-1_XXZ}\\
\hat{h}_{j,j+1}^{S=1}&\equiv\hat{D}_{j,j+1}+\hat{U}_{j,j+1}+\hat{F}_{j,j+1}+\hat{F}_{j,j+1}', \\
\hat{D}_{j,j+1}&\equiv \frac{1}{2}(\cket{0,-}-\cket{-,0})(\bra{0,-}-\bra{-,0})_{j,j+1},\label{eq:definition_of_D_Motzkin}\\
\hat{U}_{j,j+1}&\equiv \frac{1}{2}(\cket{0,+}-\cket{+,0})(\bra{0,+}-\bra{+,0})_{j,j+1},\label{eq:definition_of_U_Motzkin}\\
\hat{F}_{j,j+1}&\equiv \frac{1}{2}(\cket{0,0}-\cket{+,-})(\bra{0,0}-\bra{+,-})_{j,j+1},\label{eq:definition_of_F_Motzkin}\\
\hat{F}'_{j,j+1}&\equiv \frac{1}{2}(\cket{0,0}-\cket{-,+})(\bra{0,0}-\bra{-,+})_{j,j+1}.\label{eq:definition_of_F'_Motzkin}
\end{align}
Upon omitting $\hat{F}_{j,j+1}'$, the bulk local terms coincide with those of the Motzkin-chain Hamiltonian~\cite{Movassagh2017entanglement,Hao2023exact}. However, the full Hamiltonian and its ground-state structure differ from those of the Motzkin chain.

We next investigate the excited states of the parent Hamiltonian. We do not provide analytical expressions for the low-energy excited states in general fixed-$n$ sectors. The $n=1$ and $n=2M-1$ sectors, however, reduce to one-particle and one-hole problems, respectively. For other values of $n$, we evaluate the excitation energy variationally, as described in Sec.~\ref{subsubsec:ARQMBS_spin-S}. The parent Hamiltonian satisfies the locality conditions and Eq.~(\ref{eq:assumption_number_of_hpj}), and its local terms obey $\|4J^2\hat{h}_{j,j+1}^{S=1}\|_{\rm op}=O(1)$. In addition, the Perron--Frobenius theorem is applicable to the parent Hamiltonian~\cite{Tasaki2020physics}, because its off-diagonal matrix elements are nonpositive and the local transitions generated by $\hat D_{j,j+1}$, $\hat U_{j,j+1}$, $\hat F_{j,j+1}$, and $\hat F'_{j,j+1}$ connect all product-basis configurations within each fixed-$U(1)$-charge sector. Hence, the ground state in each fixed-charge sector is unique. Under the condition $\epsilon\leq\frac{n}{M}\leq2-\epsilon$, where $\epsilon>0$ is a system-size-independent constant, the parent-Hamiltonian gap has an $O(M^{-2})$ upper bound. Therefore, ARQMBS states exist.

Next, we consider the corresponding open quantum systems. The Hamiltonian and Lindblad operators are given by Eqs.~(\ref{eq:definition_of_Hamiltonian_spin-1/2_XXZ_example}) and (\ref{eq:definition_of_Lj_spin-1/2_XXZ_ex}), with $\hat{S}_j^{\mu}$ taken to be a spin-1 operator. A special case of this model was investigated in Ref.~\cite{Gu2025spontaneous}.

Direct calculation shows that the effective Lindbladian becomes
\begin{align}
\hat{\mathcal{L}}_{\rm eff}&=-\frac{4J^2}{\hbar^2\gamma}\sum_{j=1}^{M-1}\hat{h}_{j,j+1}^{S=1}=-\frac{1}{\hbar^2\gamma}\hat{H}_{\rm p}.\label{eq:effective_Lindbladian_S=1_XXZ}
\end{align}
As mentioned above, we do not provide analytical expressions for the low-energy excited states of this parent Hamiltonian. The variational argument gives an $O(1/M^2)$ upper bound on the Lindbladian gap. This proves gaplessness but does not establish a quadratic lower bound. 

\subsection{Spinless fermion model}\label{subsec:spinless_fermion}
Next, we consider a spinless fermion model. The Hamiltonian and operators $\{\hat{L}_j\}$ are given by
\begin{align}
\hat{H}&=-J\sum_{j=1}^{M-1}(\hat{c}^{\dagger}_{j+1}\hat{c}_j+\hat{c}^{\dagger}_j\hat{c}_{j+1})+U\sum_{j=1}^{M-1}\hat{n}_j\hat{n}_{j+1},\label{eq:definition_of_Hamiltonian_spinless_fermion_ex}\\
\hat{L}_j&\equiv \sqrt{\gamma}\hat{n}_j,\label{eq:definition_of_Lj_spinless_fermion}
\end{align}
where $J$ is the hopping amplitude, $U$ is the strength of the density-density nearest-neighbor interaction, and $\gamma$ is the strength of the inter-ladder interaction and has units of $[T]^{-1}$. 

The ladder operator, the root state, and the RQMBS states are given by
\begin{align}
\hat{Q}^{\dagger}&\equiv \sum_{j=1}^M\hat{c}_j^{\dagger}\otimes\hat{c}_j^{\dagger},\label{eq:definition_of_ladder_operator_spinless_fermions_ex}\\
\cket{S_0}&\equiv \cket{\rm vac}\otimes\cket{\rm vac},\label{eq:definition_of_root_state_spinless_fermion_ex}\\
\cket{S_n}&\equiv \frac{1}{\sqrt{W_{M,n}}}\sum_{\bm{n}:\hat{N}\cket{\bm{n}}=n\cket{\bm{n}}}\cket{\bm{n}}\otimes\cket{\bm{n}},\;n=0, 1,\ldots, M.\label{eq:definition_of_RQMBS_spinless_fermion_ex}
\end{align}
The enlarged scar subspace can be written as
\begin{align}
\mathcal{H}_{P_n}&={\rm Span}\{\cket{\bm{n}}\otimes\cket{\bm{n}}~|~n_j=0,1, \hat{N}\cket{\bm{n}}=n\cket{\bm{n}} \},\label{eq:enlarged_scar_subspace_spinless_fermion}\\
\mathcal{H}_P&\equiv \bigoplus_{n=0}^M\mathcal{H}_{P_n}.\label{eq:definition_ot_total_enlarged_scar_subspace_spinless_fermion}
\end{align}
The parent Hamiltonian becomes
\begin{align}
&\hat{H}_{\rm p}\notag \\
&=2J^2\sum_{j=1}^{M-1}\left[(\cket{1,0}\otimes\cket{1,0}-\cket{0,1}\otimes\cket{0,1})(\bra{1,0}\otimes\bra{1,0})\right.\notag \\
&\left.\hspace{1.0em}+(\cket{0,1}\otimes\cket{0,1}-\cket{1,0}\otimes\cket{1,0})(\bra{0,1}\otimes\bra{0,1})\right]_{j,j+1}.\label{eq:parent_Hamiltonian_spinless_fermion_bracket_form}
\end{align}
Using the single-ladder mapping and defining the following spin-$1/2$ operators on the effective single ladder $\hat{s}_j^+\equiv \cket{0}\bra{1}_j$, $\hat{s}_j^-\equiv \cket{1}\bra{0}_j$, $\hat{s}_j^z\equiv (1/2)(\cket{0}\bra{0}-\cket{1}\bra{1})_j$, $\hat{s}^x_j=(\hat{s}^+_j+\hat{s}^-_j)/2$, and $\hat{s}_j^y\equiv (\hat{s}_j^+-\hat{s}_j^-)/(2i)$, we obtain
\begin{align}
\hat{H}_{\rm p}=4J^2\sum_{j=1}^{M-1}\left(\frac{1}{4}-\hat{\bm{s}}_j\cdot\hat{\bm{s}}_{j+1}\right).\label{eq:parent_Hamiltonian_spinless_fermion_single_ladder_ex}
\end{align}
This is the same as Eq.~(\ref{eq:parent_Hamiltonian_spin-1/2_XXZ}) up to an overall factor. Therefore, under the condition $\epsilon\leq n/M\leq1-\epsilon$, where $\epsilon>0$ is a system-size-independent constant, we obtain low-energy gapless excitations by the same argument as in the spin-$1/2$ case.

Next, we consider the corresponding open quantum systems. The Hamiltonian and Lindblad operators are given by  Eqs.~(\ref{eq:definition_of_Hamiltonian_spinless_fermion_ex}) and (\ref{eq:definition_of_Lj_spinless_fermion}), respectively. This model has been extensively studied in various contexts~\cite{Medvedyeva2016exact,Ishiyama2025exact,Ishiyama2025exact_current,Penc2026linear,Essler2026interacting}. 

Direct calculation shows that the effective Lindbladian becomes
\begin{align}
\hat{\mathcal{L}}_{\rm eff}&=-\frac{4J^2}{\hbar^2\gamma}\sum_{j=1}^{M-1}\left(\frac{1}{4}-\hat{\bm{s}}_j\cdot\hat{\bm{s}}_{j+1}\right)=-\frac{1}{\hbar^2\gamma}\hat{H}_{\rm p}.\label{eq:effective_Lindbladian_spinless_fermion}
\end{align}
As in the spin-$1/2$ {\it XXZ} model, we can obtain the following decay rates:
\begin{align}
\lambda_k=\frac{4J^2}{\hbar^2\gamma}\left[1-\cos\left(\frac{\pi k}{M}\right)\right],\:k=1,2,\ldots,M-1.\label{eq:Lindbiadian_spectra_spinless_fermion}
\end{align}
For $k=O(1)$, we obtain $\lambda_k=O(1/M^2)$, which is characteristic of diffusive scaling.

In the absence of the nearest-neighbor interaction [i.e., $U=0$ in Eq.~\eqref{eq:definition_of_Hamiltonian_spinless_fermion_ex}], the Lindbladian is mapped to the one-dimensional Fermi--Hubbard model with a purely imaginary interaction strength. At this solvable point, a gapless spectrum with $1/M^2$ scaling was obtained by using the Bethe-ansatz method \cite{Medvedyeva2016exact}, and the infinite-temperature state corresponds to an $\eta$-pairing state of the Fermi--Hubbard model. The quadratic spectrum is closely related to excitations around the $\eta$-pairing state (see, e.g., Ref.~\cite{Nakagawa2021eta} for a calculation of the quadratic dispersion relation). Our result establishes gaplessness beyond this solvable point, albeit only in the strong-dissipation regime. We note that the Lindbladian with $U\neq 0$ no longer possesses the $\eta$-pairing symmetry but remains gapless due to the ARQMBS states.

\subsection{Soft-core boson model}\label{subsec:soft-core_boson}
Finally, we consider a soft-core boson model. The Hamiltonian and operators $\{\hat{L}_j\}$ are given by
\begin{align}
\hat{H}&=J\sum_{j=1}^{M-1}(\hat{b}^{\dagger}_{j+1}\hat{b}_j+\hat{b}^{\dagger}_j\hat{b}_{j+1})+\frac{U}{2}\sum_{j=1}^M\hat{n}_j(\hat{n}_j-1),\label{eq:definition_of_Bose-Hibbard_Hamiltonian_ex}\\
\hat{L}_j&\equiv \sqrt{\gamma}\hat{n}_j,\label{eq:definition_of_Lj_soft-core_boson_ex}
\end{align}
where $J$ is the hopping amplitude, $U$ is the strength of the onsite interaction, and $\gamma$ is the strength of the inter-ladder interaction and has units of $[T]^{-1}$. 

The ladder operator, the root state, and the RQMBS states are given by
\begin{align}
\hat{Q}^{\dagger}&\equiv \sum_{j=1}^M\hat{b}^{\dagger}_j\otimes\hat{b}_j^{\dagger},\label{eq:definition_of_ladder_operator_soft-core_boson_ex}\\
\cket{S_0}&\equiv \cket{\rm vac}\otimes\cket{\rm vac},\label{eq:definition_of_root_state_sof-core_boson_ex}\\
\cket{S_n}&=\frac{1}{\sqrt{W_{M,n}^{\rm B}}}\sum_{\bm{n}:\hat{N}\cket{\bm{n}}=n\cket{\bm{n}}}\cket{\bm{n}}\otimes\cket{\bm{n}},\label{eq:definition_of_RQMBS_softcore_boson_ex}
\end{align}
where $W_{M,n}^{\rm B}\equiv \binom{M+n-1}{n}$. The enlarged scar subspace is given by
\begin{align}
\mathcal{H}_{P_n}&={\rm Span}\{\cket{\bm{n}}\otimes\cket{\bm{n}} | n_j=0,1,\ldots,n, \hat{N}\cket{\bm{n}}=n\cket{\bm{n}} \},\label{eq:enlarged_scar_subspace_soft-core_boson}\\
\mathcal{H}_P&\equiv \bigoplus_{n=0}^{\infty}\mathcal{H}_{P_n}.\label{eq:definition_ot_total_enlarged_scar_subspace_soft-core_boson}
\end{align}
The parent Hamiltonian becomes
\begin{align}
\hat{H}_{\rm p}&=\sum_{j=1}^{M-1}\hat{h}_{j,j+1},\label{eq:definition_of_parent_Hamiltonian_soft-core_bosons}\\
\hat{h}_{j,j+1}&\equiv 2J^2\sum_{n,m=0}^{\infty}\left[(2nm+n+m)(\cket{n,m}\otimes\cket{n,m})_{j,j+1}\right.\notag \\
&-n(m+1)(\cket{n-1,m+1}\otimes\cket{n-1,m+1})_{j,j+1}\notag \\
&\left.-m(n+1)(\cket{n+1,m-1}\otimes\cket{n+1,m-1})_{j,j+1}\right]\notag \\
&\times(\bra{n,m}\otimes\bra{n,m})_{j,j+1}.\label{eq:definition_of_hjj+1_soft-core_bosons}
\end{align}
Here, any ket in Eq.~(\ref{eq:definition_of_hjj+1_soft-core_bosons}) containing a negative occupation number is defined to be zero. Using the single-ladder mapping, we obtain the operator representation of the parent Hamiltonian
\begin{align}
\hat{H}_{\rm p}&=2J^2\sum_{j=1}^{M-1}\left\{(2\hat{n}_j\hat{n}_{j+1}+\hat{n}_j+\hat{n}_{j+1})\right.\notag \\
&\left.-\left[\hat{b}_{j+1}^{\dagger}\hat{b}_j\sqrt{\hat{n}_j(\hat{n}_{j+1}+1)}+\sqrt{\hat{n}_j(\hat{n}_{j+1}+1)}\hat{b}^{\dagger}_j\hat{b}_{j+1}\right]\right\}.\label{eq:operator_representation_parent_Hamiltonian_Bose-Hubbard_ladder}
\end{align}
Here, we define the following operators:
\begin{align}
\hat{K}_j^+&\equiv \hat{b}_j^{\dagger}\sqrt{\hat{n}_j+1},\quad \hat{K}_j^-\equiv \sqrt{\hat{n}_j+1}\hat{b}_j,\quad \hat{K}^z_j\equiv \hat{n}_j+\frac{1}{2}.\label{eq:definition_of_K+K-Kz_soft_core_boson}
\end{align}
These operators satisfy the $SU(1,1)$ algebra
\begin{align}
[\hat{K}_j^+, \hat{K}_j^-]=-2\hat{K}_j^z,\quad [\hat{K}^z_j, \hat{K}_j^{\pm}]=\pm\hat{K}_j^{\pm}.\label{eq:SU(1,1)_algebra_for_K_soft-core_boson}
\end{align}
Using these operators, we can rewrite the parent Hamiltonian (\ref{eq:operator_representation_parent_Hamiltonian_Bose-Hubbard_ladder}) as
\begin{align}
\hat{H}_{\rm p}=4J^2\sum_{j=1}^{M-1}\left(-\frac{1}{4}-\hat{K}_j^x\hat{K}_{j+1}^x-\hat{K}_j^y\hat{K}_{j+1}^y+\hat{K}_j^z\hat{K}_{j+1}^z\right),\label{eq:parent_Hamitonian_su(1,1)}
\end{align}
where $\hat{K}_j^x\equiv (\hat{K}_j^++\hat{K}_j^-)/2$ and $\hat{K}_j^y\equiv (\hat{K}_j^+-\hat{K}_j^-)/(2i)$. This Hamiltonian is the ferromagnetic $SU(1,1)$-invariant Heisenberg model~\cite{Giardina2009duality,Carinci2019orthogonal,Groenevelt2019orthogonal}. As in the spin-$1/2$ ferromagnetic Heisenberg model, the low-energy gapless excitations can be obtained analytically. The corresponding $SU(1,1)$ magnon states can be written as
\begin{align}
\cket{AS_n(k)}&\propto \hat{K}^+_{\rm tot}(k)(\hat{K}_{\rm tot}^+)^{n-1}\cket{\rm vac},\quad n=1,2,\ldots,\label{eq:SU(1,1)_magnon_excited_state}\\
\hat{H}_{\rm p}\cket{AS_n(k)}&=4J^2\left[1-\cos\left(\frac{\pi k}{M}\right)\right]\cket{AS_n(k)},\label{eq:eigenenergy_for_SU(1,1)_magnon}
\end{align}
where $\hat{K}_{\rm tot}^+\equiv  \sum_{j=1}^M\hat{K}_j^+$, $\hat{K}_{\rm tot}^+(k)\equiv \sum_{j=1}^Mf_j(k)\hat{K}_j^+$, and $f_j(k)\equiv \cos[\pi k(2j-1)/(2M)]$ with $k=1,2,\ldots,M-1$. The above results follow from $[\hat{H}_{\rm p},\hat{K}_{\rm tot}^{+}]=0$ and the fact that $\hat{K}_{\rm tot}^{+}(k)|{\rm vac}\rangle$ is an exact eigenstate of the parent Hamiltonian. As in the spin-1/2 {\it XXZ} case, we obtain the gapless excited states of the parent Hamiltonian. For $0<\nu_-\leq n/M\leq\nu_+<\infty$, where $\nu_-$ and $\nu_+$ are system-size-independent constants, the soft-core boson model supports ARQMBS states.

We note that in this model, the exact $SU(1,1)$ mapping allows us to establish gaplessness directly, without using the additional bound in Eq.~(\ref{eq:assumptions_for_bosons}).

Next, we consider the corresponding open quantum systems. The Hamiltonian and Lindblad operators are given by Eqs.~(\ref{eq:definition_of_Bose-Hibbard_Hamiltonian_ex}) and (\ref{eq:definition_of_Lj_soft-core_boson_ex}). This model has been studied  theoretically~\cite{Poletti2012interaction,Poletti2013emergence,Wellnitz2022rise,Tang2026mean}, and corresponding quantum-simulation experiments have been reported~\cite{Bouganne2020anomalous,Vatre2023dynamics}.

Direct calculation shows that the effective Lindbladian becomes [see also Ref.~\cite{Wellnitz2022rise}]
\begin{align}
&\hat{\mathcal{L}}_{\rm eff}\notag \\
&=-\frac{4J^2}{\hbar^2\gamma}\sum_{j=1}^{M-1}\left(-\frac{1}{4}-\hat{K}_j^x\hat{K}_{j+1}^x-\hat{K}_j^y\hat{K}_{j+1}^y+\hat{K}_j^z\hat{K}_{j+1}^z\right)\notag \\
&=-\frac{1}{\hbar^2\gamma}\hat{H}_{\rm p}.\label{eq:effective_Lindbladian_Bose-Hubbard}
\end{align}
The Lindbladian decay rates are given by
\begin{align}
\lambda_k=\frac{4J^2}{\hbar^2\gamma}\left[1-\cos\left(\frac{\pi k}{M}\right)\right],\:k=1,2,\ldots,M-1.\label{eq:Lindbiadian_spectra_softcore_bosons}
\end{align}
For $k=O(1)$, we obtain $\lambda_k=O(1/M^2)$, which is characteristic of diffusive scaling.

\section{Summary and outlook}\label{sec:summary}
In this paper, we have proved a correspondence between the spectrum of the parent Hamiltonian of the RQMBS states and that of the second-order effective Lindbladian in the strongly dissipative regime. To prove this, we have rigorously established the existence of the ARQMBS states corresponding to the RQMBS states for spin-$S$, multicomponent fermionic, and multicomponent bosonic systems on $d$-dimensional Bravais lattices. We have proved that when restricted to the enlarged scar subspace, the parent Hamiltonian admits a local Rokhsar--Kivelson-type frustration-free decomposition. Using the SMA, we have constructed variational states that are orthogonal to all RQMBS states and obtained an $O(|\bm{k}_{\rm min}|^2)$ upper bound on the parent Hamiltonian gap, which proves the gaplessness of the parent Hamiltonian and the vanishing of the energy variance of the variational states with respect to the original Hamiltonian in the thermodynamic limit. We have also shown that the SMA variational states have the same qualitative EE scaling as the corresponding RQMBS states. From these results, we have concluded that the ARQMBS states exist under the assumptions summarized in Sec.~\ref{subsec:summary_construction_ARQMBS}.

Through the duality between doubled systems and vectorized open quantum systems~\cite{Teretenkov2024duality}, the RQMBS states in closed systems are mapped to the infinite-temperature states of the corresponding fixed-charge sectors in open quantum systems with strong $U(1)$ or $U(1)^f$ symmetry. We have evaluated the R\'enyi-2 and R\'enyi-1 correlators for spin-$S$, multicomponent fermionic, and multicomponent bosonic systems and found that these states exhibit SWSSB order in both the R\'enyi-2 and R\'enyi-1 senses~\cite{Lessa2025strong,Ziereis2025strong} under the finite-density conditions. In addition, for soft-core bosons, we have shown directly that the fidelity correlator exhibits long-range order. We have proved that the gap of the second-order effective Lindbladian in the strongly dissipative regime is bounded above and below by system-size-independent multiples of the parent-Hamiltonian gap and has an $O(|\bm k_{\min}|^2)$ upper bound. When $\Gamma_{\bm{n},\bm{m}}=\Gamma$ for all active transitions, the effective Lindbladian is exactly proportional to the parent Hamiltonian of the RQMBS states, and hence the two operators have identical eigenmodes. In models possessing exact quadratic low-energy dispersions, the corresponding low-decay-rate modes can be identified with diffusive NG modes of SWSSB~\cite{Huang2025hydrodynamics}.

Within the assumptions of the present work, the spectral argument does not depend on the detailed form of the Hamiltonian. It is essential to assume that the Lindblad operators are functions of local charge-density operators. In the strongly dissipative regime, the effective Lindbladian can be interpreted as the generator of an effective classical master equation in the state space $\mathcal{H}_P$. From this perspective, the frustration-freeness of $-\hat{\mathcal{L}}_{\rm eff}$ is related to a detailed-balance condition for the effective Markov process \cite{Masaoka2025rigorous}. If we allow a $U(1)$-symmetric Lindblad operator that is not diagonal with respect to the local charge basis, the detailed-balance condition of the effective Lindbladian could be violated. For example, the Lindbladian with stochastic asymmetric hopping reduces to the generator of the asymmetric simple exclusion process in a certain limit \cite{Haga2021liouvillian}, which does not satisfy the detailed-balance condition and exhibits a dynamical critical exponent of $3/2$ \cite{Masaoka2025rigorous,Gwa1992bethe,Kim95bethe}. Thus, the $O(|\bm{k}_{\rm min}|^2)$ upper bound on the Lindbladian gap could be violated for such Lindblad operators.

We have several open questions. First, our analysis is restricted to the strongly dissipative regime. Diffusive low-decay-rate modes are known to persist away from this regime in several models. In fact, the quadratic dispersion has been found in more general situations~\cite{Cai2013algebraic,Poletti2012interaction,Poletti2013emergence,Medvedyeva2016exact,Ishiyama2025exact,Ishiyama2025exact_current,Penc2026linear,Essler2026interacting,Wellnitz2022rise,Tang2026mean}. Our approach does not appear to be directly applicable to weak- and intermediate-dissipation regimes. Understanding the NG modes associated with SWSSB from the perspective of ARQMBS states remains a challenging problem.

A second open question is to identify conditions under which the Lindbladian spectrum is quadratic and corresponds to diffusive modes. Our calculations provide $O(|\bm k_{\min}|^2)$ upper bounds on both the parent-Hamiltonian gap and the corresponding effective Lindbladian gap. Establishing the quadratic dispersion requires matching lower bounds on the gaps, which are more difficult to derive than the corresponding upper bounds. Additional assumptions may therefore be necessary to establish these lower bounds. These upper bounds are also compatible with softer dispersions~\cite{Masaoka2024quadratic,Masaoka2025rigorous}, such as those occurring in dipole- or higher-moment-conserving systems~\cite{Feldmeier2020anomalous,Gromov2020fraction} and disordered systems~\cite{Agarwal2015anomalous,Vosk2015theory}. Dipole-conserving Hamiltonians and disorder potentials are allowed in our setup. We may obtain anomalously diffusive NG modes with dispersions proportional to $|\bm{k}|^z$ ($z>2$) in such systems. For example, dipole-moment-conserving systems exhibit $z=4$~\cite{Feldmeier2020anomalous,Gromov2020fraction}.

A third direction is to extend our analysis to systems with non-Abelian strong symmetries, such as strong $SU(N)$ or $SO(N)$ symmetries~\cite{Zhang2020stationary,Hashimoto2026construction,Li2025highly}. For non-Abelian groups, the Hilbert space decomposes into irreducible-representation spaces and multiplicity spaces, rather than simultaneous eigenspaces of commuting charges~\cite{Hall2015LieGroups}. Therefore, it is nontrivial to identify the appropriate maximally mixed sector states and their closed-system counterparts.

Finally, we discuss prospects for experimental implementation. In isolated systems, RQMBS states and their characteristic entanglement structures have been realized in a superconducting-qubit ladder~\cite{Dong2023disorder}, whereas ARQMBS states have not yet been observed. Very recently, the first experimental signatures of generic AQMBS states were reported on a trapped-ion quantum processor~\cite{Logaric2026dynamical}. In this experiment, AQMBS states corresponding to one-magnon excitations were prepared using quantum circuits. This direction is promising for observing ARQMBS states because Hamiltonians hosting RQMBS states are generally complicated and it is challenging to realize them in analog quantum simulators. We also note that the preparation of QMBS states in the domain-wall-conserving model was demonstrated on digital quantum simulators~\cite{Gustafson2023preparing}.

On the open-system side, $U(1)$ SWSSB was recently observed in an experiment on an ultracold Fermi gas in a two-dimensional optical lattice~\cite{Wang2026observation}. In this experiment, R\'enyi-1 and R\'enyi-2 correlators were reconstructed by combining experimental data with theoretical analysis, thereby demonstrating SWSSB order. Beyond observing SWSSB order, verifying the associated diffusive NG modes requires resolving long-wavelength, low-decay-rate modes. Although such measurements are experimentally challenging as they require dynamics with strong $U(1)$ symmetry and state preparation in a fixed-charge sector, promising candidates are $U(1)$-symmetric systems with dephasing. Such systems have been experimentally realized on various quantum-simulation platforms~\cite{Maier2019environment,Bouganne2020anomalous,Vatre2023dynamics,Liang2025dephasing}. With continued advances in quantum technologies, diffusive NG modes may become accessible in analog,
digital, and hybrid quantum simulators.


\begin{acknowledgments}
The authors thank D. Hashimoto, T. Nikuni, R. Kaneko, H. Yoshida, H. Katsura, and Y. Chiba for their helpful discussions. We used OpenAI’s ChatGPT (GPT-5.6 Pro and 6.0 Pro) to assist with literature organization, checks of derivations and calculations, and revision of the scientific descriptions. The authors independently verified all calculations, references, and scientific statements and take full responsibility for the final manuscript. This work was supported by JSPS KAKENHI Grants No.~JP25K00215 (M.K.), No.~JP22K13983 (T.H.), and No.~JP24K16989 (M.N.), and JST ASPIRE No.~JPMJAP24C2 (M.K.).
\end{acknowledgments}

\section*{Data Availability}
The data that support the findings of this article are openly
available~\cite{kunimi_2026_23156908}.
\vspace{1.0em}
\appendix
\section{Proof of the existence of rainbow quantum many-body scar states}\label{app:proof_of_RQMBS}
In this appendix, we prove Eq.~(\ref{eq:RQMBS_for_general_case}). The key identity, given in Ref.~\cite{Wildeboer2022quantum}, is
\begin{align}
(\hat{A}\otimes\hat{1})\sideset{}{'}\sum_{\bm{n}}\cket{\bm{n}}\otimes\cket{\bm{n}}=(\hat{1}\otimes\hat{A}^{\rm T})\sideset{}{'}\sum_{\bm{n}}\cket{\bm{n}}\otimes\cket{\bm{n}},\label{eq:identity_for_infinite_temperature_state}
\end{align}
where $\hat{A}$ is an operator acting only on one subsystem and satisfies $[\hat{A}, \hat{U}]=0$, and ${\rm T}$ denotes the transpose of the operator. Using the above identity, we obtain
\begin{align}
(\hat{H}\otimes\hat{1}-\hat{1}\otimes\hat{H}^{\ast})|\tilde{S}_n\rangle&=[\hat{H}\otimes\hat{1}-(\hat{H}^{\ast})^{\rm T}\otimes\hat{1}]|\tilde{S}_n\rangle\notag \\
&=(\hat{H}\otimes\hat{1}-\hat{H}\otimes\hat{1})|\tilde{S}_n\rangle\notag \\
&=0,\label{eq:vanishing_first_and_second_terms_for_RQMBS}
\end{align}
where we used $(\hat{H}^{\ast})^{\rm T}=\hat{H}^{\dagger}=\hat{H}$. The interaction terms reduce to
\begin{align}
&\sum_j(\hat{L}_j\otimes\hat{1}-\hat{1}\otimes\hat{L}_j)^2|\tilde{S}_n\rangle\notag \\
&=\sum_j(-2\hat{L}_j\otimes\hat{L}_j+\hat{L}_j^2\otimes \hat{1}+\hat{1}\otimes\hat{L}_j^2)|\tilde{S}_n\rangle\notag \\
&=0,\label{eq:vanishing_interaction_terms_for_RQMBS}
\end{align}
where we used the fact that $\hat{L}_j$ is diagonal in the product basis, and hence $\hat{L}_j^{\rm T}=\hat{L}_j$. Therefore, we have proved that $\hat{H}_{\rm tot}|\tilde{S}_n\rangle=0$.

\section{Details of the proofs for spin-$S$ systems}\label{app:proof_of_RQMBS_spin-S}
In this appendix, we prove some relations used in Sec.~\ref{subsec:spin-S}.

\subsection{Proof of $(\hat{Q}^{\dagger})^n\cket{S_0}=|\tilde{S}_n\rangle$}\label{app:detail_of_Qdagger_to_S0}
Here, we prove the relation $(\hat{Q}^{\dagger})^n\cket{S_0}=|\tilde{S}_n\rangle$ by induction. For $n=1$, direct calculations show
\begin{widetext}
\begin{align}
\hat{Q}^{\dagger}\cket{S_0}&=\sum_j|-S, -S,\ldots,\underbrace{-S+1}_{j},\ldots,-S\rangle\otimes|-S, -S,\ldots,\underbrace{-S+1}_j,\ldots,-S\rangle\notag \\
&=|\tilde{S}_1\rangle.\label{eq:relation_Qdagger_to_S0_spin-S_appendix}
\end{align}
We assume that $(\hat{Q}^{\dagger})^k\cket{S_0}=|\tilde{S}_k\rangle$ holds. We then calculate
\begin{align}
(\hat{Q}^{\dagger})^{k+1}\cket{S_0}&=k!\sum_{j=1}^M\sum_{m=-S}^{S-1}\sum_{\bm{n}:\hat{N}\cket{\bm{n}}=k\cket{\bm{n}}}(m+S+1)\hat{F}_{j}^{m+1,m}\cket{\bm{n}}\otimes\hat{F}_{j}^{m+1,m}\cket{\bm{n}}\notag \\
&=k!\sum_{j=1}^M\sum_{\bm{n}:\hat{N}\cket{\bm{n}}=k\cket{\bm{n}}, n_j\le S-1}(n_j+S+1)\cket{\bm{n}+\bm{e}_j}\otimes\cket{\bm{n}+\bm{e}_j}\notag \\
&=k!\sum_{j=1}^M\sum_{\bm{n}':\hat{N}\cket{\bm{n}'}=(k+1)\cket{\bm{n}'}, n_j'\ge -S+1}(n'_j+S)\cket{\bm{n}'}\otimes\cket{\bm{n}'}\notag \\
&=k!\sum_{j=1}^M\sum_{\bm{n}':\hat{N}\cket{\bm{n}'}=(k+1)\cket{\bm{n}'}}(n'_j+S)\cket{\bm{n}'}\otimes\cket{\bm{n}'}\notag \\
&=k!\sum_{j=1}^M(\hat{n}_j\otimes \hat{1})\sum_{\bm{n}:\hat{N}\cket{\bm{n}}=(k+1)\cket{\bm{n}}}\cket{\bm{n}}\otimes\cket{\bm{n}}\notag \\
&=(k+1)!\sum_{\bm{n}:\hat{N}\cket{\bm{n}}=(k+1)\cket{\bm{n}}}\cket{\bm{n}}\otimes\cket{\bm{n}}\notag \\
&=|\tilde{S}_{k+1}\rangle,\label{eq:proof_of_Qdagger_to_n_to_S0_calculation}
\end{align}
\end{widetext}
where $\bm{e}_j$ denotes the unit vector whose $j$th component is one and all other components are zero.

\subsection{Proof of the operator identity in RSGA}\label{app:proof_of_operator_identity_in_RSGA}
Here, we prove that the operator identity required by the RSGA holds at order at most $4Sr_0$:
\begin{align}
&\underbrace{[[[[\hat{H}_{\rm tot},\hat{Q}^{\dagger}],\hat{Q}^{\dagger}],\ldots],\hat{Q}^{\dagger}]}_{4Sr_0+1\text{ times}}\notag \\
&\equiv [[[[\hat{H}_{\rm tot},\hat{Q}^{\dagger}],\hat{Q}^{\dagger}],\ldots],\hat{Q}^{\dagger}]_{4Sr_0+1}\notag \\
&=0.\label{eq:RSGA-n_condition4_spin-S_appendix}
\end{align}
The proof follows from the result of Ref.~\cite{ODea2026locality}. The proof relies on the locality of the Hamiltonian and the nilpotency of the ladder operators
\begin{align}
(\hat{q}_j^{\dagger})^{2S+1}=0.\label{eq:nilpotency_of_the_ladder_operator_appendix}
\end{align}
Since $\hat{H}_{\rm tot}$ is $r_0$-local, we can write
\begin{align}
\hat{H}_{\rm tot}\equiv \sum_j\hat{h}_{{\rm tot},j},\label{eq:decomposition_of_Htot_spin-S_appendix}
\end{align}
where $|{\rm supp}(\hat{h}_{{\rm tot},j})|\le r_0=O(1)$. The commutator becomes
\begin{align}
[\hat{H}_{\rm tot}, \hat{Q}^{\dagger}]&=\sum_j[\hat{h}_{{\rm tot},j}, \hat{Q}^{\dagger}]\notag \\
&=\sum_j[\hat{h}_{{\rm tot},j},\hat{Q}^{\dagger}_j],\label{eq:calculation_commutator_Htot_and_Qdagger_appendix}\\
\hat{Q}_j^{\dagger}&\equiv \sum_{l\in T_j}\hat{q}_l^{\dagger},\label{eq:definition_of_Qj_dagger}
\end{align}
where we introduced the set
\begin{align}
T_j&\equiv \{l | (\bm{R}_l,1)\in {\rm supp}(\hat{h}_{{\rm tot},j})\text{ or }(\bm{R}_l,2)\in{\rm supp}(\hat{h}_{{\rm tot},j})\}.
\end{align}
Because $\hat{q}_j^{\dagger}$ is a linear combination of $\hat{F}_j^{m+1,m}\otimes\hat{F}_j^{m+1,m}$, we show ${\rm supp}([\hat{h}_{{\rm tot},j},\hat{Q}_j^{\dagger}])\subseteq\{(\bm{R}_l,\alpha)~|~l\in T_j, \alpha=1,2\}$. Then, the multiple commutator reduces to
\begin{align}
[[[[\hat{h}_{{\rm tot},j},\hat{Q}^{\dagger}],\hat{Q}^{\dagger}],\ldots],\hat{Q}^{\dagger}]_{n}=[[[[\hat{h}_{{\rm tot},j}, \hat{Q}^{\dagger}_j], \hat{Q}_j^{\dagger}],\ldots],\hat{Q}_j^{\dagger}]_{n}.\label{eq:multiple_commutator_reduces_local_multiple_commutator}
\end{align}
The nested commutator can be expanded as [see Ex.~14 of Chap.~3 in Ref.~\cite{Hall2015LieGroups}]
\begin{align}
&[[[[\hat{h}_{{\rm tot},j}, \hat{Q}^{\dagger}_j], \hat{Q}_j^{\dagger}],\ldots],\hat{Q}_j^{\dagger}]_{n}\notag \\
&=\sum_{l=0}^n\binom{n}{l}(-\hat{Q}_j^{\dagger})^l\hat{h}_{{\rm tot},j}(\hat{Q}^{\dagger}_j)^{n-l}.\label{eq:expansion_multiple_commutator_appendix}
\end{align}
Since $|T_j|\le |{\rm supp}(\hat{h}_{{\rm tot},j})|\le r_0$ and the operators $\hat{q}_l^{\dagger}$ at distinct positions commute, Eq.~(\ref{eq:nilpotency_of_the_ladder_operator_appendix}) implies
\begin{align}
(\hat{Q}_j^{\dagger})^{2Sr_0+1}=0.\label{eq:nilpotent_condition_Qj_dagger_appendix}
\end{align}
We obtain $[[[[\hat{h}_{{\rm tot},j}, \hat{Q}^{\dagger}_j], \hat{Q}_j^{\dagger}],\ldots],\hat{Q}_j^{\dagger}]_{n}=0$ for $n\ge 4Sr_0+1$. Therefore, the operator identity required for the RSGA holds at order at most $4Sr_0$.

\subsection{Proof of the locality of the parent Hamiltonian}\label{app:proof_locality_parent_Hamitonian_spin-S}
Here, we prove that the parent Hamiltonian $\hat{H}_{\rm p}$ admits a local frustration-free decomposition. We start from the Rokhsar--Kivelson form of the parent Hamiltonian
\begin{align}
\hat{H}_{\rm p}&=2\sum_{\bm{n},\bm{m}}|H_{\bm{n},\bm{m}}|^2\hat{P}_{\bm{n},\bm{m}},\label{eq:parent_Hamiltonian_bracket_form_spin-S_appendix}\\
H_{\bm{n},\bm{m}}&\equiv \bra{\bm{n}}\hat{H}\cket{\bm{m}},\label{eq:definition_of_H_n_m_spin-S_appendix}\\
\hat{P}_{\bm{n},\bm{m}}&\equiv \frac{1}{2}(\cket{\bm{n}}\otimes\cket{\bm{n}}-\cket{\bm{m}}\otimes\cket{\bm{m}})\notag \\
&\quad\times(\bra{\bm{n}}\otimes\bra{\bm{n}}-\bra{\bm{m}}\otimes\bra{\bm{m}}).\label{eq:definition_of_operator_P_spin-S_appendix}
\end{align}
We note that diagonal matrix elements $H_{\bm{n},\bm{n}}$ do not contribute to the parent Hamiltonian because $\hat{P}_{\bm{n},\bm{n}}=0$. In the following, we regard the parent Hamiltonian as an operator on the enlarged scar subspace $\mathcal{H}_P$. 

We write the subsystem Hamiltonian $\hat{H}$ as
\begin{align}
\hat{H}=\sum_j\hat{h}_j,\label{eq:subsystem_Hamiltonian_appendix}
\end{align}
where $\hat{h}_j$ is a Hermitian $r$-local operator of range $R$ with $r,R=O(1)$. We denote the support of $\hat{h}_j$ by
\begin{align}
S_j\equiv {\rm supp}(\hat{h}_j).\label{eq:definition_of_Sj_support_appendix}
\end{align}
We now consider a nonzero matrix element $H_{\bm{n},\bm{m}}\not=0$ with $\bm{n}\not=\bm{m}$. From Eq.~(\ref{eq:subsystem_Hamiltonian_appendix}), the matrix element becomes
\begin{align}
\bra{\bm{n}}\hat{H}\cket{\bm{m}}&=\sum_j\bra{\bm{n}}\hat{h}_j\cket{\bm{m}}\notag \\
&=\sum_{j\in I(\bm{n},\bm{m})}\bra{\bm{n}}\hat{h}_j\cket{\bm{m}},\label{eq:matrix_element_written_by_set_I_appendix}
\end{align}
where $I(\bm{n},\bm{m})$ denotes the set of indices $j$ for which $\bra{\bm{n}}\hat{h}_j\cket{\bm{m}}\not=0$. From the locality of the Hamiltonian, the matrix element of $\hat{h}_j$ for $j\in I(\bm{n},\bm{m})$ reduces to
\begin{align}
\bra{\bm{n}}\hat{h}_j\cket{\bm{m}}&=\langle\bm{n}_{S_j}|\hat{h}_j|\bm{m}_{S_j}\rangle\delta_{\bm{n}_{\bar{S}_j},\bm{m}_{\bar{S}_j}},\label{eq:matrix_element_local_Hamiltonian_appendix}
\end{align}
where $\bar{S}_j$ represents the complement of $S_j$ and $|\bm{n}_{S_j}\rangle$ denotes the restriction of the product-basis state $\cket{\bm{n}}$ to $S_j$. Thus, the two configurations $\bm{n}$ and $\bm{m}$ coincide outside the region $S_j$. Since $\bm{n}\not=\bm{m}$, we may choose a site $x$ such that $n_x\not=m_x$. For every $j\in I(\bm{n},\bm{m})$, Eq.~(\ref{eq:matrix_element_local_Hamiltonian_appendix}) then implies $x\in S_j$. For $l\in I(\bm{n},\bm{m})$ and $l\not=j$, we also obtain $x\in S_l$. Therefore, $S_j\cap S_l\not=\varnothing$ for all $j,l\in I(\bm{n},\bm{m})$.  

For each $j$, we define the neighboring-term index set and the associated region
\begin{align}
\mathcal{N}_j&\equiv\{ l~|~ S_j\cap S_l\not=\varnothing\},\label{eq:definition_of_set_Nj_appendix}\\
X_j&\equiv \bigcup_{l\in \mathcal{N}_j}S_l.\label{eq:definition_of_region_Xj_appendix}
\end{align}
For configurations $\bm{n}_{X_j}$ and $\bm{m}_{X_j}$ on $X_{j}$, we define the local active-index set by
\begin{align}
\mathcal{I}_{j}(\bm{n}_{X_j},\bm{m}_{X_j})\equiv\left\{l\in\mathcal{N}_{j}~|~ \langle\bm{n}_{X_j}|\hat{h}_{l}|\bm{m}_{X_j}\rangle\neq0\right\}.\label{eq:definition_of_local_active_set_appendix}
\end{align}
From the above discussion, we obtain $I(\bm{n},\bm{m})\subseteq \mathcal{N}_j$. Thus, for $j\in I(\bm{n},\bm{m})$, the matrix element can be written as 
\begin{align}
H_{\bm{n},\bm{m}}=\sum_{l\in\mathcal{N}_j}\bra{\bm{n}}\hat{h}_l\cket{\bm{m}},\quad j\in I(\bm{n},\bm{m}).\label{eq:rewrite_matrix_element_Nj_appendix}
\end{align}
Equation~(\ref{eq:matrix_element_local_Hamiltonian_appendix}) implies $\bm{n}_{\bar{S}_j}=\bm{m}_{\bar{S}_j}$. Since $S_j\subseteq X_j$, we also have $\bm{n}_{\bar{X}_j}=\bm{m}_{\bar{X}_j}$. Thus, $H_{\bm{n},\bm{m}}$ is determined by the restrictions of the two configurations to $X_j$.

We now rewrite the parent Hamiltonian. We define the local matrix element and the local positive-semidefinite operator on $X_j$ by
\begin{align}
H^{X_j}_{\bm{n}_{X_j},\bm{m}_{X_j}}&\equiv \sum_{l\in N_j}\bra{\bm{n}_{X_j}}\hat{h}_l\cket{\bm{m}_{X_j}},\label{eq:definition_of_local_matrix_elements_appendix}\\
\hat{P}^{X_j}_{\bm{n}_{X_j},\bm{m}_{X_j}}&\equiv \frac{1}{2}(\cket{\bm{n}_{X_j}}\otimes\cket{\bm{n}_{X_j}}-\cket{\bm{m}_{X_j}}\otimes\cket{\bm{m}_{X_j}})\notag \\
&\times (\bra{\bm{n}_{X_j}}\otimes\bra{\bm{n}_{X_j}}-\bra{\bm{m}_{X_j}}\otimes\bra{\bm{m}_{X_j}}).\label{eq:definition_of_local_operators_appendix}
\end{align}
For configurations $\bm{n}$ and $\bm{m}$ satisfying $j\in I(\bm{n},\bm{m})$, the following relations hold:
\begin{align}
I(\bm{n},\bm{m})&=\mathcal{I}_j(\bm{n}_{X_j}, \bm{m}_{X_j}),\label{eq:relation_I_and_Ij_appendix}\\
H_{\bm{n},\bm{m}}&=H^{X_j}_{\bm{n}_{X_j},\bm{m}_{X_j}}.\label{eq:relation_H_and_HXj_appendix}
\end{align}
In particular, both quantities are independent of the common configuration on $\bar{X}_j$. These relations follow because the two configurations coincide outside $X_j$. We note that $\hat{P}^{X_j}_{\bm{n}_{X_j},\bm{m}_{X_j}}$ is positive semidefinite. For every pair $(\bm{n},\bm{m})$ with $H_{\bm{n},\bm{m}}\not=0$, we use 
\begin{align}
1=\sum_{j\in I(\bm{n},\bm{m})}\frac{1}{|I(\bm{n},\bm{m})|}.\label{eq:identity_for_I_n_m_appendix}
\end{align}
This identity distributes each global matrix-element contribution equally among the local terms that generate it. Thus, we obtain
\begin{align}
\hat{H}_{\rm p}&=2\sum_{j}\hspace{-1.0em}\sum_{\substack{\bm{n}_{X_j},\bm{m}_{X_j} \\ j\in I_j(\bm{n}_{X_j},\bm{m}_{X_j})}}\hspace{-1em}\frac{|H^{X_j}_{\bm{n}_{X_j},\bm{m}_{X_j}}|^2}{|I_j(\bm{n}_{X_j},\bm{m}_{X_j})|}\hat{P}^{X_j}_{\bm{n}_{X_j},\bm{m}_{X_j}}\otimes \hat{P}^{\bar{X}_j},\label{eq:rewrite_parent_Hamiltonian_totyu1_appendix} \\
\hat{P}^{\bar{X}_j}&\equiv \sum_{\bm{n}_{\bar{X}_j}}(|\bm{n}_{\bar{X}_j}\rangle\otimes|\bm{n}_{\bar{X}_j}\rangle)(\langle\bm{n}_{\bar{X}_j}|\otimes\langle\bm{n}_{\bar{X}_j}|),\label{eq:definition_of_Q_in_parent_Hamiltonian_appendix}
\end{align}
where the factor $1/|I_j(\bm{n}_{X_j},\bm{m}_{X_j})|$ compensates for the fact that each global matrix-element contribution is assigned to every active local term. For spin-$S$ systems, the enlarged scar subspace has a tensor-product structure
\begin{align}
\mathcal{H}_P=&\bigotimes_{j=1}^M\mathcal{H}^{\rm L}_{P,j},\label{eq:enlarged_scar_subspace_tensor_product_form_appendix}\\
\mathcal{H}^{\rm L}_{P,j}&\equiv {\rm Span}\{\cket{\eta}_{j,1}\otimes\cket{\eta}_{j,2}~|~\eta=-S,-S+1,\ldots,S \}.\label{eq:definition_of_local_enlarged_scar_subspace_appendix}
\end{align}
Thus, $\hat{P}^{\bar{X}_j}$ acts as the identity on the complement of $X_j$ within $\mathcal{H}_P$. Therefore, the restriction of the parent Hamiltonian to $\mathcal{H}_P$ can be written as a sum of local operators
\begin{align}
\hat{H}_{\rm p}&=\sum_j\hat{h}_{{\rm p},j},\label{eq:parent_Hamiltonian_local_sum_form_appendix}\\
\hat{h}_{{\rm p},j}&\equiv 2\sum_{\substack{\bm{n}_{X_j},\bm{m}_{X_j} \\ j\in I_j(\bm{n}_{X_j},\bm{m}_{X_j}) }}\frac{|H^{X_j}_{\bm{n}_{X_j},\bm{m}_{X_j}}|^2}{|I_j(\bm{n}_{X_j},\bm{m}_{X_j})|}\hat{P}^{X_j}_{\bm{n}_{X_j},\bm{m}_{X_j}}.\label{eq:definition_of_hp_j_appendix}
\end{align}

We next establish the locality of $\hat{h}_{{\rm p},j}$. By definition, every support $S_l$ entering the union that defines $X_j$ intersects $S_j$. Since every $\hat{h}_l$ is $r$-local and has range $R$, with $r, R=O(1)$, and since the lattice has finite coordination number, the region $X_j$ contains $O(1)$ sites and has $O(1)$ diameter.
 
Finally, we show that the local decomposition is frustration-free. Whenever $H_{\bm{n}_{X_{j}},\bm{m}_{X_{j}}}^{X_{j}}\not= 0$, the $U(1)$ symmetry of $\hat{H}$ implies that the corresponding global configurations carry the same total charge. Since the two configurations coincide outside $X_{j}$, $\bm{n}_{X_{j}}$ and $\bm{m}_{X_{j}}$ carry the same charge within $X_{j}$. Because the rainbow QMBS state is an equal-weight superposition within each fixed-charge sector, we obtain $\hat{P}^{X_j}_{\bm{n}_{X_j},\bm{m}_{X_j}}|\tilde{S}_n\rangle=0$. Since all coefficients in Eq.~(\ref{eq:definition_of_hp_j_appendix}) are nonnegative, each $\hat{h}_{{\rm p},j}$ is positive semidefinite and satisfies $\hat{h}_{{\rm p},j}|\tilde{S}_n\rangle=0$ for all $j$. Thus, the local decomposition is frustration-free.
 
\subsection{Proof of the gaplessness of the parent Hamiltonian}\label{app:proof_gaplessness_parent_Hamitonian_spin-S}
First, we prove $\langle\tilde{S}_{n'}|\widetilde{AS}_n\rangle=0$ for all $n,n'$. Because $|\tilde{S}_{n'}\rangle$ and $|\widetilde{AS}_n\rangle$ carry charges $n'$ and $n$, respectively, $\langle\tilde{S}_{n'}|\widetilde{AS}_n\rangle=0$ unless $n=n'$. For $n=n'$, direct calculations show that
\begin{align}
\langle\tilde{S}_n|\widetilde{AS}_n\rangle&=\frac{1}{n}\langle\tilde{S}_n|\sum_{j=1}^Me^{i\bm{k}\cdot\bm{R}_j}(\hat{n}_j\otimes\hat{1})|\tilde{S}_n\rangle\notag \\
&=\frac{(n!)^2}{n}\sum_{j=1}^Me^{i\bm{k}\cdot\bm{R}_j}\sum_{\bm{n};\hat{N}\cket{\bm{n}}=n\cket{\bm{n}}}\langle\bm{n}|\hat{n}_j|\bm{n}\rangle\notag \\
&=\frac {(n!)^2}{n}\sum_{j=1}^Me^{i\bm{k}\cdot\bm{R}_j}C\notag \\
&=0,\label{eq:proof_of_orthogonality_of_AS_n_appendix}
\end{align}
where $C$ is a $j$-independent constant and we used Eq.~(\ref{eq:definition_of_crystal_momentum_spin-S}). Therefore, the variational state $|\widetilde{AS}_n\rangle$ is orthogonal to all RQMBS states. 

Next, we evaluate the denominator $f_{\rm d}(\bm{k})$ in Eq.~(\ref{eq:bound_gap_and_energy_variance}). To this end, we define
\begin{align}
W_{M,n}&\equiv \frac{1}{(n!)^2}\langle\tilde{S}_n|\tilde{S}_n\rangle.\label{eq:number_of_basis_in_RQMBS_appendix}
\end{align}
This quantity represents the number of product-basis states appearing in $|\tilde{S}_n\rangle$. A combinatorial argument gives
\begin{align}
W_{M,n}&=
\begin{cases}
\vspace{0.5em}\displaystyle{\sum_{m=0}^{2S}W_{M-1,n-m}},\quad \text{for }0\le n\le 2SM,\\
0,\quad \text{for }n<0\text{ or }n>2SM,
\end{cases}
\label{eq:expression_of_W_appendix}\\
W_{0,0}&=1,\quad W_{0,n\not=0}=0.\label{eq:reccurent_relation_W_appendix}
\end{align}
For the $S=1/2$ case, $W_{M,n}$ reduces to the binomial coefficient:
\begin{align}
W_{M,n}^{S=1/2}=\binom{M}{n}.\label{eq:W_M_n_S=1/2_case}
\end{align}

Then, we calculate $f_{\rm d}(\bm{k})$ using Eq.~(\ref{eq:definition_of_crystal_momentum_spin-S}):
\begin{align}
&f_{\rm d}(\bm{k})\notag \\
&=\frac{1}{n^2}\sum_{j,l}e^{-i\bm{k}\cdot(\bm{R}_j-\bm{R}_l)}\langle\tilde{S}_n|(\hat{n}_j\otimes\hat{1})(\hat{n}_l\otimes\hat{1})|\tilde{S}_n\rangle\notag \\
&=\frac{(n!)^2}{n^2}\sum_{j,l}e^{-i\bm{k}\cdot(\bm{R}_j-\bm{R}_l)}\sum_{\bm{n}:\hat{N}\cket{\bm{n}}=n\cket{\bm{n}}}q_jq_l\notag \\
&=\frac{(n!)^2}{n^2}\left[\sum_{j}\sum_{\bm{n}:\hat{N}\cket{\bm{n}}=n\cket{\bm{n}}}q_j^2\right.\notag \\
&\left.\hspace{5.0em}+\sum_{j,l,j\not=l}e^{-i\bm{k}\cdot(\bm{R}_j-\bm{R}_l)}\sum_{\bm{n}:\hat{N}\cket{\bm{n}}=n\cket{\bm{n}}}q_jq_l\right]\notag \\
&=\frac{(n!)^2}{n^2}M\sum_{\bm{n}:\hat{N}\cket{\bm{n}}=n\cket{\bm{n}}}\left(q_1^2-q_1q_2\right)\notag \\
&\equiv \frac{(n!)^2}{n^2}M(A_1-B_{12}),\label{eq:calculation_fd_in_SMA_Spin-S_appendix}
\end{align}
where $q_j\equiv \bra{\bm{n}}\hat{n}_j\cket{\bm{n}}\in \{0,1,\ldots, 2S\}$ and we used the fact that $\sum_{\bm{n}:\hat{N}\cket{\bm{n}}=n\cket{\bm{n}}}q_j^2$ and $\sum_{\bm{n}:\hat{N}\cket{\bm{n}}=n\cket{\bm{n}}}q_jq_l$ are independent of the site indices. Using $n=\sum_jq_j$, we obtain the relation
\begin{align}
&&\left(\sum_jq_j\right)^2&=n^2\notag \\
&\Rightarrow&\;\sum_{\bm{n}:\hat{N}\cket{\bm{n}}=n\cket{\bm{n}}}\sum_{j,l}q_jq_l&=\sum_{\bm{n}:\hat{N}\cket{\bm{n}}=n\cket{\bm{n}}}n^2\notag \\
&\Rightarrow&\; B_{12}&=\frac{n^2W_{M,n}-MA_1}{M(M-1)}.\label{eq:relation_B12_and_A1_appendix}
\end{align}
Therefore, we obtain
\begin{align}
&f_{\rm d}(\bm{k})\notag \\
&=\frac{(n!)^2}{n^2}\frac{1}{M-1}\left[M^2\sum_{m=0}^{2S}m^2W_{M-1,n-m}-n^2W_{M,n}\right].\label{eq:evaluation_of_denominator_SMA_spin-S_appendix}
\end{align}
For the $S=1/2$ case, we can obtain the closed form of $f_{\rm d}(\bm{k)}$:
\begin{align}
f^{S=1/2}_{\rm d}(\bm{k})=\frac{(n!)^2}{n}\frac{1-n/M}{1-1/M}\binom{M}{n}.\label{eq:denominator_S=1/2_case_appendix}
\end{align}
For $S>1/2$, $W_{M,n}$ does not reduce to a single binomial coefficient and a comparably simple form of $f_{\rm d}(\bm{k})$ is not available. We therefore derive a lower bound sufficient for the present purpose.

To derive the inequality for $f_{\rm d}(\bm{k})$, we define the normalized scar state $\cket{S_n}\equiv |\tilde{S}_n\rangle/\sqrt{\langle\tilde{S}_n|\tilde{S}_n\rangle}$. Using the relation
\begin{align}
v_{M,n}&\equiv \bra{S_n}\left(\hat{n}_1-\frac{n}{M}\right)^2\cket{S_n}\notag \\
&=\frac{1}{W_{M,n}}\sum_{m=0}^{2S}m^2W_{M-1,n-m}-\frac{n^2}{M^2},\label{eq:relation_n1-n/M_squared_spin-S_appendix}
\end{align}
we can rewrite Eq.~(\ref{eq:evaluation_of_denominator_SMA_spin-S_appendix}) as
\begin{align}
f_{\rm d}(\bm{k})&=\frac{1}{n^2}\frac{M^2}{M-1}\bra{S_n}\left(\hat{n}_1-\frac{n}{M}\right)^2\cket{S_n}\langle\tilde{S}_n|\tilde{S}_n\rangle\notag \\
&= \frac{1}{n^2}\frac{M^2}{M-1}v_{M,n}\langle\tilde{S}_n|\tilde{S}_n\rangle.\label{eq:rewrite_denominator_totyu1_spin-S_appendix}
\end{align}
Direct calculations show that the following relations hold:
\begin{align}
&\bra{S_n}\left(\hat{n}_1-\frac{n}{M}\right)\left(\hat{n}_2-\frac{n}{M}\right)\cket{S_n}=-\frac{v_{M,n}}{M-1},\label{eq:relation1_v_spin-S_appendix}\\
&\bra{S_n}(\hat{n}_1+\hat{n}_2)^2\cket{S_n}-(\bra{S_n}(\hat{n}_1+\hat{n}_2)\cket{S_n})^2\notag \\
&=2\frac{M-2}{M-1}v_{M,n}.\label{eq:relation2_v_spin-S_appendix}
\end{align}
From these results, we obtain
\begin{align}
&\bra{S_n}(\hat{n}_1+\hat{n}_2)(4S-\hat{n}_1-\hat{n}_2)\cket{S_n}\notag \\
&=8S\frac{n}{M}-4\frac{n^2}{M^2}-2\frac{M-2}{M-1}v_{M,n}.\label{eq:relation3_v_spin-S_Appendix}
\end{align}
In a similar manner, we obtain
\begin{align}
\bra{S_n}(\hat{n}_1-\hat{n}_2)^2\cket{S_n}&=\frac{2M}{M-1}v_{M,n}.\label{eq:relation4_v_spin-S_appendix}
\end{align}
From the combinatorics, we can write
\begin{align}
&\bra{S_n}(\hat{n}_1-\hat{n}_2)^2\cket{S_n}\notag \\
&=\frac{1}{W_{M,n}}\sum_{m_1,m_2=0}^{2S}(m_1-m_2)^2W_{M-2,n-m_1-m_2}\notag \\
&=\frac{1}{W_{M,n}}\sum_{m=0}^{4S}\sum_{a=\max(0,m-2S)}^{\min(2S,m)}(2a-m)^2W_{M-2,n-m},\label{eq:calculation_combinatorics_n1-n2_appendix}
\end{align}
where we define $m\equiv m_1+m_2$ and $a\equiv m_1$. The number of allowed values of $a$ is given by $L_m+1=\min(m,4S-m)+1$, because $0\le a\le 2S$ and $0\le m-a\le 2S$. Direct calculations show 
\begin{align}
\sum_{a=\max(0,m-2S)}^{\min(2S,m)}(2a-m)^2&=\frac{1}{3}L_m(L_m+1)(L_m+2).\label{eq:sum_n1-n2_spin-S_appendix}
\end{align}
From the definition of $L_m$, the following inequality holds:
\begin{align}
L_m(L_m+2)&\ge \frac{m}{2S}(4S-m)\notag\\
\Rightarrow\quad \frac{L_m}{3}(L_m+1)(L_m+2)&\ge \frac{m}{6S}(L_m+1)(4S-m).\label{eq:inequality_for_Lm}
\end{align}
We can also write
\begin{align}
&\bra{S_n}(\hat{n}_1+\hat{n}_2)(4S-\hat{n}_1-\hat{n}_2)\cket{S_n}\notag \\
&=\frac{1}{W_{M,n}}\sum_{m=0}^{4S}m(4S-m)(L_m+1)W_{M-2,n-m}.\label{eq:n1+n2_times_4S-n1-n2_combination_form_appendix}
\end{align}
Therefore, we obtain
\begin{align}
&\bra{S_n}(\hat{n}_1-\hat{n}_2)^2\cket{S_n}\notag \\
&\ge \frac{1}{6S}\bra{S_n}(\hat{n}_1+\hat{n}_2)(4S-\hat{n}_1-\hat{n}_2)\cket{S_n}.\label{eq:inequality_for_n1-n2_and_n1+n2_appendix}
\end{align}
From Eqs.~(\ref{eq:relation3_v_spin-S_Appendix}), (\ref{eq:relation4_v_spin-S_appendix}), and (\ref{eq:inequality_for_n1-n2_and_n1+n2_appendix}), we obtain the inequality
\begin{align}
v_{M,n}&\ge \frac{2(M-1)}{(6S+1)M-2}\dfrac{n}{M}\left(2S-\dfrac{n}{M}\right)\notag \\
&\ge \frac{1}{6S+1}\frac{n}{M}\left(2S-\frac{n}{M}\right).\label{eq:inequality_for_v_M_n_spin-S_appendix}
\end{align}
A lower bound on the denominator is given by
\begin{align}
f_{\rm d}(\bm{k})\ge \frac{1}{n}\frac{M}{M-1}\frac{1}{6S+1}\left(2S-\dfrac{n}{M}\right)\langle\tilde{S}_n|\tilde{S}_n\rangle.\label{eq:final_form_of_inequality_f_d_spin-S_appendix}
\end{align}

We now evaluate the numerator $f_{\rm n}(\bm{k})$. First, we rewrite $f_{\rm n}(\bm{k})$ as
\begin{align}
f_{\rm n}(\bm{k})&=\frac{1}{n^2}\langle\tilde{S}_n|\hat{\rho}(-\bm{k})\hat{H}_{\rm p}\hat{\rho}(\bm{k})|\tilde{S}_n\rangle,\label{eq:rewrite_numerator_f_spin-S_appendix}
\end{align}
where we have used $\hat{\rho}(\bm{k})^{\dagger}=\hat{\rho}(-\bm{k})$. For $\bm{k}=\bm{0}$, we obtain
\begin{align}
f_{\rm n}(\bm{0})&=\frac{1}{n^2}\langle\tilde{S}_n|\hat{\rho}(\bm{0})\hat{H}_{\rm p}\hat{\rho}(\bm{0})|\tilde{S}_n\rangle=0,\label{eq:numerator_at_k=0_spin-S_appendix}
\end{align}
where we used $\hat{\rho}(\bm{0})=\hat{N}\otimes\hat{1}$ and $\hat{H}_{\rm p}|\tilde{S}_n\rangle=0$. We next consider the case $\bm{k}\not=\bm{0}$. Since $\hat{h}_{{\rm p},j}$ is positive semidefinite, its square root $\hat{h}_{{\rm p},j}^{1/2}$ is well defined. The numerator reduces to
\begin{align}
f_{\rm n}(\bm{k})&=\frac{1}{n^2}\sum_j\langle\tilde{S}_n|\hat{\rho}(-\bm{k})\hat{h}_{{\rm p},j}^{1/2}\hat{h}_{{\rm p},j}^{1/2}\hat{\rho}(\bm{k})|\tilde{S}_n\rangle\notag \\
&=\frac{1}{n^2}\sum_j\|\hat{h}_{{\rm p},j}^{1/2}\hat{\rho}(\bm{k})|\tilde{S}_n\rangle \|^2,\label{eq:numerator_norm_form_spin-S_appendix}
\end{align}
where $\|\cdots\|$ denotes the Hilbert-space norm. We can rewrite the operator part in Eq.~(\ref{eq:numerator_norm_form_spin-S_appendix}) as
\begin{align}
\hat{h}_{{\rm p},j}^{1/2}\hat{\rho}(\bm{k})|\tilde{S}_n\rangle&=\hat{h}_{{\rm p},j}^{1/2}\sum_{l=1}^{M}e^{i \bm{k}\cdot\bm{R}_l}(\hat{n}_l\otimes\hat{1})|\tilde{S}_n\rangle\notag \\
&=\hat{h}_{{\rm p},j}^{1/2}\sum_{l=1}^{M}(e^{i \bm{k}\cdot\bm{R}_l}-e^{i\bm{k}\cdot\bm{R}_{l_0}})(\hat{n}_l\otimes\hat{1})|\tilde{S}_n\rangle\notag \\
&=\hat{h}_{{\rm p},j}^{1/2}\sum_{l\in S_{{\rm p},j}^{(1)}}(e^{i \bm{k}\cdot\bm{R}_l}-e^{i\bm{k}\cdot\bm{R}_{l_0}})(\hat{n}_l\otimes\hat{1})|\tilde{S}_n\rangle.\label{eq:rewrite_operator_in_numarator_spin-S_appendix}
\end{align}
Here,
\begin{align}
S_{\mathrm{p},j}^{(1)}\equiv\left\{l\,\middle|\,(\bm{R}_{l},1)\in\operatorname{supp}(\hat{h}_{\mathrm{p},j})\right\}.\label{eq:definition_of_Spj1}
\end{align}
We choose
$l_{0}\in S_{\mathrm{p},j}^{(1)}$, and $\bm{R}_{l_{0}}$ denotes its position. In deriving Eq.~(\ref{eq:rewrite_operator_in_numarator_spin-S_appendix}), we have used $\hat{h}^{1/2}_{{\rm p},j}|\tilde{S}_n\rangle=0$, $(\hat{N}\otimes\hat{1})|\tilde{S}_n\rangle=n|\tilde{S}_n\rangle$, and $[\hat{h}_{{\rm p},j}^{1/2}, \hat{n}_l\otimes\hat{1}]=0$ for $l\not\in S_{{\rm p},j}^{(1)}$. From the Cauchy--Schwarz inequality and $1-\cos(x)\le x^2/2$, we obtain
\begin{align}
&f_{\rm n}(\bm{k})\notag \\
&\le \frac{1}{n^2}\sum_j|S_{{\rm p},j}^{(1)}|\notag \\
&\quad\times\sum_{l\in S^{(1)}_{{\rm p},j}}\|(e^{i\bm{k}\cdot\bm{R}_l}-e^{i\bm{k}\cdot\bm{R}_{l_0}})\hat{h}_{{\rm p},j}^{1/2}(\hat{n}_l\otimes\hat{1})|\tilde{S}_n\rangle\|^2\notag \\
&\le \frac{1}{n^2}\sum_j|S_{{\rm p},j}^{(1)}|\sum_{l\in  S^{(1)}_{{\rm p},j}}\|\hat{h}_{{\rm p},j}^{1/2}(\hat{n}_l\otimes\hat{1})|\tilde{S}_n\rangle\|^2\bm{k}^2|\bm{R}_l-\bm{R}_{l_0}|^2\notag \\
&\le \frac{1}{n^2}\sum_j\sum_{l\in  S^{(1)}_{{\rm p},j}}\|\hat{h}_{{\rm p},j}^{1/2}(\hat{n}_l\otimes\hat{1})|\tilde{S}_n\rangle\|^2\bm{k}^2C_1,\label{eq:inequality_fn_totyu_spin-S_appendix}
\end{align}
where $C_1$ is an $O(1)$ constant. The last line follows from the assumption that $\hat{h}_{{\rm p},j}$ is $r'$-local and has range $R'\;[r',R'=O(1)]$. We note that under periodic boundary conditions, one needs to use the appropriate distance for evaluating $|\bm{R}_l-\bm{R}_{l_0}|$. To proceed with the evaluation of Eq.~(\ref{eq:inequality_fn_totyu_spin-S_appendix}), we use the operator norm. From the assumptions on $\hat{h}_{{\rm p},j}$, we obtain the following inequality
\begin{align}
\|\hat{h}_{{\rm p},j}^{1/2}(\hat{n}_l\otimes\hat{1})|\tilde{S}_n\rangle\|^2&\le  \|\hat{h}_{{\rm p},j}^{1/2}(\hat{n}_l\otimes\hat{1}) \|^2_{\rm op} \langle\tilde{S}_n|\tilde{S}_n\rangle\notag \\
&\le C_2\langle\tilde{S}_n|\tilde{S}_n\rangle,\label{eq:inequality_from_operator_norm_spin-S_appendix}
\end{align}
where $C_2$ is an $O(1)$ constant. Here, we use the fact that $\|\hat{h}_{{\rm p},j}\|_{\rm op}\le J_{\rm p}=O(1)$ and $\|\hat{n}_j\|_{\rm op}\le 2S=O(1)$. Therefore, we obtain
\begin{align}
f_{\rm n}(\bm{k})\le \frac{M}{n^2}C_1C_2C_3\bm{k}^2\langle\tilde{S}_n|\tilde{S}_n\rangle,\label{eq:inequality_final_form_f_n_spin-S_appendix}
\end{align}
where $C_3$ is an $O(1)$ constant arising from Eq.~(\ref{eq:assumption_number_of_hpj}) and the bound $|S_{{\rm p},j}^{(1)}|\le r'=O(1)$. 

Combining Eqs.~(\ref{eq:final_form_of_inequality_f_d_spin-S_appendix}) and (\ref{eq:inequality_final_form_f_n_spin-S_appendix}), we obtain the following upper bound on the parent-Hamiltonian gap:
\begin{align}
\Delta_{{\rm p},n}&\le \Delta E_n^2\notag \\
&=\frac{f_{\rm n}(\bm{k})}{f_{\rm d}(\bm{k})}\notag \\
&\le \frac{M}{n}\left(1-\frac{1}{M}\right)\frac{6S+1}{2S-n/M}C_1C_2C_3|\bm{k}|^2\notag \\
&\le \frac{M}{n}\frac{6S+1}{2S-n/M}C_1C_2C_3|\bm{k}|^2.\label{eq:inequality_for_energy_variance_spin-S_appendix}
\end{align}
For $\epsilon\le n/M\le 2S-\epsilon$, where $\epsilon>0$ is independent of the system size, we obtain $\Delta_{{\rm p},n}\le \Delta E_n^2=O(\bm{k}^2)$. Choosing $\bm{k}=\bm{k}_{\rm min}$ as defined in Eq.~(\ref{eq:definition_of_kmin}) and taking the thermodynamic limit, we prove that the parent Hamiltonian is gapless. At the same time, the energy variance of the variational state vanishes in the thermodynamic limit.

\subsection{Entanglement entropy of the RQMBS and variational states}\label{app:entanglement_entropy_of_the_variational_states}
In this subsection, we calculate the EEs of the normalized RQMBS and normalized variational states $|AS_n\rangle\equiv |\widetilde{AS}_n\rangle/\sqrt{\langle\widetilde{AS}_n |\widetilde{AS}_n\rangle}$. Before calculating the EE of the variational state, we consider the EE of the RQMBS states. For simplicity, we assume that $M$ is even. 

The normalized RQMBS state can be written as
\begin{align}
\cket{S_n}=\frac{1}{\sqrt{W_{M,n}}}\sum_{\bm{n}:\hat{N}\cket{\bm{n}}=n\cket{\bm{n}}}\cket{\bm{n}}\otimes\cket{\bm{n}}.\label{eq:normalized_RQMBS_EE_appendix}
\end{align}
Here, $W_{M,n}$ represents the number of states satisfying $\hat{N}\cket{\bm{n}}=n\cket{\bm{n}}$. First, we consider the type-I bipartition. In this case, each Schmidt coefficient $S_{\bm{n}}$ is given by
\begin{align}
S_{\bm{n}}&=\frac{1}{\sqrt{W_{M,n}}}.\label{eq:Schmidt_coefficient_typeI_cut_appendix}
\end{align}
Therefore, the von Neumann EE is given by
\begin{align}
S_{\rm vN}^{\rm I}&=-\sum_{\bm{n}:\hat{N}\cket{\bm{n}}=n\cket{\bm{n}}}S_{\bm{n}}^2\ln S_{\bm{n}}^2\notag \\
&=\ln W_{M,n}.\label{eq:EE_of_RQMBS_typeI_appendix}
\end{align}
Here, we investigate the scaling of $W_{M,n}$. We first show that the following inequality holds:
\begin{align}
W_{M,n}\ge \binom{M}{m},\quad m\equiv \lfloor n/(2S)\rfloor,\label{eq:inequality_for_W_M_n_appendix}
\end{align}
where $\lfloor\cdot\rfloor$ is the floor function. This can be derived as follows: We set $n=2Sm+l$, where $0\le l \le 2S-1$. We count configurations in which $m$ sites have spin $S$, one additional site has spin $-S+l$, and all remaining sites have spin $-S$. In this case, the total charge becomes $n$. This number must be smaller than or equal to $W_{M,n}$. For $l=0$, we obtain $W_{M,n}\ge \binom{M}{m}$. For $1\le l\le 2S-1$, we obtain $W_{M,n}\ge \binom{M}{m}(M-m)\ge \binom{M}{m}$. This proves Eq.~(\ref{eq:inequality_for_W_M_n_appendix}).

Next, we use the inequality for the binomial coefficient [see Eq.~(11.40) of Ref.~\cite{CoverThomas2006}]
\begin{align}
\binom{M}{m}&\ge \frac{e^{MH_2(m/M)}}{M+1},\label{eq:inequality_binomial_constatn_and_H2_appendix}\\
H_2(x)&\equiv -x\ln(x)-(1-x)\ln(1-x),\quad 0<x<1.\label{eq:definition_of_H2_appendix}
\end{align}
We define $H_2(0)=H_2(1)=0$ by continuity. From Eq.~(\ref{eq:inequality_binomial_constatn_and_H2_appendix}), we obtain
\begin{align}
S_{\rm vN}^{{\rm I}}\ge M H_2(m/M)-\ln(M+1).\label{eq:bound_for_S_typeI_RQMBS_appendix}
\end{align}
For a positive $M$-independent constant $\epsilon$ and $\epsilon\le n/M\le 2S-\epsilon$, the inequality $H_2(x)\ge 2\ln(2)\min(x,1-x)$ gives
\begin{align}
&S_{\rm vN}^{\rm I}\notag \\
&\ge 
\begin{cases}
\vspace{0.5em}\ln(2)\epsilon \dfrac{M}{S}-2\ln(2)-\ln(M+1),\;\text{for }m/M\le 1/2,\\
\ln(2)\epsilon \dfrac{M}{S}-\ln(M+1),\;\text{for }m/M>1/2.
\end{cases}
\label{eq:inequality_for_Svn_type_I_appendix}
\end{align}
Therefore, we obtain a volume-law scaling of the EE for the type-I bipartition.

Next, we consider the type-II bipartition. In this case, the RQMBS state can be rewritten as~\cite{Popkov2005logarithmic}
\begin{align}
\cket{S_n}&=\sum_{m=m_{\rm min}}^{m_{\rm max}}\sqrt{\lambda_m}|\Psi^A_{m}\rangle\otimes|\Psi_{n-m}^B\rangle,\label{eq:Schmidt_decomposition_RQMBS_type_II_appendix}\\
\lambda_m&\equiv\frac{W_{M/2,m}W_{M/2,n-m}}{W_{M,n}},\label{eq:definition_of_S_m_typeII_appendix}\\ 
|\Psi^{A}_{m}\rangle&\equiv \frac{1}{\sqrt{W_{M/2,m}}}\sum_{\bm{n}_A: \sum_j(n_{A,j}+S)=m}\cket{\bm{n}_A}\otimes \cket{\bm{n}_A},\label{eq:subsystemA_wave_function_spin-S_appendix}\\
|\Psi^{B}_{n-m}\rangle&\equiv \frac{1}{\sqrt{W_{M/2,n-m}}}\sum_{\bm{n}_B: \sum_j(n_{B,j}+S)=n-m}\cket{\bm{n}_B}\otimes \cket{\bm{n}_B},\label{eq:subsystemB_wave_function_spin-S_appendix}\\
m_{\rm min}&\equiv \max(0,n-SM),\label{eq:definition_of_m_min_appendix}\\
m_{\rm max}&\equiv \min(n, SM),\label{eq:definition_of_m_max_appendix}
\end{align}
where $|\Psi^A_m\rangle$ and $|\Psi^B_{n-m}\rangle$ denote normalized states of subsystems $A$ and $B$ with charges $m$ and $n-m$, respectively. From the above results, we obtain the Schmidt rank $S_{\rm rank}=\min(n,2SM-n)+1$. Because the EE is bounded by the logarithm of the Schmidt rank, we obtain
\begin{align}
S_{\rm vN}^{{\rm II}}\le\ln S_{\rm rank}\le \ln(SM+1).\label{eq:upper_bound_of_Svn_RQMBS_appendix}
\end{align}
Thus, the EE obeys a subvolume law for the type-II bipartition.

Now, we consider the variational state $|AS_n\rangle$. The state $|AS_n\rangle$ can be written as
\begin{align}
|AS_n\rangle&=\sum_{\bm{n}:\hat{N}\cket{\bm{n}}=n\cket{\bm{n}}}\frac{|F(\bm{k},\bm{n})|e^{i\theta(\bm{k},\bm{n})}}{\sqrt{\mathcal{N}(\bm{k})W_{M,n}}}\cket{\bm{n}}\otimes\cket{\bm{n}},\label{eq:expression_AS_n_Schmidt_decomposition_form_appendix}\\
F(\bm{k},\bm{n})&\equiv \sum_{j=1}^Me^{i\bm{k}\cdot\bm{R}_j}q_j\equiv |F(\bm{k},\bm{n})|e^{i\theta(\bm{k},\bm{n})},\label{eq:definition_of_F_k_n_appendix}\\
\mathcal{N}(\bm{k})&\equiv \frac{1}{W_{M,n}}\sum_{\bm{n}:\hat{N}\cket{\bm{n}}=n\cket{\bm{n}}}|F(\bm{k},\bm{n})|^2.\label{eq:definition_of_mathcal_N_k_appendix}
\end{align}
From these results, the Schmidt coefficient $S_{\bm{n}}^{AS}$ is given by
\begin{align}
S_{\bm{n}}^{AS}&=\sqrt{\frac{|F(\bm{k},\bm{n})|^2}{\mathcal{N}(\bm{k})W_{M,n}}}.\label{eq:Schmidt_coefficient_AS_n_typeI}
\end{align}
The normalization condition yields $\sum_{\bm{n}:\hat{N}\cket{\bm{n}}=n\cket{\bm{n}}}(S_{\bm{n}}^{AS})^2=1$. The entanglement entropy can be written as
\begin{align}
&S_{\rm vN}^{{\rm I}: AS}\notag \\
&=\ln W_{M,n}\notag \\
&\quad -\frac{1}{W_{M,n}}\sum_{\bm{n}:\hat{N}\cket{\bm{n}}=n\cket{\bm{n}}}\frac{|F(\bm{k},\bm{n})|^2}{\mathcal{N}(\bm{k})}\ln\left[\frac{|F(\bm{k},\bm{n})|^2}{\mathcal{N}(\bm{k})}\right]\notag \\
&=S_{\rm vN}^{\rm I}-\frac{1}{W_{M,n}}\sum_{\bm{n}:\hat{N}\cket{\bm{n}}=n\cket{\bm{n}}}\frac{|F(\bm{k},\bm{n})|^2}{\mathcal{N}(\bm{k})}\ln\left[\frac{|F(\bm{k},\bm{n})|^2}{\mathcal{N}(\bm{k})}\right].\label{eq:EE_of_AS_n_cut_typeI_appendix}
\end{align}
From the inequality $\langle x\ln x\rangle\ge 0$ for $\langle x\rangle=1$, where $\langle\cdots\rangle$ denotes the uniform average over product-basis configurations in the fixed-charge sector, we obtain $S_{\rm vN}^{{\rm I}:AS}\le S_{\rm vN}^{\rm I}$. Furthermore, using $|F(\bm{k},\bm{n})|\le n$, we obtain $(S_{\bm n}^{AS})^2\le n^2/[\mathcal{N}(\bm{k})W_{M,n}]$. The entanglement entropy is therefore bounded below by
\begin{align}
S_{\rm vN}^{{\rm I}: AS}&\ge -\sum_{\bm{n}:\hat{N}\cket{\bm{n}}=n\cket{\bm{n}}}\frac{|F(\bm{k},\bm{n})|^2}{\mathcal{N}(\bm{k})W_{M,n}}\ln\left[\frac{n^2}{\mathcal{N}(\bm{k})W_{M,n}}\right]\notag \\
&=S_{\rm vN}^{\rm I}-\ln\left[\frac{n^2}{\mathcal{N}(\bm{k})}\right].\label{eq:lower_bound_of_EE_type_I_ARQMBS_appendix}
\end{align}
Direct calculations show that
\begin{align}
\mathcal{N}(\bm{k})&=\frac{M^2}{M-1}v_{M,n}.\label{eq:relation_mathcal_N_and_v_M_n_appendix}
\end{align}
Using the inequality $\hat{n}_j^2\le 2S\hat{n}_j$, we obtain
\begin{align}
v_{M,n}\le \frac{n}{M}\left(2S-\frac{n}{M}\right).\label{eq:upper_bound_of_v_M_n_appendix}
\end{align}
From Eqs.~(\ref{eq:inequality_for_v_M_n_spin-S_appendix}) and (\ref{eq:upper_bound_of_v_M_n_appendix}), we obtain $v_{M,n}=\Theta(1)$ whenever $\epsilon\le n/M\le 2S-\epsilon$, where $\epsilon>0$ is independent of the system size. Thus, we obtain $\mathcal{N}(\bm{k})=\Theta(M)$. Equation~(\ref{eq:lower_bound_of_EE_type_I_ARQMBS_appendix}) then gives the volume-law scaling:
\begin{align}
S_{\rm vN}^{{\rm I}:AS}\ge S_{\rm vN}^{\rm I}-O(\ln M).\label{eq:proof_of_volumelaw_scaling_bipartitionI_appendix}
\end{align}

We next consider the type-II bipartition. From Eq.~(\ref{eq:Schmidt_decomposition_RQMBS_type_II_appendix}), $|AS_n\rangle$ can be written as
\begin{align}
&\cket{AS_n}\notag \\
&\propto \hat{\rho}(\bm{k})\cket{S_n}\notag \\
&=\hspace{-0.9em}\sum_{m=m_{\rm min}}^{m_{\rm max}}\hspace{-0.9em}\sqrt{\lambda_m}\left[\hat{\rho}_A(\bm{k})|\Psi^A_m\rangle\otimes|\Psi^B_{n-m}\rangle\right.\notag \\
&\left.\hspace{5.0em}+|\Psi^A_m\rangle\otimes\hat{\rho}_B(\bm{k})|\Psi^B_{n-m}\rangle\right].\label{eq:expression_of_AS_n_bipartition_II}
\end{align}
Here, we decompose the operator $\hat{\rho}(\bm{k})$ as
\begin{align}
\hat{\rho}(\bm{k})&=\hat{\rho}_A(\bm{k})+\hat{\rho}_B(\bm{k}),\label{eq:definition_of_decomposition_rho_A_and_B_appendix}\\
\hat{\rho}_{A}(\bm{k})&\equiv\sum_{j\in\Lambda_{A}}e^{i\bm{k}\cdot\bm{R}_{j}}(\hat{n}_{j}\otimes\hat{1}),\label{eq:definition_of_rho_A_k_appendix}\\
\hat{\rho}_{B}(\bm{k})&\equiv\sum_{j\in\Lambda_{B}}e^{i\bm{k}\cdot\bm{R}_{j}}(\hat{n}_{j}\otimes\hat{1})\label{eq:definition_of_rho_b_k_appendix},
\end{align}
where $\Lambda_A$ and $\Lambda_B$ denote the sets of lattice sites in physical subsystem $1$ that belong to regions $A$ and $B$, respectively. From this result, the Schmidt rank of $\cket{AS_n}$ is at most $2[{\rm min}(n,2SM-n)+1]$. Therefore, the EE satisfies the following subvolume-law bound:
\begin{align}
S_{\rm vN}^{{\rm II}: AS}\le \ln S_{\rm rank}^{\rm AS}\le \ln[2SM+2]=O(\ln M).\label{eq:subvolue_law_scaling_AS_n_bipartition_II_appendix}
\end{align}

\section{Details of the proofs for multicomponent fermionic systems}\label{app:proof_of_RQMBS_fermions}
In this appendix, we prove some relations used in Sec.~\ref{subsec:multicomponent_Fermion}.

\subsection{Proof of the local frustration-free decomposition of the parent Hamiltonian}\label{app:proof_local_FF_decomposition_parent_Hamitonian_fermions}
Here, we show that, as in the spin-$S$ case, the parent Hamiltonian for fermionic systems admits a local frustration-free decomposition. In fermionic systems, we need to treat fermionic signs carefully because even if $\hat{h}_j$ is local, the matrix element $\bra{\bm{n}}\hat{h}_j\cket{\bm{m}}$ may depend on occupations outside ${\rm supp}(\hat{h}_j)$ owing to fermionic signs. For example, let us consider $\hat{h}_j=\hat{c}^{\dagger}_1\hat{c}_3+\hat{c}_3^{\dagger}\hat{c}_1$ in a three-site spinless fermion system. The support of $\hat{h}_j$ is given by $\{1,3\}$. The matrix element becomes $\bra{1,n_2,0}\hat{h}_j\cket{0,n_2,1}=(-1)^{n_2}$, showing that the matrix element depends on the occupation outside the support. 

For a finite region $X$, let $\hat{U}_{X}$ be a unitary operator that moves all fermionic modes in $X$ before those in its complement $\overline{X}$, while preserving the relative ordering
within each set. Its action on a basis state is written as
\begin{align}
\hat{U}_{X}|\bm{n}\rangle=u_{X}(\bm{n})|\bm{n}_{X}\rangle\otimes|\bm{n}_{\overline{X}}\rangle,\label{eq:action_of_UX_to_basis_state}
\end{align}
where $u_{X}(\bm{n})=\pm1$ is the fermionic reordering sign.

Now, we consider the matrix element of $\hat{h}_j$. Owing to the $U(1)^f$ symmetry, we may choose a decomposition $\hat{H}=\sum_j\hat{h}_j$ such that each $\hat{h}_j$ has even fermion parity. For ${\rm supp}(\hat{h}_{j})\subseteq X$, we have
\begin{align}
\langle\bm{n}|\hat{h}_{j}|\bm{m}\rangle&=u_{X}(\bm{n})^{\ast}u_{X}(\bm{m})\langle\bm{n}_{X}|\hat{h}_{j}^{(X)}|\bm{m}_{X}\rangle\delta_{\bm{n}_{\overline{X}},\bm{m}_{\overline{X}}},\label{eq:relation_matrix_element_hj_for_fermion_systems}\\
\hat{U}_X\hat{h}_j\hat{U}_X^{\dagger}&\equiv \hat{h}_j^{(X)}\otimes\hat{1}_{\overline{X}}.\label{eq:definition_of_h_j^x_fermions_app}
\end{align}
Applying this relation with $X=X_{j}$, the reordering factor is common to all
$l\in\mathcal{N}_{j}$. Hence, for $H_{\bm{n},\bm{m}}\not=0$ and $j\in I(\bm{n},\bm{m})$, we obtain
\begin{align}
H_{\bm{n},\bm{m}}&=\omega_{X_{j}}(\bm{n},\bm{m})H_{\bm{n}_{X_{j}},\bm{m}_{X_{j}}}^{X_{j}},\label{eq:matrix_element_local_decomposition_fermion_app}\\
H^{X_j}_{\bm{n}_{X_j},\bm{m}_{X_j}}&\equiv \sum_{l\in\mathcal{N}_j}\langle\bm{n}_{X_j}|\hat{h}^{(X_j)}_l|\bm{m}_{X_j}\rangle,\label{eq:definition_of_HX_nXj_mXj_fermion_app}\\
I(\bm{n},\bm{m})&=I_j(\bm{n}_{X_j},\bm{m}_{X_j}),\label{eq:relation_region_I_and_Ij_fermion_app}\\
I_j(\bm{n}_{X_j}, \bm{m}_{X_j})&\equiv \left\{l\in \mathcal{N}_j~|~ \langle\bm{n}_{X_j}|\hat{h}^{(X_j)}_{l}|\bm{m}_{X_j}\rangle\not=0\right\},\label{eq:definition_of_mathcal_I_j_fermion_app}\\
|\omega_{X_{j}}(\bm{n},\bm{m})|&=1,\label{eq:absolute_value_of_omega_fermion_app}
\end{align}
and therefore
\begin{align}
|H_{\bm{n},\bm{m}}|^{2}=|H_{\bm{n}_{X_{j}},\bm{m}_{X_{j}}}^{X_{j}}|^{2}.\label{eq:relation_Hnm_and_Hnxjmxj_fermions}
\end{align}

We next consider the projector $\hat{P}_{\bm{n},\bm{m}}$ for fermionic systems. Using the reordering operator, we obtain
\begin{align}
&\hat{U}_X\otimes\hat{U}_X(\cket{\bm{n}}\otimes\cket{\bm{n}})\notag \\
&=[u_X(\bm{n})]^2(\cket{\bm{n}_X}\otimes\cket{\bm{n}_{\overline{X}}})\otimes(\cket{\bm{n}_X}\otimes\cket{\bm{n}_{\overline{X}}})\notag \\
&=(\cket{\bm{n}_X}\otimes\cket{\bm{n}_{X}})\otimes(\cket{\bm{n}_{\overline{X}}}\otimes\cket{\bm{n}_{\overline{X}}}),\label{eq:reordering_operator_to_n_n_fermion_app}
\end{align}
where we have regrouped the tensor factors in the last line. Using this result, for $\bm{n}_{\overline{X}}=\bm{m}_{\overline{X}}$, we obtain
\begin{align}
&(\hat{U}_X\otimes\hat{U}_X)\hat{P}_{\bm{n},\bm{m}}(\hat{U}_X^{\dagger}\otimes\hat{U}_X^{\dagger})\notag \\
&=\hat{P}^X_{\bm{n}_X,\bm{m}_X}\otimes(\cket{\bm{n}_{\overline{X}}}\otimes\cket{\bm{n}_{\overline{X}}})(\bra{\bm{n}_{\overline{X}}}\otimes\bra{\bm{n}_{\overline{X}}}).\label{eq:decomposition_Pnm_fermion_app}
\end{align}

The enlarged scar subspace for fermionic systems can be written as
\begin{align}
\mathcal{H}_P&\equiv \bigotimes_{j=1}^M\mathcal{H}^{\rm L}_{P,j},\label{eq:definition_of_enlarged_scar_subspace_fermion_app}\\
\mathcal{H}^{\rm L}_{P,j}&\equiv {\rm Span}\{\cket{\bm{\eta}}_{j,1}\otimes\cket{\bm{\eta}}_{j,2}~|~ \bm{\eta}\in\{0,1\}^{N_{\rm F}} \}.\label{eq:definition_of_local_enlarged_scar_subspace_fermion_app}
\end{align}

As in the spin-$S$ case, we can show that the parent Hamiltonian admits a local frustration-free decomposition for fermionic systems. In what follows, all operator equalities involving $\hat{H}_{\rm p}$ are understood as equalities on $\mathcal{H}_P$. Whenever $H_{\bm{n}_{X_j},\bm{m}_{X_j}}^{X_j}\not=0$, the $U(1)^f$ symmetry of $\hat{H}$ implies that the corresponding global configurations carry the same charges. Since the two configurations coincide outside $X_j$, $\bm{n}_{X_j}$ and $\bm{m}_{X_j}$ carry the same charges. Therefore, we obtain $\hat{P}^{X_j}_{\bm{n}_{X_j},\bm{m}_{X_j}}|\tilde{S}_{\bm{l}}\rangle=0$ and  
\begin{align}
\hat{H}_{\rm p}&=\sum_j\hat{h}_{{\rm p},j},\label{eq:parent_Hamiltonian_decomposition_fermion_app}\\
\hat{h}_{{\rm p},j}&\ge0,\quad \hat{h}_{{\rm p},j}|\tilde{S}_{\bm{l}}\rangle=0,\quad\text{for all }j,\label{eq:FF_condition_for_fermions_app}
\end{align}
where $\hat{h}_{{\rm p},j}$ can be written as
\begin{align}
\hat{h}_{{\rm p},j}&\equiv (\hat{U}_{X_j}^{\dagger}\otimes\hat{U}_{X_j}^{\dagger})\left[\hat{h}_{{\rm p},j}^{(X_j)}\otimes\hat{1}_{\bar{X}_j}\right](\hat{U}_{X_j}\otimes\hat{U}_{X_j}),\label{eq:definition_of_hpj_fermions_app}\\
\hat{h}_{{\rm p},j}^{(X_j)}&\equiv 2\sum_{\substack{\bm{n}_{X_j},\bm{m}_{X_j} \\ j\in I_j(\bm{n}_{X_j},\bm{m}_{X_j})}}\frac{|H^{X_j}_{\bm{n}_{X_j},\bm{m}_{X_j}}|^2}{| I_j(\bm{n}_{X_j},\bm{m}_{X_j})|}\hat{P}^{X_j}_{\bm{n}_{X_j},\bm{m}_{X_j}}.\label{eq:definition_of_hXj_p_j_fermions_app}
\end{align}
Transforming
$\hat{h}_{{\rm p},j}^{(X_j)}$
back to the original fermionic-mode ordering does not
generate a parity string outside $X_j$. Since $X_j$ contains $O(1)$ sites and has diameter $O(1)$, $\hat{h}_{{\rm p},j}$ is $r'$-local and has range $R'$, where $r',R'=O(1)$.

\subsection{Proof of the gaplessness of the parent Hamiltonian}\label{app:proof_gaplessness_parent_Hamitonian_fermions}
Here, we prove that the parent Hamiltonian in Eq.~(\ref{eq:local_decompotion_parent_Hamiltonian_fermions}) for multicomponent fermionic systems is gapless. First, we prove $\langle\tilde{S}_{\bm{l}'}|\widetilde{AS}_{\bm{l}}\rangle=0$ for all $\bm{l}$ and $\bm{l}'$. Because $|\tilde{S}_{\bm{l}'}\rangle$ and $|\widetilde{AS}_{\bm{l}}\rangle$ carry the charge vectors $\bm{l}'$ and $\bm{l}$, respectively, we obtain $\langle\tilde{S}_{\bm{l}'}|\widetilde{AS}_{\bm{l}}\rangle=0$ unless $\bm{l}=\bm{l}'$. For $\bm{l}=\bm{l}'$, a direct calculation shows that
\begin{align}
&\langle\tilde{S}_{\bm{l}} | \widetilde{AS}_{\bm{l}}\rangle\notag \\
&=\frac{1}{l_{s_0}}\langle\tilde{S}_{\bm{l}}|\sum_{j=1}^M\sum_{\sigma\in g_{s_0}}e^{i\bm{k}\cdot\bm{R}_j}(\hat{n}_{j,\sigma}\otimes\hat{1})|\tilde{S}_{\bm{l}}\rangle\notag \\
&=\frac{1}{l_{s_0}}\prod_{s=1}^f(l_s!)^2\sum_{j=1}^M\sum_{\sigma\in g_{s_0}}e^{i\bm{k}\cdot\bm{R}_j}\sum_{\substack{\bm{n}:\hat{\mathcal{N}}_s\cket{\bm{n}}=l_s\cket{\bm{n}} \\ n=\sum_sl_s}}\bra{\bm{n}}\hat{n}_{j,\sigma}\cket{\bm{n}}\notag \\
&=\frac{1}{l_{s_0}}\prod_{s=1}^f(l_s!)^2\sum_{j=1}^Me^{i\bm{k}\cdot\bm{R}_j}C'\notag \\
&=0,\label{eq:proof_of_orthogonality_of_variational_state_fermions}
\end{align}
where $C'$ is a constant independent of $j$ and $\sigma$;  in the last equality, we used Eq.~(\ref{eq:definition_of_crystal_momentum_spin-S}). Therefore, the variational state $|\widetilde{AS}_{\bm{l}}\rangle$ is orthogonal to all RQMBS states. 

Next, we evaluate the denominator $f_{\rm d}(\bm{k})$ in Eq.~(\ref{eq:energy_variance_for_fermions}). To this end, we evaluate $\langle\tilde{S}_{\bm{l}}|\tilde{S}_{\bm{l}}\rangle$:
\begin{align}
\langle\tilde{S}_{\bm{l}}|\tilde{S}_{\bm{l}}\rangle&=\prod_{s=1}^f(l_s!)^2\sum_{\substack{\bm{n}:\hat{\mathcal{N}}_s\cket{\bm{n}}=l_s\cket{\bm{n}} \\ n=\sum_sl_s}}1\notag \\
&=\prod_{s=1}^f(l_s!)^2\prod_{s=1}^f\binom{|g_s|M}{l_s}\notag \\
&\equiv \prod_{s=1}^f(l_s!)^2W_{\bm{l}}.\label{eq:norm_of_RQMBS_fermions}
\end{align}
Without loss of generality, we can set $s_0=1$ and assume $l_1\ge 1$. The denominator becomes
\begin{align}
&f_{\rm d}(\bm{k})\notag \\
&=\frac{1}{l_1^2}\langle\tilde{S}_{\bm{l}}|\hat{\rho}_{1}(-\bm{k})\hat{\rho}_{1}(\bm{k}) |\tilde{S}_{\bm{l}}\rangle\notag \\
&=\frac{1}{l_1^2}\sum_{j,j'}\sum_{\sigma,\sigma'\in g_{1}}e^{-i\bm{k}\cdot(\bm{R}_j-\bm{R}_{j'})}\langle\tilde{S}_{\bm{l}}|(\hat{n}_{j,\sigma}\hat{n}_{j',\sigma'}\otimes\hat{1})|\tilde{S}_{\bm{l}}\rangle\notag \\
&=\frac{1}{l_1^2}\prod_{s=1}^f(l_s!)^2\sum_{j,j'}e^{-i\bm{k}\cdot(\bm{R}_j-\bm{R}_{j'})}\sum_{\substack{\bm{n}:\hat{\mathcal{N}}_s\cket{\bm{n}}=l_s\cket{\bm{n}} \\ n=\sum_sl_s}}q_{j,1}q_{j',1}\notag \\
&=\frac{1}{l_1^2}\prod_{s=1}^f(l_s!)^2M\sum_{\substack{\bm{n}: \hat{\mathcal{N}}_s\cket{\bm{n}}=l_s\cket{\bm{n}} \\ n=\sum_sl_s}}(q_{1,1}^2-q_{1,1}q_{2,1})\notag \\
&=\frac{1}{l_1^2}\prod_{s=1}^f(l_s!)^2M\prod_{s=2}^f\binom{|g_s|M}{l_s}\sideset{}{'}\sum_{\bm{n}}(q_{1,1}^2-q_{1,1}q_{2,1}),\label{eq:calculation_fd_fermion_totyu1}
\end{align}
where $q_{j,s}\equiv \sum_{\sigma\in g_{s}}\bra{\bm{n}}\hat{n}_{j,\sigma}\cket{\bm{n}}\equiv \sum_{\sigma\in g_s}\eta_{j,\sigma}$ and the primed sum in the last line runs over all configurations of the fermionic modes belonging to $g_1$ with $l_1$ fermions. A direct calculation gives
\begin{align}
\sideset{}{'}\sum_{\bm{n}}q_{1,1}^2&=\sideset{}{'}\sum_{\bm{n}}\left(\sum_{\sigma\in g_1}\eta_{1,\sigma}+\sum_{\substack{\sigma,\sigma'\in g_1 \\ \sigma\not=\sigma' }}\eta_{1,\sigma}\eta_{1,\sigma'}\right)\notag \\
&=|g_1|\binom{|g_1|M-1}{l_1-1}\notag \\
&\quad +|g_1|(|g_1|-1)\binom{|g_1|M-2}{l_1-2},\label{eq:calculation_q11_squared_fermion_app}\\
\sideset{}{'}\sum_{\bm{n}}q_{1,1}q_{2,1}&=|g_1|^2\binom{|g_1|M-2}{l_1-2}.\label{eq:calculation_q11q21_fermion_app}
\end{align}
Using the formula $\binom{n}{k}+\binom{n}{k-1}=\binom{n+1}{k}$, we obtain
\begin{align}
f_{\rm d}(\bm{k})&=\frac{1}{l_1^2}\prod_{s=1}^f(l_s!)^2M\frac{|g_1|\binom{|g_1|M-2}{l_1-1}}{\binom{|g_1|M}{l_1}}W_{\bm{l}}\notag \\
&=\frac{1}{l_1}\frac{|g_1|M-l_1}{|g_1|M-1}\langle\tilde{S}_{\bm{l}}|\tilde{S}_{\bm{l}}\rangle.\label{eq:final_results_of_fd_k_fermion_app}
\end{align}
We note that $f_{\rm d}(\bm{k})=0$ for $l_1=|g_1|M$. Hence, we consider the case $1\le l_1\le |g_1|M-1$. 

Now, we evaluate the numerator $f_{\rm n}(\bm{k})$ in Eq.~(\ref{eq:energy_variance_for_fermions}). Following the calculation for the spin-$S$ case, we obtain
\begin{align}
&f_{\rm n}(\bm{k})\notag \\
&=\frac{1}{l_1^2}\sum_j\|\hat{h}^{1/2}_{{\rm p},j}\hat{\rho}_1(\bm{k})|\tilde{S}_{\bm{l}}\rangle\|^2\notag \\
&\le \frac{1}{l_1^2}\sum_j|S_{{\rm p},j}^{(1)}||g_1|\notag \\
&\quad\times\sum_{x\in S_{{\rm p},j}^{(1)}} \sum_{\sigma\in g_1} \|(e^{i\bm{k}\cdot\bm{R}_x}-e^{i\bm{k}\cdot\bm{R}_{x_0}})\hat{h}_{{\rm p},j}^{1/2}(\hat{n}_{x,\sigma}\otimes\hat{1})|\tilde{S}_{\bm{l}}\rangle \|^2\notag \\
&\le \frac{|g_1|M}{l_1^2}C_5\bm{k}^2\langle\tilde{S}_{\bm{l}}|\tilde{S}_{\bm{l}}\rangle,\label{eq:inequality_fn_k_appC}
\end{align}
where $x_0\in S_{{\rm p},j}^{(1)}$ and $C_5$ is an $O(1)$ constant. To obtain the above inequality, we assume that $\|\hat{h}_{{\rm p},j}\|_{\rm op}\le J_{\rm p}=O(1)$ and $\| \hat{n}_{x,\sigma}\|_{\rm op}\le 1$.

From the above results, we obtain the upper bound on the gap of the parent Hamiltonian:
\begin{align}
\Delta_{{\rm p},\bm{l}}&\le \Delta E_{\bm{l}}^2=\frac{f_{\rm n}(\bm{k})}{f_{\rm d}(\bm{k})}\notag \\
&\le \frac{|g_1|M}{l_1}\frac{|g_1|M-1}{|g_1|M-l_1}C_5\bm{k}^2.\label{eq:gap_of_parent_Hamiltonian_fermion_app}
\end{align}
For $\epsilon \le l_1/(|g_1|M)\le 1-\epsilon$, where $\epsilon>0$ is independent of the system size, we obtain $\Delta_{{\rm p},\bm{l}}\le \Delta E_{\bm{l}}^2\le (C_5/\epsilon^2)\bm{k}^2=O(\bm{k}^2)$. Following the same argument as in the spin-$S$ case, we conclude that the energy variance of the variational state $|\widetilde{AS}_{\bm{l}}\rangle$ vanishes in the thermodynamic limit. The same conclusion follows for a general $s_0$ upon replacing $\hat{\rho}_1(\bm{k})$ with $\hat{\rho}_{s_0}(\bm{k})$.

\subsection{Entanglement entropy of the RQMBS and variational states}\label{app:entanglement_entropy_of_the_variational_states_fermion}
In this subsection, we evaluate the EEs of the normalized RQMBS and normalized variational states $|AS_{\bm{l}}\rangle$ for fermionic systems. We first consider the EE of the RQMBS states and then turn to the variational state. We assume that $M$ is even.

The normalized RQMBS state can be written as
\begin{align}
|S_{\bm{l}}\rangle=\frac{1}{\sqrt{W_{\bm{l}}}}\sum_{\bm{n}:\hat{\mathcal{N}}_s\cket{\bm{n}}=l_s\cket{\bm{n}}}\cket{\bm{n}}\otimes\cket{\bm{n}}.\label{eq:normalized_RQMBS_fermions_app}
\end{align}
We first consider the type-I bipartition. In this case, the EE is given by
\begin{align}
S_{\rm vN}^{{\rm I}}=\ln W_{\bm{l}}=\sum_{s=1}^f\ln\binom{|g_s|M}{l_s}.\label{eq:EE_of_RQMBS_typeI_fermion}
\end{align}
From Eqs.~(\ref{eq:inequality_binomial_constatn_and_H2_appendix}) and (\ref{eq:definition_of_H2_appendix}), we obtain the inequality
\begin{align}
S_{\rm vN}^{\rm I}\ge M\sum_{s=1}^f|g_s|H_2\left(\frac{l_s}{|g_s|M}\right)-\sum_{s=1}^f\ln(|g_s|M+1).\label{eq:inequality_EE_typeI_RQMBS_fermions}
\end{align}
We assume that, for at least one $s$, $\epsilon\le l_s/(|g_s|M)\le 1-\epsilon$, where $\epsilon>0$ is independent of $M$. Using the bound on $H_2(\cdot)$ employed in Appendix \ref{app:entanglement_entropy_of_the_variational_states}, the right-hand side scales linearly with $M$, up to logarithmic corrections. This means that $S_{\rm vN}^{\rm I}$ obeys a volume law. 

Next, we consider the type-II bipartition. The Schmidt decomposition of the RQMBS state is given by
\begin{align}
|S_{\bm{l}}\rangle&=\sum_{\bm{m}}\sqrt{\lambda_{\bm{m}}} |\Psi^A_{\bm{m}}\rangle\otimes|\Psi^B_{\bm{l}-\bm{m}}\rangle,\label{eq:typeII_bipartition_RQMBS_fermion_app}\\
\lambda_{\bm{m}}&\equiv \frac{W^A_{\bm{m}}W^B_{\bm{l}-\bm{m}}}{W_{\bm{l}}},\label{eq:Schmidt_coefficients_RQMBS_fermion_app}\\
W_{\bm{m}}^A&\equiv \prod_{s=1}^f\binom{|g_s|M/2}{m_s},\label{eq:definition_of_WA_RQMBS_fermion_app}\\
W_{\bm{l}-\bm{m}}^B&\equiv \prod_{s=1}^f\binom{|g_s|M/2}{l_s-m_s},\label{eq:definition_of_WB_RQMBS_fermion_app}\\
|\Psi^A_{\bm{m}}\rangle&\equiv \frac{1}{\sqrt{W^A_{\bm{m}}}}\sum_{\bm{n}_A: \hat{\mathcal{N}}_s^A\cket{\bm{n}_A}=m_s\cket{\bm{n}_A}}\cket{\bm{n}_A}\otimes\cket{\bm{n}_A},\label{eq:definition_of_psi_A_RQMBS_fermion_app}\\
|\Psi^B_{\bm{l}-\bm{m}}\rangle&\equiv \frac{1}{\sqrt{W^B_{\bm{l}-\bm{m}}}}\sum_{\bm{n}_B: \hat{\mathcal{N}}_s^B\cket{\bm{n}_B}=(l_s-m_s)\cket{\bm{n}_B}}\cket{\bm{n}_B}\otimes\cket{\bm{n}_B}.\label{eq:definition_of_psi_B_RQMBS_fermion_app}
\end{align}
Here, $\hat{\mathcal{N}}_s^A$ and $\hat{\mathcal{N}}_s^B$ denote the fermion-number operators for the component group $g_s$ in regions $A$ and $B$, respectively. The vector $\bm{m}\equiv (m_1,m_2,\ldots, m_f)$ specifies the numbers of fermions in region $A$. The allowed range of $m_s$ is given by
\begin{align}
\max(0,l_s-|g_s|M/2)\le m_s\le \min\left(l_s,|g_s|M/2\right).\label{eq:range_of_allowed_ms_fermions_app}
\end{align}
We obtain the Schmidt rank
\begin{align}
&S_{\rm rank}\notag \\
&=\prod_{s=1}^f\left[\min(l_s,|g_s|M/2)-\max(0,l_s-|g_s|M/2)+1\right]\notag \\
&=\prod_{s=1}^f[\min(l_s, |g_s|M-l_s)+1].\label{eq:Schmidt_rank_RQMBS_fermions_app}
\end{align}
Therefore, the EE for the type-II bipartition obeys a subvolume law:
\begin{align}
S_{\rm vN}^{\rm II}&\le \ln S_{\rm rank}=\sum_{s=1}^f\ln[\min(l_s, |g_s|M-l_s)+1]\notag \\
&\le \sum_{s=1}^f\ln\left(\frac{|g_s|M}{2}+1\right)=O(\ln M).\label{eq:subvolume_law_RQMBS_typeII_fermion_app}
\end{align}

Now, we consider the variational state $|AS_{\bm{l}}\rangle$. The normalized variational state can be written as
\begin{align}
|AS_{\bm{l}}\rangle&=\sum_{\bm{n}:\hat{\mathcal{N}}_s\cket{\bm{n}}=l_s\cket{\bm{n}}}\frac{|F_1(\bm{k},\bm{n})|e^{i\theta(\bm{k},\bm{n})}}{\sqrt{\mathcal{N}_{\bm{l}}(\bm{k})W_{\bm{l}}}}\cket{\bm{n}}\otimes\cket{\bm{n}},\label{eq:variational_state_typeI_fermion_app}\\
F_1(\bm{k},\bm{n})&\equiv \sum_{j=1}^Me^{i\bm{k}\cdot\bm{R}_j}q_{j,1}\equiv |F_1(\bm{k},\bm{n})|e^{i\theta(\bm{k},\bm{n})},\label{eq:definition_of_F1_fermion_app}\\
\mathcal{N}_{\bm{l}}(\bm{k})&\equiv \frac{1}{W_{\bm{l}}}\sum_{\bm{n}:\hat{\mathcal{N}}_s\cket{\bm{n}}=l_s\cket{\bm{n}}}|F_1(\bm{k},\bm{n})|^2.\label{eq:definition_of_N_l_k_fermions_app}
\end{align}
The EE for the type-I bipartition is given by
\begin{align}
&S_{\rm vN}^{{\rm I}: AS}\notag \\
&=S_{\rm vN}^{\rm I}-\frac{1}{W_{\bm{l}}}\sum_{\bm{n}:\hat{\mathcal{N}}_s\cket{\bm{n}}=l_s\cket{\bm{n}}}\frac{|F_1(\bm{k},\bm{n})|^2}{\mathcal{N}_{\bm{l}}(\bm{k})}\ln\left[\frac{|F_1(\bm{k},\bm{n})|^2}{\mathcal{N}_{\bm{l}}(\bm{k})}\right].\label{eq:EE_typeI_ARQMBS_fermion_app}
\end{align}
Following the same argument as in Appendix \ref{app:entanglement_entropy_of_the_variational_states}, we obtain the upper bound: $S_{\rm vN}^{{\rm I}: AS}\le S_{\rm vN}^{\rm I}=\Theta(M)$. From the inequality $|F_1(\bm{k},\bm{n})|\le \sum_{j=1}^Mq_{j,1}=l_1$, we obtain
\begin{align}
S_{\rm vN}^{{\rm I}: AS}\ge S_{\rm vN}^{\rm I}-\ln\left[\frac{l_1^2}{\mathcal{N}_{\bm{l}}(\bm{k})}\right]=S_{\rm vN}^{\rm I}-\ln\left[\frac{l_1(|g_1|M-1)}{|g_1|M-l_1}\right],\label{eq:lower_bound_EE_ARQMBS_typeI_fermions_app}
\end{align}
where we used Eq.~(\ref{eq:final_results_of_fd_k_fermion_app}). Assuming $\epsilon\le l_1/(|g_1|M)\le 1-\epsilon$ for some positive system-size-independent constant $\epsilon$, we obtain
\begin{align}
S_{\rm vN}^{{\rm I}: AS}\ge S_{\rm vN}^{\rm I}-O(\ln M).\label{eq:lower_bound_EE_typeI_ARQMBS_fermions_app}
\end{align}
Thus, the variational state obeys a volume law for the type-I bipartition.

Next, we consider the EE of the variational state for the type-II bipartition. The variational state can be written as
\begin{align}
|AS_{\bm{l}}\rangle&\propto \hat{\rho}_1(\bm{k})|S_{\bm{l}}\rangle\notag \\
&=\sum_{\bm{m}}\sqrt{\lambda_{\bm{m}}}\left[\hat{\rho}_{1,A}(\bm{k})|\Psi^A_{\bm{m}}\rangle\otimes|\Psi^B_{\bm{l}-\bm{m}}\rangle\right.\notag \\
&\left.\hspace{8.0em}+|\Psi^A_{\bm{m}}\rangle\otimes\hat{\rho}_{1,B}(\bm{k})|\Psi^B_{\bm{l}-\bm{m}}\rangle\right],\label{eq:ARQMBS_typeII_Fermion_app}
\end{align}
where we decompose the operator $\hat{\rho}_{1}(\bm{k)}$ as
\begin{align}
\hat{\rho}_1(\bm{k})&=\hat{\rho}_{1,A}(\bm{k})+\hat{\rho}_{1,B}(\bm{k}),\label{eq:decomposition_density_operator1_fermions_app}\\
\hat{\rho}_{1,A}(\bm{k})&\equiv \sum_{j\in\Lambda_A}\sum_{\sigma\in g_1}e^{i\bm{k}\cdot\bm{R}_j}(\hat{n}_{j,\sigma}\otimes \hat{1}),\label{eq:definition_of_rho_1_A_fermion_app}\\
\hat{\rho}_{1,B}(\bm{k})&\equiv \sum_{j\in\Lambda_B}\sum_{\sigma\in g_1}e^{i\bm{k}\cdot\bm{R}_j}(\hat{n}_{j,\sigma}\otimes \hat{1}).\label{eq:definition_of_rho_1_B_fermion_app}
\end{align}
From Eq.~(\ref{eq:ARQMBS_typeII_Fermion_app}), the Schmidt rank $S_{\rm rank}^{\rm AS}$ is at most $2S_{\rm rank}=O(M^f)$. We obtain the following upper bound on the EE:
\begin{align} 
S_{\rm vN}^{{\rm II}: AS}\le \ln S_{\rm rank}^{\rm AS}=O(\ln M).\label{eq:bound_for_EE_typeII_ARQMBS_fermion_app}
\end{align}
Thus, the EE for the type-II bipartition obeys a subvolume law. 

\section{Details of the proofs for multicomponent bosonic systems}\label{app:proof_of_RQMBS_bosons}
In this appendix, we prove some relations used in Sec.~\ref{subsec:multicomponent_Boson}.

\subsection{Proof of the gaplessness of the parent Hamiltonian}\label{app:proof_gaplessness_parent_Hamitonian_bosons}
Here, we prove that the parent Hamiltonian in Eq.~(\ref{eq:local_decompotion_parent_Hamiltonian_bosons}) for multicomponent bosonic systems is gapless. For bosonic systems, we can also prove $\langle \tilde{S}_{\bm{l}'}|\widetilde{AS}_{\bm{l}}\rangle=0$. The proof is essentially the same as in the fermionic case [see Eq.~(\ref{eq:proof_of_orthogonality_of_variational_state_fermions})], and we therefore omit the details.

Next, we evaluate the denominator $f_{\rm d}(\bm{k})$ in Eq.~(\ref{eq:energy_variance_for_bosonic_systems}). To this end, we evaluate $\langle\tilde{S}_{\bm{l}}|\tilde{S}_{\bm{l}}\rangle$:
\begin{align}
\langle\tilde{S}_{\bm{l}}|\tilde{S}_{\bm{l}}\rangle&=\prod_{s=1}^f(l_s!)^2\sum_{\bm{n}:\hat{\mathcal{N}}_s\cket{\bm{n}}=l_s\cket{\bm{n}} }1\notag \\
&=\prod_{s=1}^f(l_s!)^2\prod_{s=1}^f\binom{|g_s|M+l_s-1}{l_s}\\
&\equiv \prod_{s=1}^f(l_s!)^2W_{\bm{l}}.\label{eq:norm_of_RQMBS_bosons_app}
\end{align}
Without loss of generality, we set $s_{0}=1$ and assume $l_1\ge 1$. As in Eq.~(\ref{eq:calculation_fd_fermion_totyu1}), the denominator for $\bm{k}\neq\bm{0}$ becomes
\begin{align}
&f_{\rm d}(\bm{k})\notag \\
&=\frac{M}{l_1^2}\prod_{s=1}^f(l_s!)^2\prod_{s=2}^f\binom{|g_s|M+l_s-1}{l_s}\sideset{}{'}\sum_{\bm{n}}(q_{1,1}^2-q_{1,1}q_{2,1}),\label{eq:calculation_totyu2_denominator_bosons_app}
\end{align}
where $q_{j,s}\equiv \sum_{\sigma\in g_s}\bra{\bm{n}}\hat{n}_{j,\sigma}\cket{\bm{n}}\equiv \sum_{\sigma\in g_s}\eta_{j,\sigma}$ and the primed sum runs over all configurations of bosonic modes belonging to $g_1$ with $l_1$ bosons. A direct calculation gives
\begin{widetext}
\begin{align}
&\sideset{}{'}\sum_{\bm{n}}q_{1,1}^2\notag \\
&=\sideset{}{'}\sum_{\bm{n}}\left(\sum_{\sigma\in g_1}\eta_{1,\sigma}^2+\sum_{\substack{\sigma,\sigma'\in g_1 \\ \sigma\not=\sigma'}}\eta_{1,\sigma}\eta_{1,\sigma'}\right)\notag \\
&=|g_1|\sum_{m=0}^{l_1}m^2\binom{|g_1|M+l_1-m-2}{l_1-m}+|g_1|(|g_1|-1)\sum_{m_1=0}^{l_1}\sum_{m_2=0}^{l_1}m_1m_2\binom{|g_1|M+l_1-m_1-m_2-3}{l_1-m_1-m_2}\notag \\
&=|g_1|\sum_{m=0}^{l_1}\left[\binom{m}{1}+2\binom{m}{2}\right]\binom{|g_1|M+l_1-m-2}{l_1-m}+|g_1|(|g_1|-1)\sum_{m=0}^{l_1}\sum_{m_1=0}^{m}m_1(m-m_1)\binom{|g_1|M+l_1-m-3}{l_1-m}\notag \\
&=|g_1|\left[\binom{|g_1|M+l_1-1}{l_1-1}+2\binom{|g_1|M+l_1-1}{l_1-2}\right]+|g_1|(|g_1|-1)\sum_{m=0}^{l_1}\binom{m+1}{3}\binom{|g_1|M+l_1-m-3}{l_1-m}\notag \\
&=|g_1|\left[\binom{|g_1|M+l_1-1}{l_1-1}+(|g_1|+1)\binom{|g_1|M+l_1-1}{l_1-2}\right],\label{eq:calculation_q11squared_bosons_app}
\end{align}
\end{widetext}
where we used the following variant of the Chu--Vandermonde identity [see Eq.~(5.26) of Ref.~\cite{Graham1994concrete}]:
\begin{align}
\sum_{m=0}^{l}\binom{m}{q}\binom{R+l-m}{l-m}&=\binom{R+l+1}{l-q}.\label{eq:variant_Chu-Vandermonde_indentity_version1_app}
\end{align}
Similar calculations give
\begin{align}
&\sideset{}{'}\sum_{\bm{n}}q_{1,1}q_{2,1}\notag \\
&=\sideset{}{'}\sum_{\bm{n}}\sum_{\sigma,\sigma'\in g_1}\eta_{1,\sigma}\eta_{2,\sigma'}\notag \\
&=|g_1|^2\sum_{m_1=0}^{l_1}\sum_{m_2=0}^{l_1}m_1m_2\binom{|g_1|M+l_1-m_1-m_2-3}{l_1-m_1-m_2}\notag \\
&=|g_1|^2\binom{|g_1|M+l_1-1}{l_1-2}.\label{eq:calculation_q11q21_bosons_app}
\end{align}
Thus, we obtain 
\begin{align}
&\sideset{}{'}\sum_{\bm{n}}(q_{1,1}^2-q_{1,1}q_{2,1})\notag \\
&=|g_1|\left[\binom{|g_1|M+l_1-1}{l_1-1}+\binom{|g_1|M+l_1-1}{l_1-2}\right]\notag \\
&=|g_1|\binom{|g_1|M+l_1}{l_1-1}\notag \\
&=\frac{l_1(|g_1|M+l_1)}{M(|g_1|M+1)}\binom{|g_1|M+l_1-1}{l_1}.\label{eq:calculation_q11squared-q11q21_bosons}
\end{align}
The denominator $f_{\rm d}(\bm{k})$ becomes
\begin{align}
f_{\rm d}(\bm{k})&=\frac{M}{l_1^2}\prod_{s=1}^f(l_s!)^2\frac{l_1(|g_1|M+l_1)}{M(|g_1|M+1)}W_{\bm{l}}\notag \\
&=\frac{|g_1|M+l_1}{l_1(|g_1|M+1)}\langle\tilde{S}_{\bm{l}}|\tilde{S}_{\bm{l}}\rangle.\label{eq:final_form_of_fd_k_bosons}
\end{align}

Now, we evaluate the numerator $f_{\rm n}(\bm{k})$ in Eq.~(\ref{eq:energy_variance_for_bosonic_systems}). Following the calculations for the spin-$S$ and fermionic cases, we obtain
\begin{align}
&f_{\rm n}(\bm{k})\notag \\
&=\frac{1}{l_1^2}\sum_j\| \hat{h}_{{\rm p},j}^{1/2}\hat{\rho}_1(\bm{k})|\tilde{S}_{\bm{l}}\rangle \|^2\notag \\
&\le \frac{1}{l_1^2}\sum_j|S^{(1)}_{{\rm p},j}||g_1|\sum_{x\in S^{(1)}_{{\rm p},j}}\sum_{\sigma\in g_1}\|\hat{h}_{{\rm p},j}^{1/2}(\hat{n}_{x,\sigma}\otimes\hat{1})|\tilde{S}_{\bm{l}}\rangle \|^2|\bm{k}|^2C_6,\label{eq:inequality_for_fn_k_totyu1_bosons_app}
\end{align}
where $C_6$ is an $O(1)$ constant. 

In contrast to the spin-\(S\) and fermionic cases, we cannot obtain this bound using operator norms, because neither
$\|\hat{h}_{{\rm p},j}\|_{\rm op}$ nor $\|\hat{n}_{x,\sigma}\|_{\rm op}$ is bounded by an \(O(1)\) constant. We therefore assume that
\begin{align}
\frac{\| \hat{h}^{1/2}_{{\rm p},j}(\hat{n}_{x,\sigma}\otimes\hat{1})|\tilde{S}_{\bm{l}}\rangle \|^2}{\langle\tilde{S}_{\bm{l}}|\tilde{S}_{\bm{l}}\rangle}\le J_{\rm B},\text{ for all }j,\;\;x\in S_{{\rm p},j}^{(1)},\;\;\sigma\in g_1.\label{eq:assumptions_for_bosons_app}
\end{align}
Here, $J_{\rm B}>0$ is a system-size-independent constant. Using this assumption, the locality of the parent Hamiltonian, and the bounds $|g_{1}|,|S_{{\rm p},j}^{(1)}|=O(1)$, we obtain the following upper bound on $f_{\rm n}(\bm{k})$:
\begin{align}
f_{\rm n}(\bm{k})\le \frac{C_7}{l_1^2}M|\bm{k}|^2\langle\tilde{S}_{\bm{l}}|\tilde{S}_{\bm{l}}\rangle,\label{eq:bound_for_fn_k_bosons_app}
\end{align}
where $C_7$ is an $O(1)$ constant.

From Eqs.~(\ref{eq:final_form_of_fd_k_bosons}) and (\ref{eq:bound_for_fn_k_bosons_app}), we obtain
\begin{align}
\frac{f_{\rm n}(\bm{k})}{f_{\rm d}(\bm{k})}&\le \frac{M(|g_1|M+1)}{l_1(|g_1|M+l_1)}C_7|\bm{k}|^2.\label{eq:upper_bound_of_SMA_bosons_app}
\end{align}
For $0<\nu_-\le l_1/(|g_1|M)\le \nu_+<\infty$, we obtain the following bound on the parent-Hamiltonian gap:
\begin{align}
\Delta_{{\rm p},\bm{l}}\le \Delta E_{\bm{l}}^2\le C_8|\bm{k}|^2,\label{eq:bound_for_gap_parent_Hamiltonian}
\end{align}
where $C_8$ is an $O(1)$ constant. Following the same argument as in the spin-$S$ and fermionic cases, choosing $\bm{k}=\bm{k}_{\rm min}$ as defined in Eq.~(\ref{eq:definition_of_kmin}) and taking the thermodynamic limit, we prove that the parent Hamiltonian is gapless. At the same time, the energy variance of the variational state vanishes in the thermodynamic limit. The same conclusion follows for a general $s_0$ upon replacing $\hat{\rho}_1(\bm{k})$ with $\hat{\rho}_{s_0}(\bm{k})$.

\subsection{Entanglement entropy of the RQMBS and variational states}\label{app:entanglement_entropy_of_the_variational_states_boson}
In this subsection, we evaluate the EEs of the normalized RQMBS and normalized variational states $|AS_{\bm{l}}\rangle$ for bosonic systems. We first consider the EE of the RQMBS state and then turn to the variational state. We assume that $M$ is even and $l_s=O(M)$ for all $s$.  

The normalized RQMBS state can be written as
\begin{align}
|S_{\bm{l}}\rangle&=\frac{1}{\sqrt{W_{\bm{l}}}}\sum_{\bm{n}:\hat{\mathcal{N}}_s\cket{\bm{n}}=l_s\cket{\bm{n}} }\cket{\bm{n}}\otimes \cket{\bm{n}}.\label{eq:normalized_RQMBS_bosons_app}
\end{align}
We first consider the type-I bipartition. In this case, the EE is given by
\begin{align}
S_{\rm vN}^{\rm I}=\ln W_{\bm{l}}=\sum_{s=1}^f\ln\binom{|g_s|M+l_s-1}{l_s}.\label{eq:EE_of_RQMBS_typeI_boson_app}
\end{align}
From the inequality
\begin{align}
\binom{|g_s|M+l_s-1}{l_s}\le 2^{|g_s|M+l_s-1},\label{eq:inequality_binomial_coefficient_for_bosons_app}
\end{align}
we obtain
\begin{align}
S_{\rm vN}^{\rm I}\le \sum_{s=1}^f(|g_s|M+l_s-1)\ln 2=O(M).\label{eq:upper_bound_EE_typeI_bipartition_bosons_app}
\end{align}
From Eqs.~(\ref{eq:inequality_binomial_constatn_and_H2_appendix}) and (\ref{eq:definition_of_H2_appendix}), we obtain the inequality
\begin{align}
S_{\rm vN}^{\rm I}&\ge \sum_{s=1}^f\left[(|g_s|M+l_s-1)H_2\left(\frac{l_s}{|g_s|M+l_s-1}\right)\right.\notag \\
&\left.\hspace{3.0em}-\ln(|g_s|M+l_s)\right]\notag \\
&\ge \sum_{s=1}^f\left[2\ln 2\min(l_s,|g_s|M-1)-\ln(|g_s|M+l_s)\right].\label{eq:inequality_EE_typeI_RQMBS_bosons}
\end{align}
We assume that, for at least one $s$, $0<\nu_-\le l_s/(|g_s|M)\le \nu_+$, where $\nu_{\pm}$ are independent of $M$. Using the bound on $H_2(\cdot)$ employed in Appendix \ref{app:entanglement_entropy_of_the_variational_states}, the right-hand side scales linearly with $M$, up to logarithmic corrections. This means that $S_{\rm vN}^{\rm I}$ obeys a volume law. 

Next, we consider the type-II bipartition. The Schmidt decomposition of the RQMBS state is given by
\begin{align}
|S_{\bm{l}}\rangle&=\sum_{\substack{0\le m_s\le l_s \\ s=1,\ldots, f}}\sqrt{\lambda_{\bm{m}}} |\Psi^A_{\bm{m}}\rangle\otimes|\Psi^B_{\bm{l}-\bm{m}}\rangle,\label{eq:typeII_bipartition_RQMBS_boson_app}\\
\lambda_{\bm{m}}&\equiv \frac{W^A_{\bm{m}}W^B_{\bm{l}-\bm{m}}}{W_{\bm{l}}},\label{eq:Schmidt_coefficients_RQMBS_boson_app}\\
W_{\bm{m}}^A&\equiv \prod_{s=1}^f\binom{|g_s|M/2+m_s-1}{m_s},\label{eq:definition_of_WA_RQMBS_boson_app}\\
W_{\bm{l}-\bm{m}}^B&\equiv \prod_{s=1}^f\binom{|g_s|M/2+l_s-m_s-1}{l_s-m_s},\label{eq:definition_of_WB_RQMBS_boson_app}\\
|\Psi^A_{\bm{m}}\rangle&\equiv \frac{1}{\sqrt{W^A_{\bm{m}}}}\sum_{\bm{n}_A: \hat{\mathcal{N}}_s^A\cket{\bm{n}_A}=m_s\cket{\bm{n}_A}}\cket{\bm{n}_A}\otimes\cket{\bm{n}_A},\label{eq:definition_of_psi_A_RQMBS_boson_app}\\
|\Psi^B_{\bm{l}-\bm{m}}\rangle&\equiv \frac{1}{\sqrt{W^B_{\bm{l}-\bm{m}}}}\sum_{\bm{n}_B: \hat{\mathcal{N}}_s^B\cket{\bm{n}_B}=(l_s-m_s)\cket{\bm{n}_B}}\cket{\bm{n}_B}\otimes\cket{\bm{n}_B}.\label{eq:definition_of_psi_B_RQMBS_boson_app}
\end{align}
We note that the Chu--Vandermonde identity gives~\cite{Graham1994concrete}
\begin{align}
\sum_{m_s=0}^{l_s}\binom{|g_s|M/2+m_s-1}{m_s}\binom{|g_s|M/2+l_s-m_s-1}{l_s-m_s}\notag \\
=\binom{|g_s|M+l_s-1}{l_s}.\label{eq:Chu-Vandermonde_for_bosons_app}
\end{align}
From this result, we obtain $\sum_{\bm{m}}\lambda_{\bm{m}}=1$. Here, $\hat{\mathcal{N}}_s^A$ and $\hat{\mathcal{N}}_s^B$ denote the boson-number operators for the component group $g_s$ in regions $A$ and $B$, respectively. The vector $\bm{m}\equiv (m_1,m_2,\ldots, m_f)$ specifies the numbers of bosons in region $A$. We obtain the Schmidt rank
\begin{align}
S_{\rm rank}&=\prod_{s=1}^f(l_s+1).\label{eq:Schmidt_rank_RQMBS_bosons_app}
\end{align}
Therefore, the EE for the type-II bipartition obeys a subvolume law:
\begin{align}
S_{\rm vN}^{\rm II}&\le \ln S_{\rm rank}=O(\ln M).\label{eq:subvolume_law_RQMBS_typeII_boson_app}
\end{align}

We next consider the variational state $|AS_{\bm l}\rangle$. For a nonzero lattice wave vector $\bm{k}$, the normalized variational state can be written as
\begin{align}
|AS_{\bm{l}}\rangle&=\sum_{\bm{n}:\hat{\mathcal{N}}_s\cket{\bm{n}}=l_s\cket{\bm{n}}}\frac{|F_1(\bm{k},\bm{n})|e^{i\theta(\bm{k},\bm{n})}}{\sqrt{\mathcal{N}_{\bm{l}}(\bm{k})W_{\bm{l}}}}\cket{\bm{n}}\otimes\cket{\bm{n}},\label{eq:variational_state_typeI_boson_app}\\
F_1(\bm{k},\bm{n})&\equiv \sum_{j=1}^Me^{i\bm{k}\cdot\bm{R}_j}q_{j,1}\equiv |F_1(\bm{k},\bm{n})|e^{i\theta(\bm{k},\bm{n})},\label{eq:definition_of_F1_boson_app}\\
\mathcal{N}_{\bm{l}}(\bm{k})&\equiv \frac{1}{W_{\bm{l}}}\sum_{\bm{n}:\hat{\mathcal{N}}_s\cket{\bm{n}}=l_s\cket{\bm{n}}}|F_1(\bm{k},\bm{n})|^2\notag \\
&=\frac{l_1(|g_1|M+l_1)}{|g_1|M+1}.\label{eq:definition_of_N_l_k_boson_app}
\end{align}
The EE for the type-I bipartition is given by
\begin{align}
&S_{\rm vN}^{{\rm I}: AS}\notag \\
&=S_{\rm vN}^{\rm I}-\frac{1}{W_{\bm{l}}}\sum_{\bm{n}:\hat{\mathcal{N}}_s\cket{\bm{n}}=l_s\cket{\bm{n}} }\frac{|F_1(\bm{k},\bm{n})|^2}{\mathcal{N}_{\bm{l}}(\bm{k})}\ln\left[\frac{|F_1(\bm{k},\bm{n})|^2}{\mathcal{N}_{\bm{l}}(\bm{k})}\right].\label{eq:EE_typeI_ARQMBS_boson_app}
\end{align}
Following the same argument as in Appendix \ref{app:entanglement_entropy_of_the_variational_states}, we obtain the following upper bound: $S_{\rm vN}^{{\rm I}: AS}\le S_{\rm vN}^{\rm I}=\Theta(M)$. From the inequality $|F_1(\bm{k},\bm{n})|\le \sum_{j=1}^Mq_{j,1}=l_1$, we obtain
\begin{align}
S_{\rm vN}^{{\rm I}: AS}\ge S_{\rm vN}^{\rm I}-\ln\left[\frac{l_1^2}{\mathcal{N}_{\bm{l}}(\bm{k})}\right]=S_{\rm vN}^{\rm I}-\ln\left[\frac{l_1(|g_1|M+1)}{|g_1|M+l_1}\right],\label{eq:lower_bound_EE_ARQMBS_typeI_bosons_app}
\end{align}
where we used Eq.~(\ref{eq:definition_of_N_l_k_boson_app}). It follows that
\begin{align}
S_{\rm vN}^{{\rm I}: AS}\ge S_{\rm vN}^{\rm I}-O(\ln M).\label{eq:lower_bound_EE_typeI_ARQMBS_bosonss_app}
\end{align}
Thus, the variational state obeys a volume law for the type-I bipartition.

Next, we consider the EE of the variational state for the type-II bipartition. The variational state can be written as
\begin{align}
|AS_{\bm{l}}\rangle&\propto \hat{\rho}_1(\bm{k})|S_{\bm{l}}\rangle\notag \\
&=\sum_{\bm{m}}\sqrt{\lambda_{\bm{m}}}\left[\hat{\rho}_{1,A}(\bm{k})|\Psi^A_{\bm{m}}\rangle\otimes|\Psi^B_{\bm{l}-\bm{m}}\rangle\right.\notag \\
&\left.\hspace{8.0em}+|\Psi^A_{\bm{m}}\rangle\otimes\hat{\rho}_{1,B}(\bm{k})|\Psi^B_{\bm{l}-\bm{m}}\rangle\right],\label{eq:ARQMBS_typeII_boson_app}
\end{align}
where we decompose the operator $\hat{\rho}_{1}(\bm{k)}$ as
\begin{align}
\hat{\rho}_1(\bm{k})&=\hat{\rho}_{1,A}(\bm{k})+\hat{\rho}_{1,B}(\bm{k}),\label{eq:decomposition_density_operator1_bosons_app}\\
\hat{\rho}_{1,A}(\bm{k})&\equiv \sum_{j\in\Lambda_A}\sum_{\sigma\in g_1}e^{i\bm{k}\cdot\bm{R}_j}(\hat{n}_{j,\sigma}\otimes \hat{1}),\label{eq:definition_of_rho_1_A_boson_app}\\
\hat{\rho}_{1,B}(\bm{k})&\equiv \sum_{j\in\Lambda_B}\sum_{\sigma\in g_1}e^{i\bm{k}\cdot\bm{R}_j}(\hat{n}_{j,\sigma}\otimes \hat{1}).\label{eq:definition_of_rho_1_B_boson_app}
\end{align}
From Eq.~(\ref{eq:ARQMBS_typeII_boson_app}), the Schmidt rank $S_{\rm rank}^{\rm AS}$ is at most $2S_{\rm rank}=O(M^f)$. We obtain the following upper bound on the EE:
\begin{align} 
S_{\rm vN}^{{\rm II}: AS}\le \ln S_{\rm rank}^{\rm AS}=O(\ln M).\label{eq:bound_for_EE_typeII_ARQMBS_boson_app}
\end{align}
Thus, the EE for the type-II bipartition obeys a subvolume law. 

\section{Detailed calculations of SWSSB order}\label{app:details_of_Renyi2_correlator}
In this appendix, we show the detailed calculations of the R\'enyi-2 correlators for spin-$S$, multicomponent fermionic, and multicomponent bosonic systems. In addition, we show that the R\'enyi-1 correlator of the infinite-temperature state coincides with the R\'enyi-2 correlator. For soft-core bosons, we also evaluate the fidelity correlator.

\subsection{Spin-$S$ systems}\label{app:Renyi2_correlator_spin-S}
Here, we consider the R\'enyi-2 correlator (\ref{eq:definition_of_Renyi2_correlator}) for $\hat{O}_i=\hat{S}_i^+$ in spin-$S$ systems:
\begin{align}
R_{ij}^{(2)}&=\frac{{\rm Tr}(\hat{\rho}_{{\rm s},n}\hat{S}_i^+\hat{S}_j^-\hat{\rho}_{{\rm s},n}\hat{S}_j^+\hat{S}_i^-)}{{\rm Tr}\hat{\rho}_{{\rm s},n}^2},\quad i\not=j,\label{eq:Renyi2_correlator_spinS_app}\\
\hat{\rho}_{{\rm s},n}&\equiv \frac{1}{W_{M,n}}\sum_{\bm{n}:\hat{N}\cket{\bm{n}}=n\cket{\bm{n}}}\cket{\bm{n}}\bra{\bm{n}}.\label{eq:infinite_temperature_state_spin-S_app}
\end{align}
Using the relations 
\begin{align}
\hat{S}_i^+\hat{S}_i^-&=\hat{n}_i(2S-\hat{n}_i+1),\label{eq:relation_S+S-_app}\\
\hat{S}_i^-\hat{S}_i^+&=(\hat{n}_i+1)(2S-\hat{n}_i),\label{eq:relation_S-S+_app}
\end{align}
we obtain
\begin{align}
R_{ij}^{(2)}&=\frac{1}{W_{M,n}}\sum_{\bm{n}:\hat{N}\cket{\bm{n}}=n\cket{\bm{n}}}\notag \\
&\times\bra{\bm{n}}\hat{n}_i(2S-\hat{n}_i+1)(\hat{n}_j+1)(2S-\hat{n}_j)\cket{\bm{n}}.\label{eq:Renyi2_correlator_calculation_totyu1_Spin-S_app}
\end{align}
For the $S=1/2$ case, we obtain the following analytical expression [see also Ref.~\cite{Ziereis2025strong}]:
\begin{align}
R_{ij}^{(2)}&=\frac{W_{M-2,n-1}}{W_{M,n}}=\frac{n(M-n)}{M(M-1)}.\label{eq:spin-1/2_case_Renyi2-correlator_app}
\end{align}
For $\epsilon\le n/M\le 1-\epsilon$, we obtain $R^{(2)}_{ij}\ge \epsilon^2>0$ for any $i\not=j$. Therefore, for the $S=1/2$ case, the infinite-temperature state exhibits SWSSB order.

For the $S\ge 1$ case, we derive a lower bound on $R^{(2)}_{ij}$. Using the following inequalities
\begin{align}
\hat{n}_i(2S-\hat{n}_i+1)&\ge \hat{n}_i,\label{eq:inequality_ni_and_2S-ni+1_app}\\
(\hat{n}_i+1)(2S-\hat{n}_i)&\ge 2S-\hat{n}_i,\label{eq:inequality_ni+1_and_2S-ni_app}
\end{align}
we obtain
\begin{align}
R^{(2)}_{ij}&\ge \frac{1}{W_{M,n}}\sum_{\bm{n}:\hat{N}\cket{\bm{n}}=n\cket{\bm{n}}}\bra{\bm{n}}\hat{n}_i(2S-\hat{n}_j)\cket{\bm{n}}\notag \\
&=\frac{2Sn}{M}-\frac{1}{W_{M,n}}\sum_{\bm{n}:\hat{N}\cket{\bm{n}}=n\cket{\bm{n}}}\bra{\bm{n}}\hat{n}_i\hat{n}_j\cket{\bm{n}}\notag \\
&=\frac{2Sn}{M}-\frac{n^2}{M^2}+\frac{v_{M,n}}{M-1}\notag \\
&\ge \frac{n}{M}\left(2S-\frac{n}{M}\right),\label{eq:lower_bound_of_Renyi2_correlator_spinS_app}
\end{align}
where we used Eq.~(\ref{eq:relation1_v_spin-S_appendix}) and $v_{M,n}\ge 0$. For $\epsilon\le n/M\le 2S-\epsilon$, we obtain $R^{(2)}_{ij}\ge\epsilon^2>0$. Thus, for the $S\ge1$ case, the infinite-temperature state exhibits SWSSB order.

Next, we show that the R\'enyi-1 correlator of the infinite-temperature state in Eq.~(\ref{eq:infinite_temperature_state_spin-S_app}) coincides with the R\'enyi-2 correlator. The R\'enyi-1 correlator~\cite{Liu2025diagnosing,Weinstein2025efficient} is defined by
\begin{align}
R^{(1)}_{ij}\equiv {\rm Tr}(\sqrt{\hat{\rho}_{{\rm s},n}}\hat{S}_i^+\hat{S}_j^-\sqrt{\hat{\rho}_{{\rm s},n}}\hat{S}_j^+\hat{S}_i^-).\label{eq:definition_of_Renyi-1_correlator_for_spins}
\end{align}
Following Refs.~\cite{Liu2025diagnosing,Weinstein2025efficient}, the R\'enyi-1 and fidelity correlators are equivalent diagnostics of SWSSB. From Eq.~(\ref{eq:infinite_temperature_state_spin-S_app}), we obtain $\hat{\rho}_{{\rm s},n}^{1/2}=\sqrt{W_{M,n}}\hat{\rho}_{{\rm s},n}$. Substituting this relation into Eq.~(\ref{eq:definition_of_Renyi-1_correlator_for_spins}), we obtain $R_{ij}^{(1)}=R_{ij}^{(2)}$. Therefore, this system also exhibits SWSSB order in the R\'enyi-1 sense.

\subsection{Multicomponent fermionic systems}\label{app:Renyi2_correlator_fermions}
Next, we consider multicomponent fermionic systems. Here, we choose $\hat{O}_i=\hat{c}^{\dagger}_{i,\sigma}$ and $\hat{O}_j=\hat{c}^{\dagger}_{j,\sigma'}$ for $\sigma,\sigma'\in g_s$. The R\'enyi-2 correlator becomes
\begin{align}
R_{ij}^{(2)}&=\frac{{\rm Tr}(\hat{\rho}_{{\rm s},\bm{l}}\hat{c}^{\dagger}_{i,\sigma}\hat{c}_{j,\sigma'}\hat{\rho}_{{\rm s},\bm{l}}\hat{c}^{\dagger}_{j,\sigma'}\hat{c}_{i,\sigma})}{{\rm Tr}\hat{\rho}^2_{{\rm s},\bm{l}}},\quad i\not=j,\label{eq:Renyi2-correlator_fermions_app}\\
\hat{\rho}_{{\rm s},\bm{l}}&=\frac{1}{W_{\bm{l}}}\sum_{\bm{n}:\hat{\mathcal{N}}_s\cket{\bm{n}}=l_s\cket{\bm{n}}}\cket{\bm{n}}\bra{\bm{n}}.\label{eq:definition_of_infinite_temperature_state_fermions_app}
\end{align}
Direct calculation shows that
\begin{align}
R_{ij}^{(2)}&=\frac{1}{W_{\bm{l}}}\sum_{\bm{n}:\hat{\mathcal{N}}_s\cket{\bm{n}}=l_s\cket{\bm{n}}}\bra{\bm{n}}\hat{c}^{\dagger}_{i,\sigma}\hat{c}_{j,\sigma'}\hat{c}^{\dagger}_{j,\sigma'}\hat{c}_{i,\sigma}\cket{\bm{n}}\notag \\
&=\frac{1}{W_{\bm{l}}}\sum_{\bm{n}:\hat{\mathcal{N}}_s\cket{\bm{n}}=l_s\cket{\bm{n}}}\bra{\bm{n}}\hat{n}_{i,\sigma}(1-\hat{n}_{j,\sigma'})\cket{\bm{n}}\notag \\
&=\frac{\binom{|g_s|M-2}{l_s-1}}{\binom{|g_s|M}{l_s}}\notag \\
&=\frac{l_s}{|g_s|M}\frac{|g_s|M-l_s}{|g_s|M-1}.\label{eq:Renyi2-correlator_fermions_final_form_app}
\end{align}
For $\epsilon\le l_s/(|g_s|M)\le 1-\epsilon$, we obtain $R^{(2)}_{ij}\ge \epsilon^2>0$. Thus, the infinite-temperature state of the multicomponent fermionic system exhibits SWSSB order with respect to the $U(1)$ symmetry generated by $\hat{\mathcal{N}}_s$.

Next, we calculate the R\'enyi-1 correlator. For fermionic systems, the R\'enyi-1 correlator is defined by
\begin{align} 
R^{(1)}_{ij}&={\rm Tr}(\sqrt{\hat{\rho}_{{\rm s},\bm{l}}}\hat{c}^{\dagger}_{i,\sigma}\hat{c}_{j,\sigma'}\sqrt{\hat{\rho}_{{\rm s},\bm{l}}}\hat{c}^{\dagger}_{j,\sigma'}\hat{c}_{i,\sigma} ).\label{eq:Renyi-1_correlator_for_fermions}
\end{align}
As in the case of spin-$S$ systems, the infinite-temperature state satisfies the relation $\hat{\rho}_{{\rm s},\bm{l}}^{1/2}=\sqrt{W_{\bm{l}}}\hat{\rho}_{{\rm s},\bm{l}}$. From this relation, we obtain $R_{ij}^{(1)}=R_{ij}^{(2)}$. Therefore, the fermionic systems also exhibit SWSSB order in the R\'enyi-1 sense.

\subsection{Multicomponent bosonic systems}\label{app:Renyi2_correlator_bosons}
Finally, we consider multicomponent bosonic systems. Here, we choose $\hat{O}_i=\hat{b}^{\dagger}_{i,\sigma}$ and $\hat{O}_j=\hat{b}^{\dagger}_{j,\sigma'}$ for $\sigma,\sigma'\in g_s$. The R\'enyi-2 correlator becomes
\begin{align}
R_{ij}^{(2)}&=\frac{{\rm Tr}(\hat{\rho}_{{\rm s},\bm{l}}\hat{b}^{\dagger}_{i,\sigma}\hat{b}_{j,\sigma'}\hat{\rho}_{{\rm s},\bm{l}}\hat{b}^{\dagger}_{j,\sigma'}\hat{b}_{i,\sigma})}{{\rm Tr}\hat{\rho}^2_{{\rm s},\bm{l}}},\quad i\not=j,\label{eq:Renyi2-correlator_bosons_app}\\
\hat{\rho}_{{\rm s},\bm{l}}&=\frac{1}{W_{\bm{l}}}\sum_{\bm{n}:\hat{\mathcal{N}}_s\cket{\bm{n}}=l_s\cket{\bm{n}}}\cket{\bm{n}}\bra{\bm{n}}.\label{eq:definition_of_infinite_temperature_state_bosons_app}
\end{align}
Direct calculation shows that
\begin{align}
R_{ij}^{(2)}&=\frac{1}{W_{\bm{l}}}\sum_{\bm{n}:\hat{\mathcal{N}}_s\cket{\bm{n}}=l_s\cket{\bm{n}}}\bra{\bm{n}}\hat{b}^{\dagger}_{i,\sigma}\hat{b}_{j,\sigma'}\hat{b}^{\dagger}_{j,\sigma'}\hat{b}_{i,\sigma}\cket{\bm{n}}\notag \\
&=\frac{1}{W_{\bm{l}}}\sum_{\bm{n}:\hat{\mathcal{N}}_s\cket{\bm{n}}=l_s\cket{\bm{n}}}\bra{\bm{n}}\hat{n}_{i,\sigma}(1+\hat{n}_{j,\sigma'})\cket{\bm{n}}\notag \\
&=\frac{l_s}{|g_s|M}+\frac{\binom{|g_s|M+l_s-1}{l_s-2}}{\binom{|g_s|M+l_s-1}{l_s}}\notag \\
&=\frac{l_s}{|g_s|M}\frac{|g_s|M+l_s}{|g_s|M+1},\label{eq:Renyi2-correlator_bosons_final_form_app}
\end{align}
where we used Eq.~(\ref{eq:calculation_q11q21_bosons_app}). For $0<\nu_-\le l_s/(|g_s|M)\le \nu_+<\infty$, we obtain $R^{(2)}_{ij}\ge \nu_-(1+\nu_-)/[1+1/(|g_s|M)]>0$. Thus, the infinite-temperature state of the multicomponent bosonic system exhibits SWSSB order with respect to the $U(1)$ symmetry generated by $\hat{\mathcal{N}}_s$.

Next, we consider the R\'enyi-1 correlator. The R\'enyi-1 correlator is defined by
\begin{align}
R^{(1)}_{ij}={\rm Tr}(\sqrt{\hat{\rho}_{{\rm s},\bm{l}}}\hat{b}^{\dagger}_{i,\sigma}\hat{b}_{j,\sigma'}\sqrt{\hat{\rho}_{{\rm s},\bm{l}}}\hat{b}^{\dagger}_{j,\sigma'}\hat{b}_{i,\sigma} ).\label{eq:Renyi-1_correlator_for_bosons}
\end{align}
As in the spin-$S$ and fermionic cases, the R\'enyi-1 correlator coincides with the R\'enyi-2 correlator: $R^{(1)}_{ij}=R_{ij}^{(2)}$. Thus, the bosonic systems exhibit R\'enyi-1 long-range order as well. However, because the bosonic creation and annihilation operators are not uniformly bounded in the thermodynamic limit, the equivalence between the R\'enyi-1 and fidelity correlators~\cite{Liu2025diagnosing,Weinstein2025efficient} established for bounded operators cannot be applied directly. We therefore evaluate the fidelity correlator explicitly.

Here, we evaluate the fidelity correlator, which is defined by
\begin{align}
F_{ij}&\equiv {\rm Tr}\left(\sqrt{\sqrt{\hat{\rho}_{{\rm s},\bm{l}}} \hat{b}_{i,\sigma}^{\dagger}\hat{b}_{j,\sigma'}\hat{\rho}_{{\rm s},\bm{l}}\hat{b}_{j,\sigma'}^{\dagger}\hat{b}_{i,\sigma} \sqrt{\hat{\rho}_{{\rm s},\bm{l}}}}\right),\quad i\not=j.\label{eq:definition_of_Fidelity_correlator}
\end{align}
Direct calculation shows
\begin{align}
F_{ij}&=\frac{1}{W_{\bm{l}}}\sum_{\bm{n}: \hat{\mathcal{N}}_s\cket{\bm{n}}=l_s\cket{\bm{n}}}\sqrt{n_{i,\sigma}(n_{j,\sigma'}+1)}.\label{eq:fidelity_correlator_calculation_totyu1}
\end{align}
Here, we derive a lower bound on the fidelity correlator. Using the inequality
\begin{align}
\sqrt{n_{i,\sigma}(n_{j,\sigma'}+1)}\ge 1,\quad \text{for }n_{i,\sigma}\ge 1,\label{eq:inequality_for_square_root_of_occupation_num}
\end{align}
we obtain
\begin{align}
F_{ij}&\ge \frac{W(n_{i,\sigma}\ge 1)}{W_{\bm{l}}}\notag \\
&=1-\frac{\binom{|g_s|M+l_s-2}{l_s}}{\binom{|g_s|M+l_s-1}{l_s}}\notag \\
&=\frac{l_s/(|g_s|M)}{1+l_s/(|g_s|M)-1/(|g_s| M)}\notag \\
&\ge \frac{l_s/(|g_s|M)}{1+l_s/(|g_s|M)},\label{eq:lower_bound_of_fidelity_correlator_for_bosons}
\end{align}
where $W(n_{i,\sigma}\ge1)$ represents the number of configurations for $n_{i,\sigma}\ge 1$. Under the same conditions as for the R\'enyi-2 correlator, the fidelity correlator remains nonzero in the thermodynamic limit. Therefore, the bosonic systems exhibit SWSSB order.

\bibliography{references_v2}
\end{document}